\documentclass[12pt]{article}
\usepackage{amsthm,amsmath,latexsym,amssymb,amsfonts,amscd}
\usepackage{graphics,lscape,fancyhdr,array,stmaryrd,euscript,wrapfig}
\usepackage{empheq}
\usepackage{verbatim,slashed}
\numberwithin{equation}{section}
\usepackage{hyperref,setspace}
\usepackage[numbers,sort&compress]{natbib}
\usepackage[nottoc]{tocbibind}
\usepackage{tikz-cd}
\usepackage[all]{xy}
\usepackage{tikz}
\usetikzlibrary{fit}
\usetikzlibrary{patterns}
\usetikzlibrary{arrows.meta}
\usetikzlibrary{decorations.pathmorphing}
\usepackage{multirow}

\usepackage{rotating}% provides a rotate environment

\newcommand{\sign}{\operatorname{sign}}

\numberwithin{equation}{section}
\newcommand{\pl}{\partial}

\newcommand{\Tr}{\mathrm{Tr}}

\newcommand{\fud}[2]{{}^{#1}{}_{#2}\,}
\newcommand{\fdu}[2]{{}_{#1}{}^{#2}\,}
\newcommand{\fudu}[3]{{}^{#1}{}_{#2}{}^{#3}\,}

\newcommand{\fdud}[3]{{}_{#1}{}^{#2}{}_{#3}\,}

\newcommand{\brk}{{{\bar{k}}}}

\newcommand{\besubeqs}{\begin{subequations}}
\newcommand{\esubeqs}{\end{subequations}}

\begin{document}
%%%%%%%%%%%%%%%%%%%%%%%%%%%%%%%%%%%%%%%%%%%%%%%%%%%%%%%%%%%%%
%\pagenumbering{gobble}
\hfill
\vskip 0.01\textheight
\begin{center}
{\large\bfseries 
Dirichlet, Neumann, Mixed and self-dual holography:\\
\vspace{0.2cm}
(self-dual) Yang--Mills theory}

\vspace{0.4cm}

\vskip 0.03\textheight
\renewcommand{\thefootnote}{\fnsymbol{footnote}}
Evgeny \textsc{Skvortsov}\footnote{Also affiliated with Lebedev Institute of Physics.}  \& Richard van Dongen

\vskip 0.03\textheight

{\em Service de Physique de l'Univers, Champs et Gravitation, \\ Universit\'e de Mons, 20 place du Parc, 7000 Mons, 
Belgium}\\
\vspace*{5pt}

\renewcommand{\thefootnote}{\arabic{footnote}}
\end{center}

\vskip 0.02\textheight

\begin{abstract}
Motivated by applications of self-dual theories to the AdS/CFT correspondence, we study self-dual Yang--Mills theory (SDYM) and its relation to Yang--Mills theory and to Chalmers--Siegel theory with Dirichlet, Neumann, and mixed boundary conditions. A Fefferman–Graham analysis of SDYM is performed to identify its boundary CFT data. We make a proposal for self-dual holography that defines $3d$ ``self-dual CFTs''. The bulk-to-bulk and boundary-to-bulk propagators for SDYM and for Yang--Mills/Chalmers--Siegel theory with mixed boundary conditions are derived in Feynman and axial gauges. Three- and four-point functions are computed in the spinor-helicity formalism, and the relations among the results in the various theories are clarified. The flat limit and the gauge-(in)dependence of the results are analyzed.
\end{abstract}

\newpage
\tableofcontents
\newpage

%%%%%%%%%%%%%%%%%%%%%%%%%%%%%%%%%%%%%%%%%%%%%%%%%%%%%%%%%%%%%
\section{Introduction}
\label{sec:intro}
%%%%%%%%%%%%%%%%%%%%%%%%%%%%%%%%%%%%%%%%%%%%%%%%%%%%%%%%%%%%%
AdS/CFT \cite{Maldacena:1997re,Gubser:1998bc,Witten:1998qj} is a rich tool to investigate models of quantum gravity via CFT lenses. For many reasons it would be advantageous to have a pair of dual theories that are both simple enough, potentially soluble, and nontrivial on both sides. While numerous dualities have been proposed, a relatively new idea is to study self-dual theories in this context, with the simplest examples being self-dual Yang--Mills theory (SDYM), self-dual gravity (SDGR), and supersymmetric and higher-spin extensions thereof. Indeed, self-duality has not yet been mentioned much within the AdS/CFT correspondence until recently \cite{Skvortsov:2018uru,Sharapov:2022awp,Jain:2024bza,Aharony:2024nqs,Chowdhury:2024dcy,Sharma:2025ntb}, but see \cite{Petkou:2004nu,deHaro:2007fg,deHaro:2008gp,Compere:2008us}.

Apart from admitting a formulation where the self-duality is manifest,\footnote{Self-dual theories are very handy in the light-cone gauge where their self-duality is less obvious. The light-cone gauge form of self-dual theories is significantly more complicated in (A)dS, \cite{Metsaev:2018xip,Lipstein:2023pih,Neiman:2023bkq,Neiman:2024vit,CarrilloGonzalez:2024sto,Chowdhury:2024dcy,Kozaki:2025jrj}. } self-dual theories may be characterized by several ``usual'' properties: (a) tree-level amplitudes vanish in flat space (but not in AdS); (b) loop amplitudes are very simple and, in particular, UV-finite; (c) solutions of self-dual theories are also solutions of the parent theories; (d) self-dual theories have simple formulations on twistor space; (e) self-dual theories are integrable in some sense. 

For (A)dS/CFT applications, the most important properties are that self-dual theories admit a negative or positive cosmological constant and are UV-finite. It is the UV-divergences of the pure gravity that require a complete model of quantum gravity. This either complicates the bulk dynamics, e.g. calling for a stringy completion, or encourages one to look for simpler models in lower dimensions where gravity does not have propagating degrees of freedom. Self-dual theories, e.g. self-dual gravity can provide us with a useful compromise where a theory is simple enough, yet nontrivial and free of UV-divergences. 

So far, self-dual theories have not yet been embedded fully into the web of AdS/CFT dualities. AdS/CFT duals of self-dual theories in the bulk can be called ``self-dual CFTs''. Concrete proposals were discussed in \cite{Sharapov:2022awp,Jain:2024bza,Aharony:2024nqs} with some speculations earlier in \cite{Ponomarev:2016lrm,Skvortsov:2018uru}. One can also mention a few (A)dS/CFT-type calculations in self-dual theories in the light-cone gauge in \cite{Skvortsov:2018uru,Albrychiewicz:2021ndv,Chowdhury:2024dcy}. However, there are very basic questions that have not yet been answered: (a) what is the Fefferman-Graham expansion of self-dual theories? (b) what are the CFT dual operators? (c) what are possible boundary conditions in self-dual theories? (d) what do Witten diagrams compute in self-dual theories? (e) what part of the ``parent'' theory's AdS/CFT dynamics is captured by the self-dual subsector? 

In the paper we are going to address these questions in the simplest example of self-dual theory --- self-dual Yang--Mills or SDYM. Accordingly, in Section \ref{sec:2.5} we introduce several useful, closely related, theories: Yang--Mills theory, its chiral reformulation, Chalmers--Siegel action and its SDYM limit. In Section \ref{sec:ads/cft} we address the most basic questions about the Fefferman-Graham expansion in Section \ref{sec:FG}, about boundary conditions in Section \ref{sec:boundarycond} and, finally, we propose a natural extension of the AdS/CFT lore to self-dual theories in Section \ref{sec:adscftdict}. Here, an important role is played by the standard AdS/CFT dualities that involve Dirichlet/Neumann/mixed boundary conditions developed in \cite{Klebanov:1999tb,Witten:2001ua,Witten:2003ya,Leigh:2003ez,Yee:2004ju,deHaro:2007eg,Hartman:2006dy, Giombi:2011ya,Leigh:2003gk,Petkou:2004nu,Compere:2008us,deHaro:2008gp,Giombi:2013yva}. In Section \ref{sec:correlators} we proceed to computing two-, three- and four-point correlation functions in YM, Chalmers--Siegel theory and SDYM. In particular, whenever possible, we give results for Dirichlet/Neumann/mixed boundary conditions. As different from flat space, where YM and SDYM directly share a subset of amplitudes, in the AdS/CFT setup one has to carefully deal with boundary conditions and SDYM should be approached together with tuning mixed boundary conditions to the self-dual point. We end up with a discussion of the results in Section \ref{sec:discussion}.

%%%%%%%%%%%%%%%%%%%%%%%%%%%%%%%%%%%%%%%%%%%%%%%%%%%%%%%%%%%%%
\section{Two and half theories}
\label{sec:2.5}
%%%%%%%%%%%%%%%%%%%%%%%%%%%%%%%%%%%%%%%%%%%%%%%%%%%%%%%%%%%%%
Since all theories we consider are conformally invariant, it is advantageous to transfer them from the usual Poincare coordinates on $\text{AdS}_4$ to the Euclidean space $\mathbb{R}^4$ with $x^i\in \mathbb{R}^3$ and $z>0$, with the boundary still being at $z=0$. Therefore, except for Section \ref{sec:FG}, we will always work in flat Euclidean space.

For AdS/CFT applications that we consider, Yang--Mills theory should be supplemented with a theta-term. To facilitate the transition to the Chalmers–Siegel formulation and to SDYM, it is convenient to express the action in the two-component spinor language:\footnote{Indices $A,B,...=1,2$, $A',B',...=1,2$ are the indices of the two two-dimensional representations of the Lorentz algebra, which is $sl(2,\mathbb{C})$ or $su(2)\oplus su(2)$, depending on the signature. The indices are raised by $\epsilon_{AB}=-\epsilon_{BA}$ and $\epsilon_{A'B'}$, e.g. $\xi^A=\epsilon^{AB}\xi_B$, $\xi^A\epsilon_{AB}=\xi_B$. For example, for the field strength we have $F_{\mu\nu} \equiv F_{AA',BB'} = F_{AB}\epsilon_{A'B'} + \bar{F}_{A'B'}\epsilon_{AB}$ and $F_{AB}=\tfrac12(\pl_{AB'} A\fdu{B,}{B'}+\pl_{BB'} A\fdu{A,}{B'}+A_{A,B'}A\fdu{B,}{B'}+A_{B,B'}A\fdu{A,}{B'})$. }
\begin{align}\label{YMthetaAction}
    S_{\text{YM},\theta}&=\frac{a}4\,\Tr\int \left(F_{AB}F^{AB} +\bar{F}_{A'B'} \bar{F}^{A'B'}\right)+\frac{b}4 \,\Tr\int \left(F_{AB}F^{AB} -\bar{F}_{A'B'} \bar{F}^{A'B'}\right)\,.
\end{align}
Here, $a$ and $b$ are the factors in front of the Yang--Mills Lagrangian $\Tr\, F_{\mu\nu} F^{\mu\nu}$ and the theta-term $\Tr\, F\wedge F$, respectively. We find it very useful to rearrange the terms as
\begin{align}\label{chiralYM}
    S_{\text{cYM},\theta}&= \frac{a}2\,\Tr\int F_{AB}F^{AB} +\frac{(b-a)}4 \,\Tr\int \left(F_{AB}F^{AB} -\bar{F}_{A'B'} \bar{F}^{A'B'}\right)\,.
\end{align}
The bulk interactions are now confined to $(F_{AB})^2$, which is much simpler to deal with. The rest is in the topological term that disappears in flat space and leads to simple contributions in $\text{AdS}_4$. We call this intermediate step chiral Yang--Mills theory, cYM. Following Chalmers and Siegel \cite{Chalmers:1996rq}, by introducing an auxiliary field $\Psi_{AB}$, one can rewrite cYM as 
\begin{align}\label{CSthetaActionB}
    S_{\text{Ch.Si.},\theta}&= a\epsilon\,\Tr\int \left(\Psi_{AB}F^{AB} - \frac{\epsilon}2 \Psi_{AB} \Psi^{AB}\right) + \tfrac{(b-a)}4 \,\Tr\int \left(F_{AB}F^{AB} -\bar{F}_{A'B'} \bar{F}^{A'B'}\right)\,.
\end{align}
SDYM emerges when the $\Psi^2$-term, which is gauge invariant on its own, is dropped. Formally, one has to take the limit $\epsilon\rightarrow 0$ and ensure that $a \epsilon$ stays finite, say $\tilde{a}=1$, and $b-a$ goes to $\varkappa$:
\begin{align}\label{SDYMthetaAction}
    S_{\text{SDYM},\theta}&= \tilde{a}\Tr\int \Psi_{AB}F^{AB} + \frac{\varkappa}4 \,\Tr\int \left(F_{AB}F^{AB} -\bar{F}_{A'B'} \bar{F}^{A'B'}\right)\,.
\end{align}
We would like to stress that $\epsilon$ is not a genuine coupling constant as long as $\epsilon\neq0$. In addition, SDYM is not a simple smooth limit of YM. It might be better to think of SDYM as of a closed subsector of YM: all solutions and amplitudes of SDYM are solutions and (a subset of) amplitudes of YM. Also, the connection between YM and SDYM requires Chalmers--Siegel theory. If one just imposes $F_{AB}=0$ in the original YM action, or better in its cYM form \eqref{chiralYM}, the bulk interactions vanish.\footnote{In Minkowski signature, $F_{AB}$ and $\bar{F}_{A'B'}$ isolate negative and positive helicity components of $A_\mu$. If we set $F_{AB}=0$, we are left with a single helicity and there cannot be any Lorentz-invariant action for this one. }

%%%%%%%%%%%%%%%%%%%%%%%%%%%%%%%%%%%%%%%%%%%%%%%%%%%%%%%%%%%%%
\section{AdS/CFT (self)-duality}
\label{sec:ads/cft}
%%%%%%%%%%%%%%%%%%%%%%%%%%%%%%%%%%%%%%%%%%%%%%%%%%%%%%%%%%%%%
The key steps to embed SDYM into AdS/CFT duality are: (a) to work out the Fefferman-Graham expansion to reveal the boundary data; (b) to investigate possible boundary conditions and the relation of those to YM; (c) as a result, there is a natural extrapolation of the AdS/CFT dictionary from YM to SDYM. We follow these steps below.

%%%%%%%%%%%%%%%%%%%%%%%%%%%%%%%%%%%%%%%%%%%%%%%%%%%%%%%%%%%%%
\subsection{Fefferman-Graham expansion}
\label{sec:FG}
%%%%%%%%%%%%%%%%%%%%%%%%%%%%%%%%%%%%%%%%%%%%%%%%%%%%%%%%%%%%%
\paragraph{YM.} The Fefferman-Graham expansion of free YM, i.e. Maxwell theory, is well-known \cite{Mueck:1998iz,Bianchi:2001kw}. Let us summarize the main results. The equations of motion\footnote{We are temporarily back to Poincare coordinates to have the standard $z$-dependence. We will use $\Phi_{A,A'}$ for a gauge field in the flat space model, $A_{A,A'}=z \Phi_{A,A'}$, and not to have too many $A$s like in $A_{A,A'}$.}
\begin{align} \label{Maxwell}
    \nabla^{BB'}\nabla_{BB'}A_{A,A'}-\nabla^{BB'}\nabla_{AA'}A_{B,B'}=0 \,,
\end{align}
can be solved asymptotically near the boundary $z=0$ by plugging in the expansion\footnote{One notational improvement is to denote a group of symmetric indices or the indices to be symmetrized by the same letter, e.g. $\Phi_{AA}\equiv \Phi_{A_1A_2}\equiv \Phi_{A_2A_1}$. Also, in this paper we work in a hybrid position-momentum space: the radial coordinate $z$ plus a $3d$ momentum $k_i\equiv k_{AA}$ along the boundary, e.g. $\Phi_{A,A'}=\Phi_{A,A'}(k,z)$.}
\begin{align}
    A_{AA}=z^\Delta\Phi_{AA}=z^\Delta (\phi_{AA}^{(0)}+z\phi^{(1)}_{AA}+z^2\phi^{(2)}_{AA}+\dots) \,.
\end{align}
Here, we imposed the axial gauge $\Phi_z=0$ and $z$ is along $\epsilon_{AA'}$. Therefore, it means that $\Phi_{A,B}$ is symmetric, hence, the comma can be dropped. To make a shortcut, we in addition impose the Coulomb condition $k_i \Phi^i=0$ on the boundary. Since the equation is second order, there are two independent fall-offs: $\Delta=1$ and $\Delta=2$. The complete solution reads
\begin{align}
    \Phi^{AA}=\cosh(kz) \Phi_0^{AA} + \tfrac{1}{k}\sinh(kz) \Phi_1^{AA}= \Phi_0^{AA}+ z \Phi_1^{AA}+...\,,
\end{align}
where $\Phi_0^i\equiv \Phi^i|_{z=0}$ and $\Phi_1^i\equiv \pl_z\Phi^i|_{z=0}$ are Dirichlet and Neumann data, respectively. In a more standard notation we have $A_i=z(a_i + z E_i)+...$, where $E_i=\pl_z \Phi_i|_{z=0}$ is the electric field and $a_i$ is a $3d$ gauge potential. In the AdS/CFT context one is interested in regular solutions, which relates the two boundary data to each other:
\begin{align}
    \Phi^{AA}=\exp[-kz] \Phi_0^{AA}=\Phi_0^{AA}-z k \Phi_0^{AA}+...
\end{align}
In other words, the regularity implies $E_i=-k \Pi _{ij}a^j$, where $\Pi_{ij}=\delta_{ij}-k_ik_j/k^2$ (if we drop the transversality condition imposed on $\Phi_i|_{z=0}\equiv a_i$). In the physical (transverse and $\epsilon_{AA'}$-traceless) gauge, the Fefferman-Graham expansion of Maxwell theory is that of a conformally-coupled scalar field $\phi=0$ that carries an additional index $i$.

\paragraph{SDYM: negative helicity.} The $\Psi$-equation of motion is, in fact, four equations on a three-component $\Psi_{AB}$ (the right conformal factor $z^2$ is extracted):
\begin{align} \label{eom}
    \nabla\fdu{B}{A'}(z^2\Psi^{AB})=0\,.
\end{align}
They imply that the boundary data $\psi_0^{AB}=\Psi^{AB}|_{z=0}$ is transverse $k_{AA} \psi_0^{AA}=0$ and the complete expansion reads
\begin{align}
    \Psi^{AA}&= \cosh(kz) \psi_0^{AA} +\tfrac{1}{k}\sinh(kz) k\fud{A}{B} \psi_0^{AB}\,.
\end{align}
Since the equation is first order, the boundary data is only Dirichlet. In order to find regular solutions, we need to split $\psi_0^{AB}$ further and the only way to do it is via a helicity decomposition. Since the helicity is a momentum space notion, the decomposition would appear nonlocal in position space. 

Following \cite{Maldacena:2011nz}, let us introduce a ``fake'' $4d$ light-like momentum $p_{AA'}=k_{AA'}+k\epsilon_{AA'}$.\footnote{We define $p_\mu p^\mu=-\tfrac12 p_{AA'}p^{AA'}$, $k_i k^i= -\tfrac12 k_{AB}k^{AB}=k^2$. $\slashed k$ is defined as $\slashed k \xi= k\fud{A}{B}\xi^B$, $\slashed k \psi= k\fud{A}{B}\psi^{AB}$. In particular, $\slashed k k \equiv k\fud{A}{B} k^B=-k k^A $, $\slashed k \brk \equiv k\fud{A}{B} \brk^B=+k \brk^A $.  } Being null, $p_{AA'}$ can be factorized as $p_{AA'}=k_A\bar{k}_{A'}$. One can define two $3d$ vectors $k_Ak_B$ and $\bar{k}_{A}\bar{k}_{B}$ that are both $3d$-null and orthogonal to $k^{AB}$. This idea is behind the spinor-helicity formalism for $3d$ CFTs and $\text{(A)dS}_4$, see e.g. \cite{Maldacena:2011nz,Jain:2021vrv}.\footnote{There are more spinor-helicity formalisms for $\text{AdS}_4/\text{CFT}_3$, see e.g. \cite{Maldacena:2011nz,Nagaraj:2018nxq,Nagaraj:2019zmk,Nagaraj:2020sji,Skvortsov:2022wzo,Baumann:2024ttn}. } Now, the decomposition of the solution space into (ir)regular solutions can be found to be
\begin{align} \label{PsiSol}
    \Psi^{AA} &= \psi_{0}^-e^{-kz}k^{A}k^A\,, &
    \Psi^{AA} &= \psi_{0}^+e^{+kz}\bar{k}^{A}\bar{k}^A\,.
\end{align}
Here, $\psi_{0}^{\pm}$ is the decomposition of the boundary data $\psi_{0}^{AA}$ into the definite helicity components.  

\paragraph{SDYM: positive helicity.} In this case there are fewer equations than fields, but there is also a gauge symmetry, (recall that $A_{A,A'}=z\Phi_{A,A'}$)
\begin{align}
    \nabla\fdu{A}{B'}A_{A,B'} =0\,, && \delta A_{AA'}=\nabla_{AA'}\xi\,.
\end{align}
With inspiration from the $\Psi$-case, (ir)regular solutions have the form
\begin{align} \label{PhiProp}
        \Phi_{A,A'}&=\phi^{0}_-e^{+kz} q_Ak_{A'}\,,\qquad &
        \Phi_{A,A'}&=\phi^{0}_+e^{-kz} q_A\bar{k}_{A'}\,,
\end{align}
where $q_A$ is an arbitrary reference spinor. In the axial gauge $\Phi_{A,A'}=\Phi_{AA'}$ is symmetric, i.e. $q_A$ is either $k_A$ or $\brk_A$, and similar solutions read
\begin{align}
        \Phi_{AA}&=\phi^{0}_-e^{+kz} k_Ak_{A}\,, &
        \Phi_{AA}&=\phi^{0}_+e^{-kz} \bar{k}_A\bar{k}_{A}\,.
\end{align}
In the physical gauge, $k_{AA}\phi_0^{AA}=0$, the Fefferman-Graham expansion can be summarized as
\begin{align}
    \Phi^{AA}&= \cosh(kz) \phi_0^{AA} -\tfrac{1}{k}\sinh(kz) \slashed k \phi_0^{AA}\,.
\end{align}

\paragraph{Helicity decomposition.} To make the above more self-evident, let us adopt the physical gauge and decompose $\Psi$ and $\Phi$ as
\begin{align}
\begin{aligned}
           \Phi^{AA}(\vec k,z)&= \epsilon_+^{AA}(\vec k) \Phi_+(\vec k,z)+\epsilon_-^{AA}(\vec k) \Phi_-(\vec k,z)\,, \\
        \Psi^{AA}(\vec k,z)&=\epsilon_+^{AA}(\vec k) \Psi_+(\vec k,z)+\epsilon_-^{AA}(\vec k) \Psi_-(\vec k,z) \,,
\end{aligned}
\end{align}
where the polarization vectors are defined as
\begin{align}\label{polarizationspinors}
    \epsilon_+^A(\vec k)&=  \frac{\brk^{A}}{\sqrt{2k}}\,, & \epsilon_-^A(\vec k)&=  \frac{k^{A}}{\sqrt{2k}}\,, & \epsilon_\pm^{AA}(\vec k)&=\epsilon_\pm^A\epsilon_\pm^A \,.
\end{align}
With $D_\pm=\pl_z \pm k$ the equations of motion are simply (note that $D_+D_-=\pl^2_z-k^2$)
\begin{align}\label{eqhel}
    D_+ \Phi_+&=0\,, & D_- \Psi_+&=0\,, & D_- \Phi_-&=0\,, & D_+ \Psi_-&=0 \,.
\end{align}
The Fefferman-Graham expansions for these two types of equations are 
\begin{align}
    D_+ f&=0 &&\rightarrow && f=\exp(-kz) f_0\,, &
    D_- g&=0 && \rightarrow && g=\exp(+kz)g_0\,.
\end{align}
The helicity clearly separates solutions into (ir)regular ones. Imposing regularity in SDYM leaves us with $\Phi_+$ and $\Psi_-$ as the fields that can have free boundary data. In the Chalmers--Siegel theory $\Psi_{\pm}$ obeys the same equations as in SDYM, but one also has $\epsilon\Psi_\pm=D_\pm \Phi_\pm$.

%%%%%%%%%%%%%%%%%%%%%%%%%%%%%%%%%%%%%%%%%%%%%%%%%%%%%%%%%%%%%
\subsection{Boundary conditions}
\label{sec:boundarycond}
%%%%%%%%%%%%%%%%%%%%%%%%%%%%%%%%%%%%%%%%%%%%%%%%%%%%%%%%%%%%%
The Fefferman-Graham decomposition allows us to identify the boundary data. In YM we have $\Phi^0_\pm \leftrightarrow \Phi_i^0\equiv \Phi_i|_{z=0}$, $\Phi^1_\pm \leftrightarrow \Phi_i^1\equiv \pl_z\Phi_i|_{z=0}$, which, in the usual terms, are a gauge field $a_i$ and the electric field $E_i$, $\pl_i E^i=0$, respectively. More invariantly, $a_i$ modulo gauge symmetry is characterized by its magnetic field $B_i\sim \epsilon_{ijk} F^{jk}$. Therefore, the standard boundary conditions are: Dirichlet, where $a_i$ or $B_i$ are fixed on the boundary; Neumann, where $E_i$ is fixed on the boundary. They are also called magnetic and electric. 

More generally, boundary conditions select a Lagrangian submanifold of the Fefferman-Graham boundary data that is consistent with having regular solutions in the bulk. The symplectic structure of Maxwell/Yang--Mills theory at the free level,
\begin{align}
    \Omega= \delta^{ij}\,\delta A^0_i \wedge \delta A^1_j= \delta \Phi_+^0\wedge \delta \Phi_+^1+\delta \Phi_-^0\wedge \delta \Phi_-^1\,,
\end{align}
tolerates so-called mixed (conformally-invariant) boundary conditions \cite{Witten:2003ya,Marolf:2006nd}
\begin{align}\label{mixedbc}
    F^{AA} e^{i\gamma}+ \bar F^{AA} e^{-i\gamma}&=\text{fixed} && \Longleftrightarrow && B_i \cos \gamma +i E_i\sin \gamma=\text{fixed} \,.
\end{align}
Here, $\gamma=0$ and $\gamma=\pi/2$ correspond to Dirichlet and Neumann conditions, respectively. The Yang--Mills action is already stationary with Dirichlet boundary conditions $\delta S_{\text{YM}}=\int_{\pl M} E_i \delta A^i$. The mixed boundary conditions require a boundary Chern--Simons term since $\delta S_{\text{CS}}=i\kappa \int _{\pl M} B_i \delta A^i$, where $\kappa$ is the non-normalized level. With the sources turned off the action $S_{\text{YM}}+S_{\text{CS}}$ is stationary provided that
\begin{align}\label{mixedbcEB}
    i\kappa B_i + E_i&=0 \,, && \kappa = -\cot(\gamma) \,.
\end{align}
In terms of the helicity components the mixed boundary conditions with sources read
\begin{align}\label{mixedbcphi}
    k \Phi^0_+ \cos \gamma +\Phi^1_+ i\sin \gamma&= \text{fixed}_+\,, &
    -k \Phi^0_- \cos \gamma +\Phi^1_- i\sin \gamma&= \text{fixed}_-\,.
\end{align}
It is obvious that SDYM should be approached as $\epsilon\rightarrow0$ in the Chalmers--Siegel action, but on a manifold with boundaries it has to be done simultaneously with imposing the self-dual (SD) boundary condition (otherwise, the system is over-constrained). The SD boundary condition is $F_{AA}|_{z=0}=0$. But it is more than a boundary condition since $F_{AA}|_{z=0}=0$ implies $F_{AA}=0$ in the bulk as well.\footnote{Indeed, in terms of the boundary data a general solution for $\Phi_{\pm}$ gives for $F_{AA}$:  $F_+=e^{+kz}(\Phi^1_++k\Phi^0_+)$, $F_-=e^{-kz}(\Phi^1_--k\Phi^0_-)$. We see that $F_\pm=0$ everywhere if $F_{\pm}|_{z=0}=0$. Even simpler: the first order operator $\pl_z\pm k$ gives a vanishing solution if we start from $0$ on the boundary.}  The SD condition can be reached at $\gamma=-i\infty$ in \eqref{mixedbc}, or $\kappa=-i$ in \eqref{mixedbcEB}. Another feature of the SD limit is that the initial source gets suppressed.\footnote{In a more standard situation of the mixed boundary conditions $a \phi+ b\pl_z \phi|_{z=0}=Q$ for a scalar field $\phi$, one can reach Dirichlet either as $b=0$ or as $a\rightarrow\infty$. The latter forces one to amplify the source $Q\rightarrow aQ$, otherwise we lose degrees of freedom. This is impossible in the SD case since $F_{AA}=0$ does not allow for a source.} 

One should not impose the SD boundary condition directly in YM. Indeed, the regular branch of $\Phi_-$ gets eliminated. Therefore, we are left with a single degree of freedom $\Phi_+$ and there cannot be any Lorentz-invariant action for this one, i.e. all bulk interactions are gone as well. In the Chalmers--Siegel theory the limit $\epsilon=0$ preserves the number of degrees of freedom: one still has two regular solutions, $\Phi_+$ and $\Psi_-$. Therefore, it is $\Phi_+$ and $\Psi_-$ that should have sources at the boundary:
\begin{align}
    \Phi_+^0&=\text{fixed} \,, & \Psi_-^0=\text{fixed}\,,
\end{align}
which is compatible with the SDYM symplectic form
\begin{align}
    \Omega&= \delta \Psi_{AA}\wedge \delta \Phi^{AA}=\delta \Psi_+ \wedge \delta \Phi_++\delta \Psi_- \wedge \delta \Phi_-\,.
\end{align}
It seems that the limit is discontinuous in that the boundary data, the sources, jump from \eqref{mixedbc} to $\Phi_+$ and $\Psi_-$. This can be made look ``less discontinuous'' by noticing that fixing $\Phi_+^0$ in SDYM means the Dirichlet condition on $\Phi_+$ in YM, i.e. the ``same'' $\Phi^0_+$ gets fixed, while fixing $\Psi_-$ in SDYM can be understood as the Neumann condition on $\Phi_-$. Indeed, with $\Phi_-^1=\text{fixed}$, the regularity implies $\Phi_-=-k^{-1} e^{-kz} \Phi_-^1$. Therefore, $\epsilon\Psi_-= D_-\Phi_-=2 e^{-kz}\Phi_-^1$ and we can identify $\epsilon\Psi_-^0=2\Phi_-^1$. 

%%%%%%%%%%%%%%%%%%%%%%%%%%%%%%%%%%%%%%%%%%%%%%%%%%%%%%%%%%%%%
\subsection{AdS/CFT dictionary}
\label{sec:adscftdict}
%%%%%%%%%%%%%%%%%%%%%%%%%%%%%%%%%%%%%%%%%%%%%%%%%%%%%%%%%%%%%
Let us recall the standard lore about Dirichlet, Neumann and mixed holography. The path integral in AdS with Dirichlet boundary conditions for the gauge field computes the generating functional $W_{\text{D}}[A]$ of correlators of a conserved current in the boundary CFT:
\begin{align}
    \left\langle \exp \int d^3 x\, J^m(x) A_m(x) \right\rangle_{\text{CFT}}&=\int D\Phi\Big|_{\Phi^0_i\rightarrow A_i} \exp{S[\Phi]} \,.
\end{align}
With the Neumann boundary conditions imposed,    the path integral computes the generating functional $W_{\text{N}}[E]$ of correlation functions of a gauge field $A_m$ on the boundary
\begin{align}\label{Neumanndual}
    \int DA\, \left\langle \exp \int d^3 x\, J^m(x) A_m(x)+E^m A_m(x) +...\right\rangle_{\text{CFT}}&=\int D\Phi\Big|_{\Phi^1_i\rightarrow E_i} \exp{S[\Phi]} \,.
\end{align}
Here, the dual theory is intimately related to the one revealed via the Dirichlet boundary conditions. If $S_\text{CFT}$ is the action of the ``Dirichlet'' CFT, then one can couple the current $J_m$ to a background gauge field $A_m$
\begin{align}
    S[A]&= S_{\text{CFT}}+ \int d^3 x\, J^m(x) A_m(x) +\text{Seagulls} 
\end{align}
and integrate over $A_m$, which was called the S-operation in \cite{Witten:2003ya}. This is what is already understood in \eqref{Neumanndual}. The Dirichlet and Neumann dualities are closely related in the bulk as well \cite{Hartman:2006dy,Giombi:2011ya} and, in principle, one does not have to recompute the partition function $W_\text{N}[E]$ diagram by diagram if $W_\text{D}[A]$ has already been computed. In the Neumann case a better observable is given by the dual current $\tilde{J}=*(dA+g AA)$, which is the magnetic field of the boundary gauge field. For $U(1)$ it is just $\tilde{J}=*dA$ and it is conserved. For a nonabelian symmetry $\tilde{J}$ is only covariantly conserved and its correlation functions depend on the gauge.

Going from Dirichlet to Neumann boundary conditions is about upgrading the source $A_m$ to a dynamical field by integrating over it in the path integral. Another operation, which is closely related to the T-operation \cite{Witten:2003ya}, is to add a Chern--Simons term (more generally, any local action $S_\text{b}[A]$ of the boundary value $A=\Phi^0$, \cite{Marolf:2006nd}) to the CFT partition function $W[A]$. This operation adds local counterterms to the CFT correlators, e.g. adding the Chern--Simons term modifies the coefficients of the parity-odd structures in two- and three-point functions.\footnote{A lightning introduction into the spinor-helicity language useful for $3d$ CFTs can be found in Appendix \ref{app:cft}. We also discuss there the structure of two-point functions of conserved currents.} The bulk point of view is quite simple as well: one just adds the boundary term $S_\text{b}$ to the bulk action $S[\Phi]$. This does not modify the boundary conditions, e.g. if we add a Chern--Simons term to the YM action the variation is
\begin{align}
    \delta S=\delta A^m( i\kappa B_m + E_m)&=0 \,,  && S[A]=S_{\text{YM}}[A]+i\kappa S_{\text{CS}}[A] \,.
\end{align}
The combined action is still stationary with the Dirichlet boundary condition $\delta A_i=0$. Another option is to fix the second factor $Q_m=( i\kappa B_m + E_m)$ on the boundary. This can be achieved via a Legendre transform. Instead, one can add $-\int_{\pl M} A^m Q_m^0$, where $Q_m^0$ is a fixed boundary data. Once the action is stationary under the mixed boundary conditions, the CFT dual operator is a gauge field on the boundary as for the Neumann case. The Dirichlet point can still be reached as $\gamma=0$ point or $\kappa=\infty$. In the latter case one has to amplify the source, otherwise it is suppressed.

If we want to make the action stationary with the mixed-boundary conditions \eqref{mixedbc}, we should add the Chern--Simons action at level $\kappa=-\cot(\gamma)$. At the self-dual point, the value of the theta-term becomes such that $a=b$, see Section \ref{sec:2.5}. In other words, the YM action approaches the cYM action with vanishing theta-term, which is equivalent to the Chalmers--Siegel action without the theta-term as well. The latter is the best starting point to make contact with SDYM.  

In order to formulate a proposal for the self-dual AdS/CFT correspondence, let us recall that (a) the only boundary data in SDYM can be assigned to $\Phi_+$ and $\Psi_-$, which from the YM vantage point corresponds to Dirichlet data for $\Phi_+$ and Neumann for $\Phi_-$; (b) $\Phi$ is a gauge field both in YM and SDYM (but in SDYM it carries a single polarization) and the Dirichlet data for $\Phi$ is a gauge field on the boundary, which couples to a conserved current on the boundary; (c) $\Psi$'s boundary data is a conserved current with a single polarization in SDYM, whereas the Neumann data for $\Phi$ in YM is also a conserved current, which couples to a gauge field on the boundary. Therefore, a natural extrapolation of the AdS/CFT dictionary is that in SDYM $\Phi$ couples to the positive helicity component $J_+$ of a conserved current on the boundary and $\Psi$ couples to the negative helicity component $A_-$ of a gauge field on the boundary. The latter can be promoted to the dual current (its magnetic field) $J_-$ that would complete $J_+$ to a full current $(J_+,J_-)$.  

Lastly, as it was mentioned, Dirichlet and Neumann results are closely related, which can be seen at the level of bulk-to-bulk $G$ and boundary-to-bulk $K$ propagators. With the help of the bulk-to-bulk propagators given in Appendix \ref{app:propagators} one can verify the following relations\footnote{Here, $\Pi^\text{e}$ and $\Pi^\text{o}$ are projectors onto the parity-even and parity-odd structures of the current's two-point functions, see Appendix \ref{app:cft}.} 
\begin{align}
    \frac{\pl}{\pl \gamma} G&= \frac{1}{k}e^{-k(z+z')} P \,, & \frac{\pl}{\pl \gamma}  K&= \frac{1}{k}e^{-kz} P\,, && P=[- \sin (2\gamma) \Pi^\text{e} +i \cos(2\gamma) \Pi^\text{o}]\,,
\end{align}
that determine what happens to the holographic correlation functions once $\gamma$ is slightly changed. Here, we omit the indices. These relations allow one to follow the ``flow'' of mixed boundary conditions as $\gamma$ changes. 

%%%%%%%%%%%%%%%%%%%%%%%%%%%%%%%%%%%%%%%%%%%%%%%%%%%%%%%%%%%%%
\section{Correlators}
\label{sec:correlators}
%%%%%%%%%%%%%%%%%%%%%%%%%%%%%%%%%%%%%%%%%%%%%%%%%%%%%%%%%%%%%
In this Section we compute correlators in YM (understood as cYM) and find a perfect match with Chalmers--Siegel theory, which is not surprising, but there are a few subtleties that are absent in flat space.  Three- and four-point correlators are computed for Dirichlet/Neumann/mixed boundary conditions both in Feynman and axial gauge. We also compute correlators in SDYM. Chalmers--Siegel's correlators approach smoothly those of SDYM. Since for non-Dirichlet boundary conditions we have a gauge field on the boundary, there is may be some gauge dependence in the correlators we compute. 

Essential ingredients are bulk-to-bulk propagators, which we collect in Appendix \ref{app:propagators}.\footnote{Propagators with mixed boundary conditions in Feynman and axial gauges in YM and Chalmers--Siegel theory seem new, from which SDYM's propagators follow. In the gauge $A_{3,0}=0$ the spin-one propagator for mixed boundary conditions was found in \cite{Chang:2012kt}. } External lines in Feynman-Witten diagrams are saturated by boundary-to-bulk propagators, which can be derived from the bulk-to-bulk ones. The kinematics of the boundary-to-bulk propagators is fixed and is visible already in the Fefferman-Graham expansion of Section \ref{sec:FG}. However, there are simple prefactors that depend on boundary conditions. Concretely, $\Phi$ and $\Psi$ external lines in Witten diagrams are to be replaced by
\besubeqs
\begin{align}
    \Phi_{AA}(k,z) &=-e^{-kz} \epsilon^{\pm}_{AA}
    \begin{cases}
        \frac{1}{2k}(1-e^{ \pm 2i\gamma})\,, & \gamma\neq0\,,\\
        1\,, & \gamma =0 \,,
    \end{cases}\\
    \Psi_{AA}(k,z) &=e^{-kz}k_Ak_A
    \begin{cases}
        +\frac{1}{2k}(1-e^{-2i\gamma}) \,, & \gamma\neq0\,,\\
        1\,, & \gamma =0 \,.
    \end{cases}
\end{align}
\esubeqs
These propagators are obtained from the boundary limits $\epsilon^{AA}(k)\langle \Phi_{AA}(-k,0)\,\, \bullet(k,z)\rangle$, where $\bullet=\Phi \text{ or } \Psi$. Note that as far as the relation between the polarization $\epsilon$ on the boundary and of the bulk field goes, we have $\epsilon_\pm(k)\langle \Phi(-k,0) \bullet(k,z') \rangle \sim \epsilon_{\pm}$. Having such external lines corresponds to computing the boundary limit of $\langle \Phi_{AA}(k_1,0)\ldots \rangle$ correlators. However, one can replace the external lines by $F_{AA}$, $\Psi_{AA}$ or $\bar{F}_{A'A'}$. These options do not change the helicity structure, but can add a factor of $k$ and modify the numerical factor, which can make the SD limit smoother. In what follows we do not discuss the purely numerical $\gamma$-dependent factors on external lines as they can always be rescaled away by changing the sources. In other words, two expressions that agree modulo numerical factors on external lines are considered equal. 

To simplify the presentation we will always use the Dirichlet external lines, which also facilitates comparison in the flat space limit. In the flat limit the leading energy pole has to reproduce the amplitude in the same theory but in flat space \cite{Maldacena:2011nz}. Concretely, for Feynman gauge we will use
\begin{align}
    \Phi_+^{A,A'}(k_i,z)&=\frac{q_i^{A}\bar{i}^{A'}}{\langle q_ii \rangle}e^{-kz}\,, & \Phi_-^{A,A'}(k_i,z)&=\frac{i^A\bar{q}_i^{A'}}{\langle \bar{q}\bar{i}\rangle} e^{-k_iz}\,,
\end{align}
and for axial gauge the external lines are ($\Psi$ is the same in all gauges)
\begin{align}
    \Phi^{AA}_+(k_i,z)&=\frac{\bar{i}^A\bar{i}^{A}}{2k_i}e^{-kz}\,, &\Phi^{AA}_-(k_i,z)&=-\frac{{i}^A{i}^{A}}{2k_i}e^{-kz}\,, 
    &  \Psi^{AA}_-(k_i,z)=i^Ai^{A} e^{-kz}\,.
\end{align}
One issue with using the same external lines for all $\gamma$ is that for non-Dirichlet boundary conditions $\gamma\neq0$ we compute the correlators of a gauge field $a_i$ on the boundary. The dual current\footnote{\label{compost}There is a nonabelian piece in the dual current $*(da+g\,a\wedge a)$. It is a composite operator and to the order we are at it is just a product of lower order expressions and we do not discuss it separately. We are not going to consider the composite operator contributions in this paper, see \cite{Richard2026}, since the main accent is on the amplitudes and having a gauge (color) symmetry in the bulk corresponding to non-abelian currents on the boundary. According to \cite{Sharapov:2022awp,Jain:2024bza,Aharony:2024nqs} that would require a Chern--Simons matter theory with a leftover global non-abelian symmetry, while the simplest examples of duality do not need that. } $*(da)$ leads to $\slashed k$ and $\slashed k (\epsilon_{\pm})=\pm k \epsilon_{\pm}$, i.e. taking the magnetic field just adds a factor of $\pm k$, which is also a factor that the Dirichlet and non-Dirichlet boundary-to-bulk propagators differ by. Therefore, using the Dirichlet propagator with $\Phi$ on the boundary computes the magnetic field automatically up to a numerical factor.

%%%%%%%%%%%%%%%%%%%%%%%%%%%%%%%%%%%%%%%%%%%%%%%%%%%%%%%%%%%%%
\subsection{Two-point functions}
\label{sec:twopoint}
%%%%%%%%%%%%%%%%%%%%%%%%%%%%%%%%%%%%%%%%%%%%%%%%%%%%%%%%%%%%%
Two-point functions can be extracted either from the action or from the bulk-to-bulk propagators. Two-point functions appear to be the most subtle ones for a number of reasons. (1) Two-point functions have a non-analytic parity-even piece and a parity-odd contact term. However, in the helicity basis these two get easily mixed with each other, see Appendix \ref{app:cft};\footnote{For example, in the light-cone gauge used in \cite{Skvortsov:2018uru,Chowdhury:2024dcy} two-point functions are off-diagonal, i.e. $\langle\pm \pm\rangle=0$ and $\langle\pm \mp\rangle\neq0$. In axial gauge it is the opposite.\label{ftn:LC}} (2) in SDYM and Chalmers--Siegel theory, fields obey first order equations with $\pl_z$ and the propagator contains $\sign(z-z')$. Therefore, the two-point function as a boundary limit of the bulk-to-bulk propagator depends on the limit, the ambiguity being a contact term.\footnote{While this is a feature of the first order systems, even for a scalar field $\langle \pl_z \phi(z) \phi(z')\rangle$ has the same ambiguity. Such terms also appear for fermions. However, for nonvanishing mass $m$ the conformal dimension is $\Delta=3/2\pm m$ and there is a nonanalytic term that ``overshadows'' the contact one.\label{ftn:fermions} } Below, we concentrate mostly on the non-Dirichlet cases, $\gamma\neq0$, as the Dirichlet results are standard and are also far away from what happens for $\gamma\neq0$. 

\paragraph{Two-point functions from propagators.} With the help of Appendix \ref{app:propagators}, one finds 
\begin{align} \label{PhiPhi2}
    \begin{aligned}
        \epsilon^{+AA}(k)\langle \Phi_{AA}(-k,0)\Phi_{BB}(k,0)\rangle\epsilon^{+BB}(-k) &= -\frac{1}{2k}(1-e^{+2i\gamma})\,,\\
        \epsilon^{-AA}(k)\langle \Phi_{AA}(-k,0)\Phi_{BB}(k,0)\rangle\epsilon^{-BB}(-k) &= -\frac{1}{2k}(1-e^{-2i\gamma})\,.
    \end{aligned}
\end{align}
We write the expressions here-above in full detail for the first time, but later abbreviate these and similar ones to just 
\begin{align}
     \langle ++\rangle &= -\frac{1}{2k}(1-e^{+2i\gamma})\,, &
        \langle --\rangle &= -\frac{1}{2k}(1-e^{-2i\gamma})\,.
\end{align}
This should be compared with the boundary-to-bulk propagators
\begin{align} \label{PhiPhi1}
    \begin{aligned}
        \epsilon^{AA}_{\pm}(k)\langle \Phi_{AA}(-k,0)\Phi_{BB}(k,z')\rangle &= -\frac{1}{2k}\epsilon_{BB}^{\pm}(k)(1-e^{\pm 2i\gamma})e^{-kz'} \,.
    \end{aligned}
\end{align}
To have a finite result in the SD limit one should rescale every positive helicity external leg by a divergent factor $f\sim (1-e^{+2i\gamma})$. Negative helicity legs can be left untouched as the prefactor remains finite. The two-point function $\langle ++\rangle $ will have to be rescaled by $f^2$, which completely suppresses it. Therefore, in the SD-limit we find 
\begin{align}
     \langle --\rangle &= -\frac{1}{2k}\,, &
        \langle ++\rangle &=0\,.
\end{align}
In the Chalmers--Siegel theory and SDYM, the bulk-to-bulk propagators projected onto the states with definite helicity give (note that the SD limit is smooth here)
\begin{align} \label{PsiPhiprops}
    \begin{aligned}
        \langle \Psi_-(z) \Phi_-(z')\rangle &= \frac12 e^{-k|z-z'|}[\sign(z-z')+1] -e^{-2i\gamma} e^{-k(z+z')}\,,\\
        \langle \Psi_+(z) \Phi_+(z')\rangle &=  \frac12 e^{-k|z-z'|}[\sign(z-z')-1] \,.
    \end{aligned}
\end{align}
This shows that the two-point functions depend on the order of limits, but the difference (as well as the result itself) is a contact term, $\delta^3(x-y)$. In general, one needs additional arguments to justify a concrete value for a contact term.

\paragraph{Two-point functions from the action.} Let us begin with the YM action. On-shell it has the same value as the on-shell Chalmers--Siegel  action and it reduces to
\begin{align}\label{YMactionOnshell}
    -S_{\text{YM},\theta}\Big|_{\text{on-shell}}&= \tfrac14\int_{\pl M}(a+b) F^{AA}\Phi_{AA}-(a-b) \bar{F}^{AA}\Phi_{AA} \,. 
\end{align}
With $F=B+E$ and $\bar{F}=B-E$, we have 
\begin{align}
     -S_{\text{YM},\theta}\Big|_{\text{on-shell}}&= \tfrac12\int_{\pl M} (b B + a E )^i\Phi_i \,.
\end{align}
With Dirichlet boundary conditions this leads to the most general two-point function
\begin{align}
    \langle J_{AA} J_{BB}\rangle&= a k \Pi^\text{e}_{AA,BB}+b k\Pi^\text{o}_{AA,BB} \,.
\end{align}
For mixed boundary conditions the YM action is not stationary and we need to add a boundary term
\begin{align}
    -S_{\text{bndry}}&= -a \Tr\int _{\pl M}\left[ E\cdot \Phi +\frac{\cos(\gamma)}{2i \sin(\gamma)} \Phi \cdot B\right]\,.
\end{align}
Here, $\Phi \cdot B$ can be recognized as the free Chern--Simons term. The first term $E\cdot \Phi$ is needed to make the action stationary with Neumann boundary conditions $\delta E_i=0$ and the Chern--Simons term corrects it to extend the action from the Neumann point $\gamma=\pi/2$ to any $\gamma\neq0$. We evaluate the on-shell action by stepping $\epsilon$ into the bulk and using the boundary-to-bulk propagators. This gives a two-point function $\langle A^{AC} A^{BM}\rangle$ of a gauge field on the boundary. The two-point function of the dual current $\tilde{J}=B=*(dA)=\slashed k A$ is
\begin{align}
    \langle J^{AA} J^{BB}\rangle&= k\fud{A}{C} k\fud{B}{M} \langle A^{AC} A^{BM}\rangle=  k \sin(\gamma)[ - \sin(\gamma)\Pi_\text{e}^{AA,BB}+i\cos (\gamma) \Pi_\text{o}^{AA,BB}] \,.
\end{align}
It is the same result as from the bulk-to-bulk propagator before. For future reference, note that $(b-a)=-a(1+i \cot(\gamma))$. 

\paragraph{Variational problem and two-point function in SDYM.} Let us write the SDYM and Chalmers--Siegel actions in terms of the definite helicity components 
\begin{align}
    S&= \int_M (\Psi_+, D_+ \Phi_+)+(\Psi_-, D_- \Phi_-)-\tfrac{\epsilon}2[(\Psi_+,\Psi_+)+(\Psi_-,\Psi_-)]\,,
\end{align}
where $(f,g)\equiv f(-k,z) g(+k,z) $. The variation of the action gives a presymplectic potential
\begin{align}
   \delta S&= -(\Psi_+, \delta \Phi_+)-(\Psi_-, \delta \Phi_-)\,.
\end{align}
However, as we discussed in Section \ref{sec:FG}, fields satisfying $D_- f=0$ do not have regular solutions in the bulk and cannot vary on the boundary, i.e. $\Phi_-=\Psi_+=0$ on the boundary. The dynamical fields are $\Phi_+$ and $\Psi_-$, i.e. $\delta \Phi_+=\delta \Psi_-=0$ on the boundary. The SDYM action is stationary as is! Nevertheless, one can try to add various $(\Psi_\pm, \Phi_\pm)$ boundary terms. Since at the free level $\Phi_-(z)=\Psi_+(z)=0$ (when the other leg is on the boundary), the only boundary term to give a contribution in the $\epsilon\rightarrow0$ limit is $(\Psi_-(\epsilon),\Phi_+(\epsilon))$. One can add it with a free coefficient.\footnote{Ignoring the helicity decomposition, a covariant version looks close to the BF-term, i.e. $\Tr\int_{\pl M} \Psi \wedge F$, i.e. $\Tr\int_{\pl M}\Psi^{AA} F_{AA}$ in components, which requires treating $\Psi$ as a one-form on the boundary. The weight of $\Psi$ is, however, $2$, which is inconsistent with the BF interpretation. } It is a contact term since it is a local functional of the boundary data. Note that $\Phi$ and $\Psi$ have dual dimensions $\Delta$, $d-\Delta$ and in this case there is a possibility for a contact two-point function. One can think of this term as Chern--Simons-like. In the standard Dirichlet case, the coefficient of the Chern--Simons term is a free parameter which contributes to the two-point function. Likewise, the coefficient of $(\Psi_-,\Phi_+)$ is free.

As different from the standard examples in AdS/CFT, the two-point function of SDYM appears to be a contact term and lacks the nonlocal part. This is also seen from the bulk-to-bulk propagators \eqref{PsiPhiprops} since the boundary limit depends on the order, but is a contact term anyway. Two-point functions of fermions have similar features, i.e. $\sign(z-z')$ in the propagator that leads to a contact term, but there is also a non-local part, cf. footnote \ref{ftn:fermions}. Since the concept of helicity is tied to momentum space, it blurs the difference between contact and non-local terms, which is especially pronounced for two-point functions, see \eqref{twopointab} and footnote \eqref{ftn:LC}. 

Lastly, let us note that one can add a Chern--Simons term to the SDYM action. The free or $U(1)$ Chern--Simons term respects the helicity decomposition, $A dA \sim k(A_+,A_+)-k(A_-,A_-)$, but the $A^3$-term does not. Therefore, there does not seem to be any way to make the action stationary without constraining $F$ on the boundary. Indeed, the SD limit of YM leads to SDYM/Chalmers--Siegel theory without any theta-term.

%%%%%%%%%%%%%%%%%%%%%%%%%%%%%%%%%%%%%%%%%%%%%%%%%%%%%%%%%%%%%
\subsection{Three-point functions}
\label{sec:threepoint}
%%%%%%%%%%%%%%%%%%%%%%%%%%%%%%%%%%%%%%%%%%%%%%%%%%%%%%%%%%%%%
As we will see, holographic correlators are very much reminiscent of flat space amplitudes. What prevents them from being the same is that in $\text{AdS}_4$ only the momentum along the boundary is conserved, but the energy component $E$ is not 
\begin{align} \label{AdSMomCons}
    &\sum_{i=1}^n \vec{k}^i_{AA'}=0 \,, & &1^{A}\bar{1}^{A'}+2^A\bar{2}^{A'}+3^A\bar{3}^{A'}= (k_1+k_2+k_3)\epsilon^{AA'}\equiv E\epsilon^{AA'}\,.
\end{align}
Since our goal is to relate YM and SDYM, we concentrate on the $(++-)$-amplitude/correlation function that is shared by both theories. 

%%%%%%%%%%%%%%%%%%%%%%%%%%%%%%%%%%%%%%%%%%%%%%%%%%%%%%%%%%%%%
\subsubsection{SDYM}
\label{sec:SDYM}
\begin{figure}[h!]
\centering
\begin{tabular}{c c}
    % S-channel Witten diagram
    \begin{tikzpicture}[scale=0.7]
        % Define the boundary circle
        \draw[thick] (0,0) circle (2.1cm);
        
        % Define the vertices inside the bulk
        \coordinate (v) at (0,0);
        
        % Define the external legs on the boundary
        \coordinate (a) at (-1.48,1.48);
        \coordinate (b) at (1.48,1.48);
        \coordinate (c) at (0,-2.1);
        
        % Draw the gauge boson lines
        \draw[decorate,decoration={snake},thick] (a) -- (v);
        \draw[decorate,decoration={snake},thick] (b) -- (v);
        \draw[decorate,decoration={snake},thick] (c) -- (v);
        
        % Draw parallel momentum arrows
        \draw[-{Latex},thick] (-1.2,0.8) -- (-0.6,0.2) node[midway,xshift=-4pt,yshift=-4pt] {\( k_1 \)};
        \draw[-{Latex},thick] (0.3,-1.4) -- (0.3,-0.55) node[midway, xshift=8pt] {\( k_2 \)};
        \draw[{Latex}-,thick]  (0.6,0.2) -- (1.2,0.8) node[midway,xshift=6pt,yshift=-6pt] {\( k_3 \)};
        
        % Add labels for fields next to endpoints
        \node[xshift=-5pt,yshift=5pt] at (a) {\( \epsilon^+_1 \)};
        \node[xshift=5pt,yshift=5pt] at (b) {\( \epsilon^-_3 \)};
        \node[yshift=-7pt] at (c) {\( \epsilon^+_2 \)};
        
        % Add labels for vertices
        \node[xshift=6pt,yshift=-4pt] at (v) {\( z \)};        
    \end{tikzpicture}
\end{tabular}
\caption{SDYM Witten diagram for the three-point correlator.}
\label{fig:SDYMthreepoint}
\end{figure}
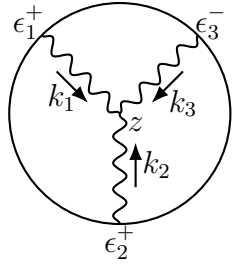

%%%%%%%%%%%%%%%%%%%%%%%%%%%%%%%%%%%%%%%%%%%%%%%%%%%%%%%%%%%%%
\noindent In SDYM the answer for the three-point correlator is simple (we are not writing down the vertex factors $V$ explicitly in the paper since they are very simple in SDYM and standard in YM). The cubic diagram on Figure \ref{fig:SDYMthreepoint} leads to
\begin{align}
    \begin{aligned}\label{SDYM3pt}
        \mathcal{W}_3^{\text{SDYM}} &= \int_0^{\infty}dz\, V_{AA',BB',CC}\Phi_+^{AA'}(k_1)\Phi_+^{BB'}(k_2)\Psi_-^{CC}(k_3)=\frac{1}{E}\mathcal{A}_3=\\
        &=-\frac{2g}{E}\frac{\langle q_13 \rangle \langle q_23 \rangle}{\langle q_11 \rangle \langle q_22 \rangle} \langle \bar{1}\bar{2} \rangle \,.
    \end{aligned}
\end{align}
It is clear that the energy pole agrees with the flat space amplitude $\mathcal{A}_3$. In flat space the momentum is fully conserved and with the help of some simple identities one can show that the amplitude does not depend on the reference spinors. In $\text{AdS}_4$, with the help of the modified identities
\begin{align} \label{momConsCorrections}
    \langle q_13 \rangle &= \frac{\langle q_11 \rangle \langle \bar{1}\bar{2} \rangle-E\langle q_1\bar{2} \rangle}{\langle \bar{2}\bar{3} \rangle} \,, & \langle q_23 \rangle &= -\frac{\langle q_22\rangle \langle \bar{1}\bar{2} \rangle+E\langle q_2\bar{1} \rangle}{\langle \bar{1}\bar{3} \rangle}\,,
\end{align}
one can rewrite the correlator as the standard flat space amplitude plus energy corrections
\begin{align}\label{corrections3}
    \mathcal{W}_3^{\text{SDYM}} &= \frac{2g}{E}\frac{\langle \bar{1}\bar{2} \rangle}{\langle \bar{1}\bar{3}\rangle\langle \bar{2}\bar{3} \rangle}\Big(\langle \bar{1}\bar{2}\rangle^2 + E \langle \bar{1}\bar{2} \rangle \frac{\langle q_11\rangle \langle q_2\bar{1}\rangle - \langle q_1 \bar{2} \rangle \langle q_2 2 \rangle}{\langle q_11 \rangle \langle q_2 2 \rangle}-E^2\frac{\langle q_1\bar{2} \rangle \langle q_2\bar{1} \rangle}{\langle q_11 \rangle \langle q_22 \rangle}\Big) \,.
\end{align}
With the reference spinors chosen as in the axial gauge, which projects onto definite helicity components, we find
\begin{align}\label{SDYM3ptAxial}
    \mathcal{W}_3^{\text{SDYM}} = -\frac{g}{2E}\frac{\langle \bar{1}3 \rangle \langle \bar{2}3 \rangle}{k_1 k_2} \langle \bar{1}\bar{2} \rangle\,.
\end{align}

%%%%%%%%%%%%%%%%%%%%%%%%%%%%%%%%%%%%%%%%%%%%%%%%%%%%%%%%%%%%%
\subsubsection{(Chiral) Yang--Mills}
\label{sec:cYMAdS}
%%%%%%%%%%%%%%%%%%%%%%%%%%%%%%%%%%%%%%%%%%%%%%%%%%%%%%%%%%%%%
We rewrite the YM action with a theta-term as the cYM action plus a theta-term
\begin{align}\notag
   \tfrac{a+b}4\,\int F_{AB}F^{AB} +\tfrac{a-b}4 \,\int \bar{F}_{A'B'} \bar{F}^{A'B'}= \tfrac{a}2\,\int F_{AB}F^{AB} +\tfrac{(b-a)}4 \,\int \left(F_{AB}F^{AB} -\bar{F}_{A'B'} \bar{F}^{A'B'}\right)\,.
\end{align}
This leads to two sets of Feynman rules depicted in Figure \ref{fig:YMthreepoint}. The key simplifications are: $(F_{AB})^2$ and $(\bar{F}_{A'B'})^2$ have much simpler interactions, especially when some of the legs have definite helicity; the topological vertex (Chern--Simons term) resides on the boundary, which leads to less singular contributions. Let us compute three-point functions in two different ways, corresponding to the two decompositions in Figure \ref{fig:YMthreepoint}. The SD vertex gives 
\begin{align}
    \begin{aligned}
        \mathcal{W}_3^{\text{SD}} &= \int_0^\infty dz V^{\text{SD}}_{AA',BB',CC'}\Phi_+^{AA'}(k_1)\Phi_+^{BB'}(k_2)\Phi_-^{CC'}(k_3)=-\frac{g}{E}\frac{\langle q_13\rangle \langle q_23 \rangle}{\langle q_11 \rangle \langle q_22 \rangle} \langle \bar{1}\bar{2} \rangle\,,
    \end{aligned}
\end{align}
This is just $\tfrac12W^{\text{SDYM}}$, \eqref{SDYM3pt}. The ASD vertex gives a more complicated expression
\begin{align}
    \begin{aligned}
                \mathcal{W}_3^{\text{ASD}} &= \int_0^\infty dz V^{\text{ASD}}_{AA',BB',CC'}\Phi_+^{AA'}(k_1)\Phi_+^{BB'}(k_2)\Phi_-^{CC'}(k_3)=\\
                &=g\frac{\langle \bar{1}\bar{2} \rangle}{E}\frac{\langle q_11\rangle \langle q_23 \rangle \langle \bar{q}_3\bar{1} \rangle + \langle q_13 \rangle \langle q_22 \rangle \langle \bar{q}_3\bar{2} \rangle}{\langle q_11\rangle \langle q_22\rangle\langle\bar{q}_3\bar{3}\rangle} \,.
    \end{aligned}
\end{align}
Both the SD and ASD contributions can be massaged to reveal the relation to the flat space amplitude
\begin{align}
    \begin{aligned}
        \mathcal{W}_3^{\text{SD}} &= \frac{g}{E}\frac{\langle \bar{1}\bar{2} \rangle}{\langle \bar{1}\bar{3}\rangle\langle \bar{2}\bar{3} \rangle}\Big(\langle \bar{1}\bar{2}\rangle^2 + E \langle \bar{1}\bar{2} \rangle \frac{\langle q_11\rangle \langle q_2\bar{1}\rangle - \langle q_1 \bar{2} \rangle \langle q_2 2 \rangle}{\langle q_11 \rangle \langle q_2 2 \rangle}-E^2\frac{\langle q_1\bar{2} \rangle \langle q_2\bar{1} \rangle}{\langle q_11 \rangle \langle q_22 \rangle}\Big)\,,\\
        \mathcal{W}_3^{\text{ASD}} &= \frac{g}{E}\frac{\langle \bar{1}\bar{2} \rangle}{\langle \bar{1}\bar{3} \rangle \langle \bar{2}\bar{3} \rangle}\Big(\langle \bar{1}\bar{2} \rangle^2 - E\frac{\langle q_11 \rangle \langle q_2\bar{1} \rangle \langle \bar{q}_3\bar{1}\rangle \langle \bar{2}\bar{3} \rangle + \langle q_1\bar{2} \rangle \langle q_22 \rangle \langle \bar{q}_3\bar{2} \rangle \langle \bar{1}\bar{3} \rangle}{\langle q_11 \rangle \langle q_22 \rangle \langle \bar{q}_3 \bar{3} \rangle }\Big) \,.
    \end{aligned}
\end{align}
Now, let us compare to the second way of representing the YM interactions. The SD-vertex is the same and the topological vertex gives 
\begin{align}\label{Wtop}
    \begin{aligned}
        \mathcal{W}_3^{\text{top}} &=g\frac{\langle q_13\rangle \langle \bar{1}\bar{2} \rangle \langle q_2\bar{q}_3 \rangle - \langle q_1q_2 \rangle \langle \bar{q}_3\bar{1}\rangle \langle \bar{2}3 \rangle}{\langle q_11\rangle \langle q_22 \rangle \langle \bar{q}_3\bar{3} \rangle}=-\frac{g}{4}\frac{\langle \bar{1}\bar{2} \rangle \langle \bar{1}3 \rangle \langle \bar{2}3 \rangle}{k_1k_2k_3}\,,
    \end{aligned}
\end{align}
where the last form is obtained in the axial gauge. As a consistency check, with the help of Fierz identities one confirms that $\mathcal{W}^{\text{SD}}_3-\mathcal{W}^{\text{ASD}}_3=\mathcal{W}^{\text{top}}_3$. Finally, the two ways to represent the same correlator are
\begin{align}\label{twoway}
        \mathcal{W}_3 &= (a+b)\mathcal{W}_3^{\text{SD}} + (a-b)\mathcal{W}_3^{\text{ASD}} = 2a\mathcal{W}_3^{\text{SD}}+(b-a)\mathcal{W}_3^{\text{top}}\,.
\end{align}
The second one being, obviously, simpler. Also, we see that $\mathcal{W}_3^{\text{SD}}=\tfrac12\mathcal{W}^{\text{SDYM}}$ with our normalization of SDYM interactions. The topological term has no energy pole and the flat space limit agrees with the standard amplitude.

\begin{figure}[h!]
\centering
\begin{tabular}{c c}
    % S-channel Witten diagram
    \begin{tikzpicture}[scale=0.7]
        
            \begin{scope}[xshift=-8cm,yshift=-6cm]
            \node at (-4,0) {$\big(a+b\big)\times$};
                % Define the boundary circle
            \draw[thick] (0,0) circle (2.1cm);
            
            % Define the vertices inside the bulk
            \coordinate (v) at (0,0);
            
            % Define the external legs on the boundary
            \coordinate (a) at (-1.48,1.48);
            \coordinate (b) at (1.48,1.48);
            \coordinate (c) at (0,-2.1);
            
            % Draw the gauge boson lines
            \draw[decorate,decoration={snake},thick] (a) -- (v);
            \draw[decorate,decoration={snake},thick] (b) -- (v);
            \draw[decorate,decoration={snake},thick] (c) -- (v);
            
            % Draw parallel momentum arrows
        \draw[-{Latex},thick] (-1.2,0.8) -- (-0.6,0.2) node[midway,xshift=-4pt,yshift=-4pt] {\( k_1 \)};
        \draw[-{Latex},thick] (0.3,-1.4) -- (0.3,-0.55) node[midway, xshift=8pt] {\( k_2 \)};
        \draw[{Latex}-,thick]  (0.6,0.2) -- (1.2,0.8) node[midway,xshift=6pt,yshift=-6pt] {\( k_3 \)};
        
        % Add labels for fields next to endpoints
        \node[xshift=-5pt,yshift=5pt] at (a) {\( \epsilon^+_1 \)};
        \node[xshift=5pt,yshift=5pt] at (b) {\( \epsilon^-_3 \)};
        \node[yshift=-7pt] at (c) {\( \epsilon^+_2 \)};

            \node at (4,0) {$+\big(a-b\big)\times$};
            \fill (0,0) circle (5pt);
            \end{scope}
            \begin{scope}[yshift=-6cm]
            % Define the boundary circle
            \draw[thick] (0,0) circle (2.1cm);
            
            % Define the vertices inside the bulk
            \coordinate (v) at (0,0);
            
            % Define the external legs on the boundary
            \coordinate (a) at (-1.48,1.48);
            \coordinate (b) at (1.48,1.48);
            \coordinate (c) at (0,-2.1);
            
            % Draw the gauge boson lines
            \draw[decorate,decoration={snake},thick] (a) -- (v);
            \draw[decorate,decoration={snake},thick] (b) -- (v);
            \draw[decorate,decoration={snake},thick] (c) -- (v);
            
            % Draw parallel momentum arrows
        \draw[-{Latex},thick] (-1.2,0.8) -- (-0.6,0.2) node[midway,xshift=-4pt,yshift=-4pt] {\( k_1 \)};
        \draw[-{Latex},thick] (0.3,-1.4) -- (0.3,-0.55) node[midway, xshift=8pt] {\( k_2 \)};
        \draw[{Latex}-,thick]  (0.6,0.2) -- (1.2,0.8) node[midway,xshift=6pt,yshift=-6pt] {\( k_3 \)};
        
        % Add labels for fields next to endpoints
        \node[xshift=-5pt,yshift=5pt] at (a) {\( \epsilon^+_1 \)};
        \node[xshift=5pt,yshift=5pt] at (b) {\( \epsilon^-_3 \)};
        \node[yshift=-7pt] at (c) {\( \epsilon^+_2 \)};
            \fill[white] (0,0) circle (5pt);
            \draw[thick] (0,0) circle (5pt);
            \node at (3,0) {$=$};
            \end{scope}

            \begin{scope}[yshift=-6cm]
                \begin{scope}[xshift=-8cm,yshift=-6cm]
            \node at (-3,0) {$=2a\times$};
                % Define the boundary circle
            \draw[thick] (0,0) circle (2.1cm);
            
            % Define the vertices inside the bulk
            \coordinate (v) at (0,0);
            
            % Define the external legs on the boundary
            \coordinate (a) at (-1.48,1.48);
            \coordinate (b) at (1.48,1.48);
            \coordinate (c) at (0,-2.1);
            
            % Draw the gauge boson lines
            \draw[decorate,decoration={snake},thick] (a) -- (v);
            \draw[decorate,decoration={snake},thick] (b) -- (v);
            \draw[decorate,decoration={snake},thick] (c) -- (v);
            
            % Draw parallel momentum arrows
        \draw[-{Latex},thick] (-1.2,0.8) -- (-0.6,0.2) node[midway,xshift=-4pt,yshift=-4pt] {\( k_1 \)};
        \draw[-{Latex},thick] (0.3,-1.4) -- (0.3,-0.55) node[midway, xshift=8pt] {\( k_2 \)};
        \draw[{Latex}-,thick]  (0.6,0.2) -- (1.2,0.8) node[midway,xshift=6pt,yshift=-6pt] {\( k_3 \)};
        
        % Add labels for fields next to endpoints
        \node[xshift=-5pt,yshift=5pt] at (a) {\( \epsilon^+_1 \)};
        \node[xshift=5pt,yshift=5pt] at (b) {\( \epsilon^-_3 \)};
        \node[yshift=-7pt] at (c) {\( \epsilon^+_2 \)};

            \node at (4,0) {$+\big(b-a\big)\times$};
            \fill (0,0) circle (5pt);
            \end{scope}
            \begin{scope}[yshift=-6cm]
            % Define the boundary circle
            \draw[thick] (0,0) circle (2.1cm);

            \filldraw[fill=gray!20, draw=black] 
            (1.97,-0.72) arc[start angle=40, end angle=140, radius=2.58cm];
            \filldraw[fill=gray!20, draw=black] 
            (-1.97,-0.72) arc[start angle=-160, end angle=-20, radius=2.1cm];
            
            % Define the vertices inside the bulk
            \coordinate (v) at (0,-0.5);
            
            % Define the external legs on the boundary
            \coordinate (a) at (-1.61,-1.35);
            \coordinate (b) at (0,-2.1);
            \coordinate (c) at (1.61,-1.35);
            
            % Draw the gauge boson lines
            \draw[decorate,decoration={snake},thick] (a) -- (v);
            \draw[decorate,decoration={snake},thick] (b) -- (v);
            \draw[decorate,decoration={snake},thick] (c) -- (v);
            
            % Add labels for fields next to endpoints
            \node[xshift=-6pt,yshift=-5pt] at (a) {\( \epsilon^+_1 \)};
            \node[yshift=-7pt] at (b) {\( \epsilon^+_2 \)};
            \node[xshift=4pt,yshift=-4pt] at (c) {\( \epsilon^-_3 \)};

             % Draw parallel momentum arrows
            \draw[-{Latex},thick] (-1.3,-0.9) -- (-0.55,-0.5) node[midway,xshift=-5pt,yshift=5pt] {\( k_1 \)};
            \draw[-{Latex},thick] (0.3,-1.8) -- (0.3,-0.95) node[midway,xshift=7pt,yshift=-2pt] {\( k_2 \)};
            \draw[-{Latex},thick] (1.3,-0.9) -- (0.55,-0.5) node[midway,xshift=5pt,yshift=5pt] {\( k_3 \)};

            \end{scope}
            \end{scope}
    \end{tikzpicture}
\end{tabular}
\caption{The YM three-point vertex is the sum of the SD and ASD vertices (top), which come from $(F_{AB})^2$ and $(\bar{F}_{A'B'})^2$, respectively. It can be rewritten as a vertex coming from $(F_{AB})^2$ and a topological one. The topological vertex is on the boundary, which corresponds to the gray region.}
\label{fig:YMthreepoint}
\end{figure}
\noindent Let us project both the SD and ASD contributions onto the definite helicity structure, 
\begin{align}
        \mathcal{W}_3^{\text{SD}} &= -\frac{g}{4E}\frac{k_3}{k_1k_2} \langle \bar{1}\bar{2} \rangle\langle \bar{1}3\rangle \langle \bar{2}3 \rangle\,, &
        \mathcal{W}_3^{\text{ASD}} &=\frac{g}{4E}\frac{k_1+k_2}{k_1k_2k_3}\langle \bar{1}\bar{2} \rangle \langle \bar{1}3\rangle \langle \bar{2}3 \rangle \,.
\end{align}
We see that the difference is exactly the topological piece \eqref{Wtop}
\begin{align}
    \mathcal{W}_3^{\text{SD}}-\mathcal{W}_3^{\text{ASD}} = -\frac{g}{4}\frac{\langle \bar{1}\bar{2}\rangle \langle \bar{1}3\rangle \langle \bar{2}3 \rangle}{k_1k_2k_3} \,.
\end{align}
The relation between SDYM and YM is more transparent from the second half of \eqref{twoway}. In the self-dual limit $(a-b)$ goes to zero and the contribution from the topological vertex disappears. On the other hand, the SD vertex gives exactly the same correlator as SDYM.

%%%%%%%%%%%%%%%%%%%%%%%%%%%%%%%%%%%%%%%%%%%%%%%%%%%%%%%%%%%%%
\subsection{Four-point functions}
\label{sec:fourpoint}
%%%%%%%%%%%%%%%%%%%%%%%%%%%%%%%%%%%%%%%%%%%%%%%%%%%%%%%%%%%%%
Four-point functions are more subtle and contain many contributions in YM theory. We mostly look at the $s$-channel diagram with $+++-$ helicity structure. Other channels can be obtained by simple permutations. While the leading energy pole has to give the flat space amplitude, which vanishes in this case, there is not ``much talking'' between different channels beyond the limit. We use Feynman and axial gauges (Feynman and Lorenz gauges in YM lead to the same expressions for Chalmers--Siegel theory and SDYM). The energy is even less conserved than for the three-point case:
\begin{align}
1^A\bar{1}^{A'}+2^A\bar{2}^{A'}+3^A\bar{3}^{A'}+4^A\bar{4}^{A'}=E\epsilon^{AA'}
\end{align}
and we also define
\begin{align}
    E_\text{L} &= k_1+k_2+k \,, & E_\text{R} &= k_3 +k_4 +k \,,  && k=|\vec k_1+\vec k_2|=|\vec k_3+\vec k_4|\,,
\end{align}
where $k^2\equiv -\frac{1}{2}(k_1+k_2)_{AA'}(k_1+k_2)^{AA'}$ and $k_{AA'}+k\epsilon_{AA'} =(k_1+k_2)_{AA}+k\epsilon_{AA'} \equiv k_A\bar{k}_{A'}$ belong to the internal line of the $s$-channel diagram.

%%%%%%%%%%%%%%%%%%%%%%%%%%%%%%%%%%%%%%%%%%%%%%%%%%%%%%%%%%%%%
\subsubsection{SDYM Lorenz}
\label{sec:4ptSDYM}
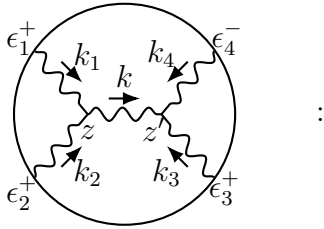
\begin{figure}[h!]
\centering
\begin{tabular}{c c}
    % S-channel Witten diagram
    \begin{tikzpicture}[scale=0.7]
        % Define the boundary circle
        \draw[thick] (0,0) circle (2.1cm);
        
        % Define the vertices inside the bulk
        \coordinate (v1) at (-0.7,0);
        \coordinate (v2) at (0.7,0);
        
        % Define the external legs on the boundary
        \coordinate (a) at (-1.7,1.2);
        \coordinate (b) at (-1.7,-1.2);
        \coordinate (c) at (1.7,1.2);
        \coordinate (d) at (1.7,-1.2);
        
        % Draw the gauge boson lines
        \draw[decorate,decoration={snake},thick] (a) -- (v1);
            \draw[decorate,decoration={snake},thick] (b) -- (v1);
        \draw[decorate,decoration={snake},thick] (c) -- (v2);
        \draw[decorate,decoration={snake},thick] (d) -- (v2);
        
        % Draw the internal propagator 
        \draw[decorate,decoration={snake},thick] (v1) -- (v2);
        
        % Draw parallel momentum arrows
        \draw[-{Latex},thick] (-1.2,1.0) -- (-0.8,0.6) node[midway,xshift=6pt,yshift=6pt] {\( k_1 \)};
        \draw[-{Latex},thick] (-1.2,-1.0) -- (-0.8,-0.6) node[midway,xshift=6pt,yshift=-6pt] {\( k_2 \)};
        \draw[{Latex}-,thick] (0.8,0.6) -- (1.2,1.0) node[midway,xshift=-6pt,yshift=7pt] {\( k_4 \)};
        \draw[{Latex}-,thick] (0.8,-0.6) -- (1.2,-1.0) node[midway,xshift=-4pt,yshift=-6pt] {\( k_3 \)};
        \draw[-{Latex},thick] (-0.3,0.3) -- (0.3,0.3) node[midway,above] {\( k \)};
        
        % Add labels for fields next to endpoints
        \node[xshift=-5pt,yshift=5pt] at (a) {\( \epsilon^+_1 \)};
        \node[xshift=-5pt,yshift=-6pt] at (b) {\( \epsilon^+_2 \)};
        \node[xshift=5pt,yshift=5pt] at (c) {\( \epsilon^-_4 \)};
        \node[xshift=4pt,yshift=-4pt] at (d) {\( \epsilon^+_3 \)};
        
        % Add labels for vertices
        \node[below] at (v1) {\( z \)};
        \node[shift={(-1mm,-2mm)}] at (v2) {\( z' \)};
        
    \end{tikzpicture}
    & \raisebox{14mm}{\(\quad : \quad\)}
\end{tabular}

\vspace{0.5cm}

\[
\begin{aligned}
\mathcal{W}_s = \int_{0}^{\infty} dz\int_0^\infty dz' & \Phi_+^{A,A'}(k_1,z)\Phi_+^{B,B'}(k_2,z)V\fdu{AA',BB',}{EE} \, \langle \Psi_{EE}(-k,z)\Phi_{F,F'}(k,z')\rangle_\text{L} \times\, \\
&\times V\fud{FF'}{,CC',DD}\Phi_+^{C,C'}(k_3,z')\Psi_-^{DD}(k_4,z') 
\end{aligned}
\]

\caption{SDYM Lorenz gauge $\text{AdS}_4$ $s$-channel Witten diagram.}
\label{fig:schannelAdS}
\end{figure}

%%%%%%%%%%%%%%%%%%%%%%%%%%%%%%%%%%%%%%%%%%%%%%%%%%%%%%%%%%%%%
SDYM is the simplest case since there is just one diagram, see Figure \ref{fig:schannelAdS}. It is convenient to isolate the part where the bulk-to-bulk times four boundary-to-bulk propagators are integrated from the rest to find\footnote{Here $\sigma=+1$ is for the Dirichlet-like propagator and $\sigma=-1$ is for the Neumann-like propagator, see also Appendix \ref{subsec:SDYMF}. Two values of $\sigma$ differ by boundary conditions on the unphysical components. Being pure gauge, the $\sigma$-term does not affect the flat limit.}
\begin{align}\label{int2}
    \begin{aligned}
        &\int_0^\infty dz \int_0^\infty dz'  \langle \Psi_{AA}(-k,z)\Phi_{B,B'}(k,z')\rangle_\text{L} e^{-(k_1+k_2)z}e^{-(k_3+k_4)z'} =\\
        &= \frac{\epsilon_{AB}}{EE_\text{L}E_\text{R}}(3_A\bar{3}_{B'}+4_A\bar{4}_{B'})-\frac{1}{4k^2E_\text{L}E_\text{R}}\Big(k_Ak_A\bar{k}_B\bar{k}_{B'}-(1+\sigma)k_A\bar{k}_Ak_B\bar{k}_{B'}\Big)\,.
    \end{aligned}
\end{align}
With the help of (note that, if we were in flat space, $-\langle 12 \rangle \langle \bar{1}\bar{2} \rangle$ would be the Mandelstam $s$)
\begin{align} \label{ELER}
    E_\text{L}E_\text{R}=-\langle 12 \rangle \langle \bar{1}\bar{2} \rangle +EE_\text{L} \,,
\end{align}
we see that the leading energy pole, where the $4$-momentum conservation is effectively restored, reduces to the flat space SDYM propagator
\begin{align}
    \lim_{E \rightarrow 0} E\int e^{-(k_1+k_2)z} \langle \Psi_{AA}(-k,z)\Phi_{B,B'}(k,z')\rangle_{\text{L}} e^{-(k_3+k_4)z'} = \langle \Psi_{AA}(-p)\Phi_{B,B'}(p)\rangle_{\text{flat}}\,.
\end{align}
Now, the vertices and polarizations are exactly the same as in flat space. Therefore, the leading pole does reproduce the flat space amplitude, which is known to vanish after adding the $s$- and $t$-channels. 

\noindent Beyond the leading energy pole approximation we get for the $s$-channel
\begin{align} \label{WSDYM}
    \begin{aligned}
        \mathcal{W}_{s,\text{L}}^{\text{SDYM}} &= -\frac{4g^2}{EE_\text{L}E_\text{R}}\frac{\langle \bar{1}\bar{2}\rangle \langle q_34 \rangle}{\langle q_11\rangle \langle q_2 2\rangle \langle q_3 3\rangle}\Big[\langle q_14 \rangle \langle q_24 \rangle \langle \bar{3}\bar{4} \rangle +\\
        &+\frac{E}{4k^2}\Big(\langle q_1q_2|kk\bar{k}\bar{k}|\bar{3}4\rangle-\frac{1+\sigma}{2}\big(\langle q_1q_2|k\bar{k}\bar{k}k|\bar{3}4\rangle+\langle q_1q_2|\bar{k}k\bar{k}k|\bar{3}4\rangle\big)\Big)\Big] \,,
    \end{aligned}
\end{align}
where $\langle ab|mnrs|cd\rangle=a^Ab^Bc^Cd^Dm_An_Br_Cs_D$. Projecting onto the definite helicity structure we get a somewhat simpler expression
\begin{align}\label{SDYM4ptLorenzAx}
    \begin{aligned}
        \mathcal{W}_{s,\text{L}}^{\text{SDYM}} &= -\frac{g^2}{2}\frac{1}{EE_\text{L}E_\text{R}}\frac{\langle \bar{1}\bar{2}\rangle \langle \bar{3}4 \rangle}{k_1k_2k_3}\Big[\langle \bar{1}4 \rangle \langle \bar{2}4 \rangle \langle \bar{3}\bar{4} \rangle +\frac{E}{4k^2}\langle \bar{1}\bar{2}|kk\bar{k}\bar{k}|\bar{3}4\rangle\Big] +\\
        &-(1+\sigma)\frac{g^2}{8}\frac{k_1-k_2}{k_1k_2k_3}\frac{\langle \bar{1}\bar{2} \rangle^2\langle \bar{3}4\rangle^2}{k^2E_\text{L}}\,.
    \end{aligned}
\end{align}
If we consider color-ordered amplitudes, $s$- and $t$-channels need to be added up to give
\begin{align}
    \begin{aligned}
       & \mathcal{W}_{s,\text{L}}^{\text{SDYM}}+\mathcal{W}_{t,\text{L}}^{\text{SDYM}} =\\
        &= \frac{4g^2}{E_\text{L}^sE_\text{R}^sE_\text{L}^tE_\text{R}^t}\frac{1}{\langle q_11\rangle \langle q_2 2\rangle \langle q_3 3\rangle}\Big[\langle q_14 \rangle \langle q_24 \rangle \langle q_34 \rangle \Big(\langle \bar{1}\bar{2} \rangle \langle \bar{2}\bar{3} \rangle \frac{\langle 23 \rangle \langle \bar{3}\bar{4} \rangle - \langle 12 \rangle \langle \bar{1}\bar{4} \rangle}{E}+\\
        &\qquad\qquad\qquad\qquad\qquad\qquad\qquad\qquad\qquad\qquad\quad+E_\text{L}^s\langle \bar{1}\bar{4} \rangle \langle \bar{2}\bar{3} \rangle -E_\text{L}^t\langle \bar{1}\bar{2} \rangle \langle \bar{3}\bar{4} \rangle  \Big)+\\
        &+\frac{E_\text{L}^sE_\text{R}^s}{4k_t^2}\langle q_14 \rangle \langle \bar{2}\bar{3} \rangle \Big(\langle q_2q_3|k_tk_t\bar{k}_t\bar{k}_t|\bar{1}4\rangle-\frac{1+\sigma}{2}\big(\langle q_2q_3|\bar{k}_tk_t\bar{k}_tk_t|\bar{1}4\rangle+\langle q_2q_3|k_t\bar{k}_t\bar{k}_tk_t|\bar{1}4\rangle\big)\Big)+\\
        &-\frac{E_\text{L}^tE_\text{R}^t}{4k_s^2}\langle \bar{1}\bar{2} \rangle \langle q_34 \rangle\Big( \langle q_1q_2|k_sk_s\bar{k}_s\bar{k}_s|\bar{3}4\rangle-\frac{1+\sigma}{2} \big(\langle q_1q_2|k_s\bar{k}_s\bar{k}_sk_s|\bar{3}4 \rangle +\langle q_1q_2|\bar{k}_sk_s\bar{k}_sk_s|\bar{3}4 \rangle \big)\Big)\Big]\,.
    \end{aligned}
\end{align}
The first term contains the energy pole and clearly vanishes in the flat limit due to the recovered four-momentum conservation. However, the leading term does not vanish beyond the flat limit and gives terms subleading in $E$ that get mixed with other subleading terms that are already present: 
\begin{align}\label{SDYM4ptComplete}
    \begin{aligned}
       & \mathcal{W}_{s,\text{L}}^{\text{SDYM}}+\mathcal{W}_{t,\text{L}}^{\text{SDYM}} =\\
        &= \frac{4g^2}{E_\text{L}^sE_\text{R}^sE_\text{L}^tE_\text{R}^t}\frac{1}{\langle q_11\rangle \langle q_2 2\rangle \langle q_3 3\rangle}\Big[\langle q_14 \rangle \langle q_24 \rangle \langle q_34 \rangle \Big(\langle \bar{1}\bar{2} \rangle \langle \bar{2}\bar{3} \rangle \langle 2\bar{4} \rangle+E_\text{L}^s\langle \bar{1}\bar{4} \rangle \langle \bar{2}\bar{3} \rangle -E_\text{L}^t\langle \bar{1}\bar{2} \rangle \langle \bar{3}\bar{4} \rangle  \Big)+\\
        &+\frac{E_\text{L}^sE_\text{R}^s}{4k_t^2}\langle q_14 \rangle \langle \bar{2}\bar{3} \rangle \Big(\langle q_2q_3|k_tk_t\bar{k}_t\bar{k}_t|\bar{1}4\rangle-\frac{1+\sigma}{2}\big(\langle q_2q_3|\bar{k}_tk_t\bar{k}_tk_t|\bar{1}4\rangle+\langle q_2q_3|k_t\bar{k}_t\bar{k}_tk_t|\bar{1}4\rangle\big)\Big)+\\
        &-\frac{E_\text{L}^tE_\text{R}^t}{4k_s^2}\langle \bar{1}\bar{2} \rangle \langle q_34 \rangle\Big( \langle q_1q_2|k_sk_s\bar{k}_s\bar{k}_s|\bar{3}4\rangle-\frac{1+\sigma}{2} \big(\langle q_1q_2|k_s\bar{k}_s\bar{k}_sk_s|\bar{3}4 \rangle +\langle q_1q_2|\bar{k}_sk_s\bar{k}_sk_s|\bar{3}4 \rangle \big)\Big)\Big] \,.
    \end{aligned}
\end{align}
This form does not feature the energy pole, which makes its vanishing in the flat limit obvious. The leftover of the leading pole piece is the first term. The second and third terms are due to \eqref{ELER}. The final form for the definite helicity structure is
\begin{align}
    \begin{aligned}
       & \mathcal{W}_{s,\text{L}}^{\text{SDYM}}+\mathcal{W}_{t,\text{L}}^{\text{SDYM}} =\\
        &= \frac{g^2}{2E_\text{L}^sE_\text{R}^sE_\text{L}^tE_\text{R}^t}\frac{1}{k_1k_2k_3}\Big[\langle \bar{1}4 \rangle \langle \bar{2}4 \rangle \langle \bar{3}4 \rangle \Big(\langle \bar{1}\bar{2} \rangle \langle \bar{2}\bar{3} \rangle \langle 2\bar{4}\rangle+E_\text{L}^s\langle \bar{1}\bar{4} \rangle \langle \bar{2}\bar{3} \rangle -E_\text{L}^t\langle \bar{1}\bar{2} \rangle \langle \bar{3}\bar{4} \rangle  \Big)+\\
        &+\frac{E_\text{L}^sE_\text{R}^s}{4k_t^2}\langle \bar{1}4 \rangle \langle \bar{2}\bar{3} \rangle \Big(\langle \bar{2}\bar{3}|k_tk_t\bar{k}_t\bar{k}_t|\bar{1}4\rangle+(1+\sigma)E_\text{R}^t\langle\bar{2}\bar{3}\rangle\langle \bar{1}4\rangle(k_2-k_3)\Big)+\\
        &-\frac{E_\text{L}^tE_\text{R}^t}{4k_s^2}\langle \bar{1}\bar{2} \rangle \langle \bar{3}4 \rangle\Big( \langle \bar{1}\bar{2}|k_sk_s\bar{k}_s\bar{k}_s|\bar{3}4\rangle+(1+\sigma) E_\text{R}^s\langle \bar{1}\bar{2} \rangle \langle \bar{3}4 \rangle(k_1-k_2) \Big)\Big]\,.
    \end{aligned}
\end{align}

%%%%%%%%%%%%%%%%%%%%%%%%%%%%%%%%%%%%%%%%%%%%%%%%%%%%%%%%%%%%%
\subsubsection{SDYM Axial}
\label{subsec:SDYMaxial}
%%%%%%%%%%%%%%%%%%%%%%%%%%%%%%%%%%%%%%%%%%%%%%%%%%%%%%%%%%%%%
Since in the SDYM holography ``one half of the dual field'' is a gauge field on the boundary, its correlators are potentially gauge dependent. Therefore, it is interesting to check the result with another gauge, the axial one. It is not necessary to recompute everything since the propagators differ by a pure gauge term
\begin{align} \label{SDYMPropDecomp}
    \begin{aligned}
        \langle \Psi_{AA}(-k,z) \Phi_{BB'}(k,z') \rangle^{\text{SDYM}}_{\text{A}} &= \langle \Psi_{AA}(-k,z)\Phi_{B,B'}(k,z')\rangle_{\text{L}}^{\text{SDYM}}+\nabla_{BB'}^{k,z'}\Delta\eta^\text{L}_{AA}\,,
    \end{aligned}
\end{align}
with $\Delta\eta^\text{L}_{AA}$ defined in \eqref{etaL}. It is only this difference that we compute, i.e.
\begin{align}\label{SDYMAxialcorr}
    \mathcal{W}^{\text{SDYM}}_{s,\text{A}} = \mathcal{W}_{s,\text{L}}^{\text{SDYM}} + \mathcal{W}_{s,\text{gauge}}^{\text{SDYM}}\,,
\end{align}
where the first term is \eqref{SDYM4ptLorenzAx}, i.e. the Feynman gauge answer with axial polarizations plugged in. For the ``gauge variation'' we find 
\begin{align} \label{SDYMaxialcorr}
        \mathcal{W}_{s,\text{gauge}}^{\text{SDYM}} &= -(1-\sigma)\frac{g^2}{8E_\text{L}}\frac{\langle \bar{1}\bar{2} \rangle^2 \langle \bar{3}4 \rangle^2}{k_1k_2k_3}\frac{k_1-k_2}{k^2}\,.
\end{align}
We observe that there is a gauge dependence for $\sigma=-1$, i.e. between the axial gauge and the SD-limit of the Feynman propagator with mixed boundary conditions. This gauge-dependent piece does not have a genuine bulk $1/E$-singularity. It also does not have a $1/(E_\text{L} E_\text{R})$-singularity, only the $1/E_\text{L}$-one, which results from integration by parts of the pure gauge term in \eqref{SDYMPropDecomp}. While the free equations of motion can be used after the integration by parts, it is the boundary effect $\eta|_{z'=0}\neq0$ that leads to \eqref{SDYMaxialcorr}. The same gauge variation is obtained near \eqref{gaugedep} in a different way. The axial gauge is ``more reliable'' as the boundary components $\Phi_i$, $\Psi_i$ have a direct CFT interpretation. As explained in Appendix \ref{subsec:SDYMF}, there is a discrete ambiguity in the SDYM propagator and for $\sigma=+1$ the Feynman and axial gauges agree, as is seen from \eqref{SDYMaxialcorr}.  

%%%%%%%%%%%%%%%%%%%%%%%%%%%%%%%%%%%%%%%%%%%%%%%%%%%%%%%%%%%%%
\subsubsection{cYM Feynman}
\label{subsec:cYM}
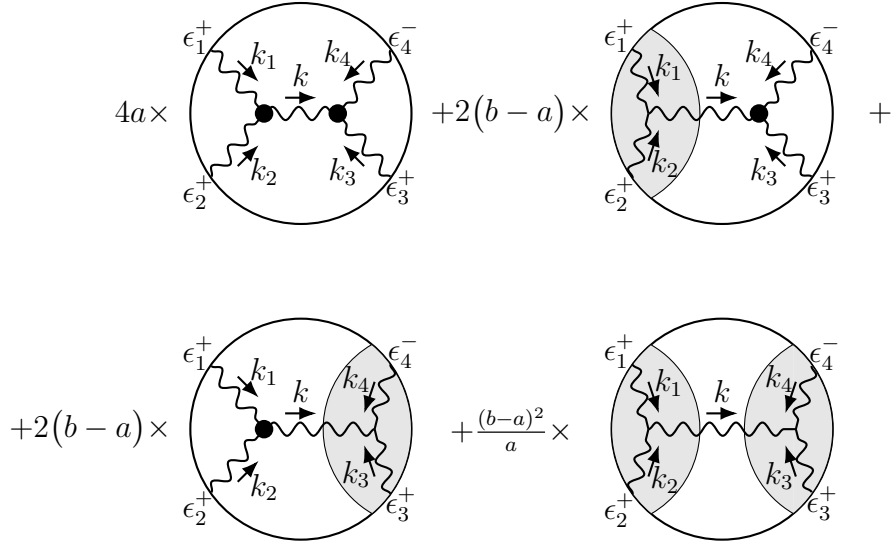
\begin{figure}[h!]
\centering
    % S-channel Witten diagram
    \begin{tikzpicture}[scale=0.7]
        \begin{scope}[xshift=-8cm]
        \node at (-3,0) {$4a\times$};
        % Define the boundary circle
        \draw[thick] (0,0) circle (2.1cm);
        
        % Define the vertices inside the bulk
        \coordinate (v1) at (-0.7,0);
        \coordinate (v2) at (0.7,0);
        
        % Define the external legs on the boundary
        \coordinate (a) at (-1.7,1.2);
        \coordinate (b) at (-1.7,-1.2);
        \coordinate (c) at (1.7,1.2);
        \coordinate (d) at (1.7,-1.2);
        
        % Draw the gauge boson lines
        \draw[decorate,decoration={snake},thick] (a) -- (v1);
        \draw[decorate,decoration={snake},thick] (b) -- (v1);
        \draw[decorate,decoration={snake},thick] (c) -- (v2);
        \draw[decorate,decoration={snake},thick] (d) -- (v2);
        
        % Draw the internal propagator 
        \draw[decorate,decoration={snake},thick] (v1) -- (v2);
        
        % Draw parallel momentum arrows
        \draw[-{Latex},thick] (-1.2,1.0) -- (-0.8,0.6) node[midway,xshift=6pt,yshift=6pt] {\( k_1 \)};
        \draw[-{Latex},thick] (-1.2,-1.0) -- (-0.8,-0.6) node[midway,xshift=6pt,yshift=-6pt] {\( k_2 \)};
        \draw[{Latex}-,thick] (0.8,0.6) -- (1.2,1.0) node[midway,xshift=-6pt,yshift=7pt] {\( k_4 \)};
        \draw[{Latex}-,thick] (0.8,-0.6) -- (1.2,-1.0) node[midway,xshift=-4pt,yshift=-6pt] {\( k_3 \)};
        \draw[-{Latex},thick] (-0.3,0.3) -- (0.3,0.3) node[midway,above] {\( k \)};
        
        % Add labels for fields next to endpoints
        \node[xshift=-5pt,yshift=5pt] at (a) {\( \epsilon^+_1 \)};
        \node[xshift=-5pt,yshift=-6pt] at (b) {\( \epsilon^+_2 \)};
        \node[xshift=5pt,yshift=5pt] at (c) {\( \epsilon^-_4 \)};
        \node[xshift=4pt,yshift=-4pt] at (d) {\( \epsilon^+_3 \)};

        \fill (v1) circle (5pt);
        \fill (v2) circle (5pt);
        \end{scope}

        \begin{scope}
        \node at (-4,0) {$+2\big(b-a\big)\times$};

            % Define the boundary circle
        \draw[thick] (0,0) circle (2.1cm);
        
        % Define the vertices inside the bulk
        \coordinate (v1) at (-1.4,0);
        \coordinate (v2) at (0.7,0);
        
        % Define the external legs on the boundary
        \coordinate (a) at (-1.7,1.23);
        \coordinate (b) at (-1.7,-1.23);
        \coordinate (c) at (1.7,1.2);
        \coordinate (d) at (1.7,-1.2);

        \filldraw[fill=gray!20, draw=black,rotate=-90] 
                    (1.61,-1.35) arc[start angle=30, end angle=150, radius=1.855cm];
                \filldraw[fill=gray!20, draw=black,rotate=-90] 
                (-1.61,-1.35) arc[start angle=220, end angle=320, radius=2.1cm];

                % Draw parallel momentum arrows
            \draw[-{Latex},thick,rotate=-90] (-0.9,-1.4) -- (-0.3,-1.2) node[midway,xshift=0.2 cm,yshift=0.2 cm] {\( k_1 \)};
            \draw[-{Latex},thick,rotate=-90] (0.9,-1.4) -- (0.3,-1.2) node[midway,xshift=0.15 cm,yshift=-0.25 cm] {\( k_2 \)};
        
        % Draw the gauge boson lines
        \draw[decorate,decoration={snake},thick] (a) -- (v1);
        \draw[decorate,decoration={snake},thick] (b) -- (v1);
        \draw[decorate,decoration={snake},thick] (c) -- (v2);
        \draw[decorate,decoration={snake},thick] (d) -- (v2);
        
        % Draw the internal propagator 
        \draw[decorate,decoration={snake},thick] (v1) -- (v2);
        
        % Draw parallel momentum arrows
        \draw[{Latex}-,thick] (0.8,0.6) -- (1.2,1.0) node[midway,xshift=-6pt,yshift=7pt] {\( k_4 \)};
        \draw[{Latex}-,thick] (0.8,-0.6) -- (1.2,-1.0) node[midway,xshift=-4pt,yshift=-6pt] {\( k_3 \)};
        \draw[-{Latex},thick] (-0.3,0.3) -- (0.3,0.3) node[midway,above] {\( k \)};
        
        % Add labels for fields next to endpoints
        \node[xshift=-5pt,yshift=5pt] at (a) {\( \epsilon^+_1 \)};
        \node[xshift=-5pt,yshift=-6pt] at (b) {\( \epsilon^+_2 \)};
        \node[xshift=5pt,yshift=5pt] at (c) {\( \epsilon^-_4 \)};
        \node[xshift=4pt,yshift=-4pt] at (d) {\( \epsilon^+_3 \)};

        \fill (v2) circle (5pt);
            \node at (3,0) {$+$};
        \end{scope}
        
        \begin{scope}[yshift=-6cm]
            \begin{scope}[xshift=-8cm,rotate=180]
                    \node at (-4,0) {$+\frac{(b-a)^2}{a}\times$};
        
                    % Define the boundary circle
                \draw[thick] (0,0) circle (2.1cm);
                
                % Define the vertices inside the bulk
                \coordinate (v1) at (-1.4,0);
                \coordinate (v2) at (0.7,0);
                
                % Define the external legs on the boundary
                \coordinate (a) at (-1.7,1.23);
                \coordinate (b) at (-1.7,-1.23);
                \coordinate (c) at (1.7,1.2);
                \coordinate (d) at (1.7,-1.2);

                \filldraw[fill=gray!20, draw=black,rotate=-90] 
                    (1.61,-1.35) arc[start angle=30, end angle=150, radius=1.855cm];
                \filldraw[fill=gray!20, draw=black,rotate=-90] 
                (-1.61,-1.35) arc[start angle=220, end angle=320, radius=2.1cm];
        
                        % Draw parallel momentum arrows
                    \draw[-{Latex},thick,rotate=-90] (-0.9,-1.4) -- (-0.3,-1.2) node[midway,xshift=-5pt,yshift=-5pt] {\( k_3 \)};
                    \draw[-{Latex},thick,rotate=-90] (0.9,-1.4) -- (0.3,-1.2) node[midway,xshift=-5pt,yshift=7pt] {\( k_4 \)};
                
                % Draw the gauge boson lines
                \draw[decorate,decoration={snake},thick] (a) -- (v1);
                \draw[decorate,decoration={snake},thick] (b) -- (v1);
                \draw[decorate,decoration={snake},thick] (c) -- (v2);
                \draw[decorate,decoration={snake},thick] (d) -- (v2);
                
                % Draw the internal propagator 
                \draw[decorate,decoration={snake},thick] (v1) -- (v2);
                
                % Draw parallel momentum arrows
                \begin{scope}[rotate=180]
                    \draw[-{Latex},thick] (-1.2,1.0) -- (-0.8,0.6) node[midway,xshift=6pt,yshift=6pt] {\( k_1 \)};
                    \draw[-{Latex},thick] (-1.2,-1.0) -- (-0.8,-0.6) node[midway,xshift=6pt,yshift=-6pt] {\( k_2 \)};
                    \draw[-{Latex},thick] (-0.3,0.3) -- (0.3,0.3) node[midway,above] {\( k \)};  
                \end{scope}

                % Add labels for fields next to endpoints
        \node[xshift=4pt,yshift=-4pt] at (a) {\( \epsilon^+_3 \)};
        \node[xshift=5pt,yshift=5pt] at (b) {\( \epsilon^-_4 \)};
        \node[xshift=-5pt,yshift=-6pt] at (c) {\( \epsilon^+_2 \)};
        \node[xshift=-5pt,yshift=5pt] at (d) {\( \epsilon^+_1 \)};

                \fill (v2) circle (5pt);
                    \node at (4,0) {$+ 2\big(b-a\big)\times$};
                \end{scope}
                \begin{scope}
                                % Define the boundary circle
                    \draw[thick] (0,0) circle (2.1cm);

                    \filldraw[fill=gray!20, draw=black,rotate=90] 
                    (1.61,-1.35) arc[start angle=30, end angle=150, radius=1.855cm];
                \filldraw[fill=gray!20, draw=black,rotate=90] 
                (-1.61,-1.35) arc[start angle=220, end angle=320, radius=2.1cm];
                    
                    % Define the vertices inside the bulk
                    \coordinate (v1) at (-1.4,0);
                    \coordinate (v2) at (1.4,0);
                    
                    % Define the external legs on the boundary
                    \coordinate (a) at (-1.7,1.23);
                    \coordinate (b) at (-1.7,-1.23);
                    \coordinate (c) at (1.7,1.2);
                    \coordinate (d) at (1.7,-1.2);

                    \filldraw[fill=gray!20, draw=black,rotate=-90] 
                    (1.61,-1.35) arc[start angle=30, end angle=150, radius=1.855cm];
                \filldraw[fill=gray!20, draw=black,rotate=-90] 
                (-1.61,-1.35) arc[start angle=220, end angle=320, radius=2.1cm];
            
                            % Draw parallel momentum arrows
                        \draw[-{Latex},thick,rotate=-90] (-0.9,-1.4) -- (-0.3,-1.2) node[midway,xshift=0.2 cm,yshift=0.2 cm] {\( k_1 \)};
                        \draw[-{Latex},thick,rotate=-90] (0.9,-1.4) -- (0.3,-1.2) node[midway,xshift=0.15 cm,yshift=-0.25 cm] {\( k_2 \)};
                    
                    % Draw the gauge boson lines
                    \draw[decorate,decoration={snake},thick] (a) -- (v1);
                    \draw[decorate,decoration={snake},thick] (b) -- (v1);
                    \draw[decorate,decoration={snake},thick] (c) -- (v2);
                    \draw[decorate,decoration={snake},thick] (d) -- (v2);
                    
                    % Draw the internal propagator 
                    \draw[decorate,decoration={snake},thick] (v1) -- (v2);
                    
                    % Draw parallel momentum arrows
                    \draw[-{Latex},thick,rotate=90] (-0.9,-1.4) -- (-0.3,-1.2) node[midway,xshift=-0.15 cm,yshift=-0.2 cm] {\( k_3 \)};
                    \draw[-{Latex},thick,rotate=90] (0.9,-1.4) -- (0.3,-1.2) node[midway,xshift=-0.15 cm,yshift=0.25 cm] {\( k_4 \)};
                    \draw[-{Latex},thick] (-0.3,0.3) -- (0.3,0.3) node[midway,above] {\( k \)}; 
                    
                    % Add labels for fields next to endpoints
                    \node[xshift=-5pt,yshift=5pt] at (a) {\( \epsilon^+_1 \)};
                    \node[xshift=-5pt,yshift=-6pt] at (b) {\( \epsilon^+_2 \)};
                    \node[xshift=5pt,yshift=5pt] at (c) {\( \epsilon^-_4 \)};
                    \node[xshift=4pt,yshift=-4pt] at (d) {\( \epsilon^+_3 \)};
                \end{scope}
        \end{scope}
        
    \end{tikzpicture}
\caption{The full exchange diagram splits into SD-SD ($\mathcal{W}^{\text{SD}}$), SD-Top ($\mathcal{T}^{\text{R}}$), Top-SD ($\mathcal{T}^{\text{R}}$) and Top-Top ($\mathcal{W}^{\text{L,R}}$) components.}
\label{fig:exchangeDecomp}
\end{figure}

%%%%%%%%%%%%%%%%%%%%%%%%%%%%%%%%%%%%%%%%%%%%%%%%%%%%%%%%%%%%%
There are many diagrams in cYM that contribute to the four-point function and it is important to represent them in a clever way to establish its relation to Chalmers--Siegel theory and, then, to SDYM. As before, we take cYM plus a theta-term representation of the action. There are two types of cubic vertices: the SD vertex and the topological one. Accordingly, there are $4$ different exchange diagrams, see Figure \ref{fig:exchangeDecomp}. There is also a single quartic vertex, see Figure \ref{fig:contactAdS}.
Therefore, the $s$-channel receives five contributions\footnote{The factors $a$, $b$ in front of the YM and theta-terms are made explicit everywhere. In particular, they do not enter the propagators and vertices on the diagrams. There is $a^{-1}$ coming the bulk-to-bulk propagators. On external lines $a^{-1}$ is simply dropped as it is the same for all lines. } 
\begin{align}\label{W4Decomp}
    \mathcal{W}_s = 4a\mathcal{W}_s^{\text{SD}}+2\big(b-a\big)\Big(\mathcal{T}_s^\text{L}+\mathcal{T}_s^\text{R}\Big)+\frac{\big(b-a\big)^2}{a}\mathcal{T}_s^{\text{L,R}}+a\mathcal{W}_{s,\text{contact}}\,.
\end{align}
We first try to massage the diagrams in such a way as to identify them with diagrams coming for the Chalmers--Siegel action. The SD-vertex contains one derivative that can hit any of the three lines. We observe that whenever it is an external line saturated by $\Phi_+$, the derivative annihilates it; whenever the derivative acts on a $\Phi_-$ external line, it produces the $\Psi_-$-line; whenever one derivative acts on the $\langle \Phi \Phi\rangle$-propagator it gives the $\langle \Psi \Phi\rangle$-propagator of the Chalmers--Siegel theory; whenever both derivatives hit the $\langle \Phi \Phi\rangle$-propagator a two-point function $\langle F_{AA}(-k,z) F_{BB}(k,z')\rangle$ is produced. 

\begin{figure}[h!]
\centering
    % S-channel Witten diagram
    \begin{tikzpicture}[scale=0.7]
        \begin{scope}[xshift=-12cm]
        % Define the boundary circle
        \draw[thick] (0,0) circle (2.1cm);
        
        % Define the vertices inside the bulk
        \coordinate (v1) at (-0.7,0);
        \coordinate (v2) at (0.7,0);
        
        % Define the external legs on the boundary
        \coordinate (a) at (-1.7,1.2);
        \coordinate (b) at (-1.7,-1.2);
        \coordinate (c) at (1.7,1.2);
        \coordinate (d) at (1.7,-1.2);
        
        % Draw the gauge boson lines
        \draw[decorate,decoration={snake},thick] (a) -- (v1);
        \draw[decorate,decoration={snake},thick] (b) -- (v1);
        \draw[decorate,decoration={snake},thick] (c) -- (v2);
        \draw[decorate,decoration={snake},thick] (d) -- (v2);
        
        % Draw the internal propagator 
        \draw[decorate,decoration={snake},thick] (v1) -- (v2);
        
        % Draw parallel momentum arrows
        \draw[-{Latex},thick] (-1.2,1.0) -- (-0.8,0.6) node[midway,xshift=6pt,yshift=6pt] {\( k_1 \)};
        \draw[-{Latex},thick] (-1.2,-1.0) -- (-0.8,-0.6) node[midway,xshift=6pt,yshift=-6pt] {\( k_2 \)};
        \draw[{Latex}-,thick] (0.8,0.6) -- (1.2,1.0) node[midway,xshift=-6pt,yshift=7pt] {\( k_4 \)};
        \draw[{Latex}-,thick] (0.8,-0.6) -- (1.2,-1.0) node[midway,xshift=-4pt,yshift=-6pt] {\( k_3 \)};
        \draw[-{Latex},thick] (-0.3,0.3) -- (0.3,0.3) node[midway,above] {\( k \)};
        
        % Add labels for fields next to endpoints
        \node[xshift=-5pt,yshift=5pt] at (a) {\( \epsilon^+_1 \)};
        \node[xshift=-5pt,yshift=-6pt] at (b) {\( \epsilon^+_2 \)};
        \node[xshift=5pt,yshift=5pt] at (c) {\( \epsilon^-_4 \)};
        \node[xshift=4pt,yshift=-4pt] at (d) {\( \epsilon^+_3 \)};

        \fill (v1) circle (5pt);
        \fill (v2) circle (5pt);
        \end{scope}
        
        \begin{scope}[xshift=-6cm]
            % Define the boundary circle
        \draw[thick] (0,0) circle (2.1cm);
        
        % Define the vertices inside the bulk
        \coordinate (v1) at (-0.7,0);
        \coordinate (v2) at (0.7,0);
        
        % Define the external legs on the boundary
        \coordinate (a) at (-1.7,1.2);
        \coordinate (b) at (-1.7,-1.2);
        \coordinate (c) at (1.7,1.2);
        \coordinate (d) at (1.7,-1.2);
        
        % Draw the gauge boson lines
        \draw[decorate,decoration={snake},thick] (a) -- (v1);
        \draw[decorate,decoration={snake},thick] (b) -- (v1);
        \draw[decorate,decoration={snake},thick] (c) -- (v2);
        \draw[decorate,decoration={snake},thick] (d) -- (v2);
        
        % Draw the internal propagator 
        \draw[decorate,decoration={snake},thick] (v1) -- (v2);
        
        % Draw parallel momentum arrows
        \draw[-{Latex},thick] (-1.2,1.0) -- (-0.8,0.6) node[midway,xshift=6pt,yshift=6pt] {\( k_1 \)};
        \draw[-{Latex},thick] (-1.2,-1.0) -- (-0.8,-0.6) node[midway,xshift=6pt,yshift=-6pt] {\( k_2 \)};
        \draw[{Latex}-,thick] (0.8,0.6) -- (1.2,1.0) node[midway,xshift=-6pt,yshift=7pt] {\( k_4 \)};
        \draw[{Latex}-,thick] (0.8,-0.6) -- (1.2,-1.0) node[midway,xshift=-4pt,yshift=-6pt] {\( k_3 \)};
        \draw[-{Latex},thick] (-0.3,0.3) -- (0.3,0.3) node[midway,above] {\( k \)};
        
        % Add labels for fields next to endpoints
        \node[xshift=-5pt,yshift=5pt] at (a) {\( \epsilon^+_1 \)};
        \node[xshift=-5pt,yshift=-6pt] at (b) {\( \epsilon^+_2 \)};
        \node[xshift=5pt,yshift=5pt] at (c) {\( \epsilon^-_4 \)};
        \node[xshift=4pt,yshift=-4pt] at (d) {\( \epsilon^+_3 \)};

        \fill[blue] (-0.4,0) circle (5pt);
        \fill[blue] (0.9,0.2) circle (5pt);
        
        \node at (3,0) {$+$}; 
        \node at (-3,0) {$=$}; 
        
        \end{scope}
        \begin{scope}[yshift=0cm]
            % Define the boundary circle
        \draw[thick] (0,0) circle (2.1cm);
        
        % Define the vertices inside the bulk
        \coordinate (v1) at (-0.7,0);
        \coordinate (v2) at (0.7,0);
        
        % Define the external legs on the boundary
        \coordinate (a) at (-1.7,1.2);
        \coordinate (b) at (-1.7,-1.2);
        \coordinate (c) at (1.7,1.2);
        \coordinate (d) at (1.7,-1.2);
        
        % Draw the gauge boson lines
        \draw[decorate,decoration={snake},thick] (a) -- (v1);
        \draw[decorate,decoration={snake},thick] (b) -- (v1);
        \draw[decorate,decoration={snake},thick] (c) -- (v2);
        \draw[decorate,decoration={snake},thick] (d) -- (v2);
        
        % Draw the internal propagator 
        \draw[decorate,decoration={snake},thick] (v1) -- (v2);
        
        % Draw parallel momentum arrows
        \draw[-{Latex},thick] (-1.2,1.0) -- (-0.8,0.6) node[midway,xshift=6pt,yshift=6pt] {\( k_1 \)};
        \draw[-{Latex},thick] (-1.2,-1.0) -- (-0.8,-0.6) node[midway,xshift=6pt,yshift=-6pt] {\( k_2 \)};
        \draw[{Latex}-,thick] (0.8,0.6) -- (1.2,1.0) node[midway,xshift=-6pt,yshift=7pt] {\( k_4 \)};
        \draw[{Latex}-,thick] (0.8,-0.6) -- (1.2,-1.0) node[midway,xshift=-4pt,yshift=-6pt] {\( k_3 \)};
        \draw[-{Latex},thick] (-0.3,0.3) -- (0.3,0.3) node[midway,above] {\( k \)};
        
        % Add labels for fields next to endpoints
        \node[xshift=-5pt,yshift=5pt] at (a) {\( \epsilon^+_1 \)};
        \node[xshift=-5pt,yshift=-6pt] at (b) {\( \epsilon^+_2 \)};
        \node[xshift=5pt,yshift=5pt] at (c) {\( \epsilon^-_4 \)};
        \node[xshift=4pt,yshift=-4pt] at (d) {\( \epsilon^+_3 \)};

        \fill[blue] (-0.4,0) circle (5pt);
        \fill[blue] (0.5,0) circle (5pt);
        \end{scope}
        
    \end{tikzpicture}

\caption{The YM $s$-channel diagram splits into two diagrams, depending on where the derivative (blue bullet) lands. The left diagram also belongs to SDYM/Chalmers--Siegel.}
\label{fig:cYM1}
\end{figure}
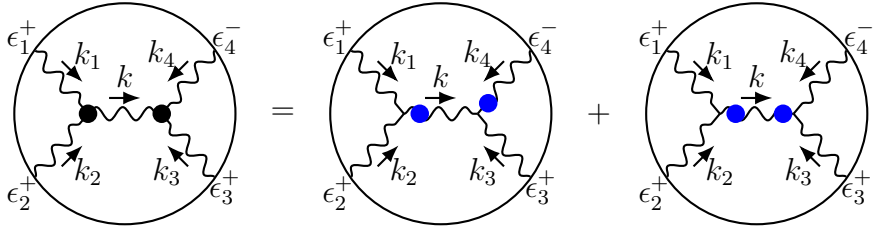

Taking the rules above into account the SD-SD exchange diagram can be further subdivided into two contributions
\begin{align} \label{Wdecomp}       \mathcal{W}_s^{\text{SD}}=\mathcal{W}_s^{1}+\mathcal{W}_s^{2} \,,
\end{align}
see also Figure \ref{fig:cYM1}. Diagram $\mathcal{W}_s^{1}$ is exactly the one present in SDYM and Chalmers--Siegel theory with the $\langle \Psi \Phi\rangle$ bulk-to-bulk propagator. The second diagram $\mathcal{W}_s^2$ is absent in SDYM and has the $\langle F_{AA}(-k,z) F_{BB}(k,z')\rangle$ two-point function instead of the propagator. The expressions for the two subdiagrams are
\begin{align}\notag% \label{w1w2}
    \begin{aligned}
        \mathcal{W}_s^{1} &= -g^2\int\Phi_{+}^{A,A'}(k_1,z)\Phi_+\fud{A,}{A'}(k_2,z)\langle \Psi_{AA}(-k,z)\Phi_{B,B'}(k,z') \rangle \Phi\fdu{+B,}{B'}(k_3,z')\Psi_-^{BB}(k_4,z')\,,\\
        \mathcal{W}_s^{2} &= g^2\int\Phi_{+}^{AA'}(k_1,z)\Phi_+\fud{A,}{A'}(k_2,z)\langle F_{AA}(-k,z)F_{BB}(k,z') \rangle \Phi_+^{BB'}(k_3,z')\Phi\fdud{-}{B,}{B'}(k_4,z') \,.
    \end{aligned}
\end{align}
In Chalmers--Siegel theory it turns out that for non-self-dual boundary conditions there is a non-vanishing $\langle \Psi \Psi\rangle$-propagator that solves a homogeneous equation. The $\langle F F\rangle$ two-point function has both the homogeneous and the contact pieces, the relation being 
\begin{align}\label{FFvsPsiPsi}
    \langle F^{AA}(-k,z) F^{BB}(k,z')\rangle&= -\delta(z-z')\epsilon^{AB}\epsilon^{AB} +\langle \Psi^{AA}(-k,z) \Psi^{BB}(k,z')\rangle\,.
\end{align}
The contact diagram is simply 
\begin{align}
    \begin{aligned}
        \mathcal{W}_{s,\text{contact}}&=\int \Phi_+^{A,A'}(k_1,z)\Phi_+^{B,B'}(k_2,z)V^{\text{quartic}}_{AA',BB',CC',DD'}\Phi_+^{C,C'}(k_3,z)\Phi_-^{D,D'}(k_4,z) \,.
    \end{aligned}
\end{align}

\paragraph{Dirichlet boundary conditions.} Let us for now forget about the diagrams containing the topological vertices. Indeed, since the Dirichlet bulk-to-bulk propagator vanishes on the boundary, the topological diagrams will not make any contributions. For the Dirichlet boundary condition, the Chern--Simons term modifies only two- and three-point functions, but not beyond. Including the coefficients provided in \eqref{W4Decomp}, the exchange diagrams give
\begin{align}
    \begin{aligned}
        4a\mathcal{W}^1_{s,\text{D}} &= -\frac{ag^2}{2EE_\text{L}E_\text{R}}\frac{\langle \bar{1}\bar{2}\rangle \langle \bar{3}4\rangle}{k_1k_2k_3}\Big(\langle \bar{1}4\rangle\langle\bar{2}4\rangle\langle \bar{3}\bar{4}\rangle-\frac{E}{4k}\big(\langle \bar{1}4 \rangle \langle \bar{2}|k\bar{k}|\bar{3}\rangle+\langle \bar{2}4 \rangle \langle \bar{1}|k\bar{k}|\bar{3}\rangle+\\
        &\qquad\qquad\qquad\qquad\qquad\qquad\qquad\qquad\qquad\qquad-\langle \bar{1}\bar{3} \rangle \langle 4|k\bar{k}|\bar{2}\rangle-\langle \bar{2}\bar{3} \rangle \langle 4|k\bar{k}|\bar{1}\rangle\big)\Big) \,,\\
        4a\mathcal{W}^2_{s,\text{D}} &= \frac{ag^2}{8}\frac{\langle \bar{1}\bar{2}\rangle \langle \bar{3}4\rangle}{k_1k_2k_3k_4}\Big(\underbrace{\frac{\langle \bar{1}\bar{3} \rangle \langle \bar{2}4\rangle +\langle \bar{1}4 \rangle\langle \bar{2}\bar{3}\rangle}{E}}_{\text{contact}}  \underbrace{-\frac{\langle \bar{1}\bar{2}|\bar{k}\bar{k}kk|\bar{3}4\rangle}{kE_\text{L}E_\text{R}}}_{\langle\Psi\Psi\rangle}\Big) \,.
    \end{aligned}
\end{align}
Here we defined $\langle a|mn|b\rangle = a^Am_Am_Bb^B$. In the last line we have split the $\langle FF \rangle$ contribution into $\langle \Psi\Psi\rangle$ and a contact part. The reason is that the contact contribution is canceled by the contact diagram 
\begin{align}
    a\mathcal{W}_{s}^{\text{contact}} &=-\frac{ag^2}{8}\frac{\langle \bar{1}\bar{2}\rangle \langle \bar{3}4\rangle}{k_1k_2k_3k_4}\frac{\langle \bar{1}\bar{3} \rangle \langle \bar{2}4\rangle +\langle \bar{1}4 \rangle\langle \bar{2}\bar{3}\rangle}{E} \,.
\end{align}
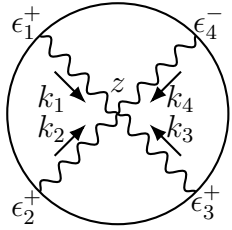
\begin{figure}[h!]
\centering
\begin{tabular}{c c}
    % S-channel Witten diagram
    \begin{tikzpicture}[scale=0.7]
        % Define the boundary circle
        \draw[thick] (0,0) circle (2.1cm);
        
        % Define the vertices inside the bulk
        \coordinate (v) at (0,0);
        
        % Define the external legs on the boundary
        \coordinate (a) at (-1.48,1.48);
        \coordinate (b) at (-1.48,-1.48);
        \coordinate (c) at (1.48,1.48);
        \coordinate (d) at (1.48,-1.48);
        
        % Draw the gauge boson lines
        \draw[decorate,decoration={snake},thick] (a) -- (v);
        \draw[decorate,decoration={snake},thick] (b) -- (v);
        \draw[decorate,decoration={snake},thick] (c) -- (v);
        \draw[decorate,decoration={snake},thick] (d) -- (v);

        % Draw parallel momentum arrows
        \draw[-{Latex},thick] (-1.2,0.8) -- (-0.6,0.2) node[midway,xshift=-0.25cm,yshift=-0.15cm] {\( k_1 \)};
        \draw[-{Latex},thick] (-1.2,-0.8) -- (-0.6,-0.2) node[midway,xshift=-0.25cm,yshift=0.15cm] {\( k_2 \)};
        \draw[{Latex}-,thick] (0.6,0.2) -- (1.2,0.8) node[midway,xshift=0.2cm,yshift=-0.15cm] {\( k_4 \)};
        \draw[{Latex}-,thick] (0.6,-0.2) -- (1.2,-0.8) node[midway,xshift=0.2cm,yshift=0.15cm] {\( k_3 \)};
        
        % Add labels for fields next to endpoints
        \node[xshift=-5pt,yshift=5pt] at (a) {\( \epsilon^+_1 \)};
        \node[xshift=-5pt,yshift=-6pt] at (b) {\( \epsilon^+_2 \)};
        \node[xshift=5pt,yshift=5pt] at (c) {\( \epsilon^-_4 \)};
        \node[xshift=4pt,yshift=-4pt] at (d) {\( \epsilon^+_3 \)};
        
        % Add labels for vertices
        \node[above] at (0,0.2) {\( z \)};
        
    \end{tikzpicture}
\end{tabular}

\caption{YM four-point contact diagram.}
\label{fig:contactAdS}
\end{figure}%

\noindent The rest gives exactly the result in the Chalmers--Siegel theory where both $\Psi\Phi$ and $\Psi\Psi$ bulk-to-bulk propagators are present. One can also interpret the answer as a result of computing the boundary limit of the $\langle \Psi \Psi \Psi \Phi\rangle $ four-point function. We also showed that the Dirichlet result is gauge independent. Indeed, it turns out that the ``gauge parameters'' that determine the difference between the Feynman and axial gauge propagators vanish on the boundary. Therefore, there are no boundary terms resulting from integration by parts. The Dirichlet results can be shown to agree with \cite{Armstrong:2020woi}.

\begin{figure}[h!]
\centering
    \begin{tikzpicture}[scale=0.7]        
                    % Define the boundary circle
                \draw[thick] (0,0) circle (2.1cm);
                
                % Define the vertices inside the bulk
                \coordinate (v1) at (-0.7,0);
                \coordinate (v2) at (0.7,0);

                \coordinate (v3) at (-0.4,1);
                \coordinate (v4) at (0.4,1);
                
                % Define the external legs on the boundary
                \coordinate (a) at (-1.7,1.23);
                \coordinate (b) at (-1.7,-1.23);
                \coordinate (c) at (1.7,1.2);
                \coordinate (d) at (1.7,-1.2);

                \filldraw[fill=gray!20, draw=black,rotate=180] 
                    (1.61,-1.35) arc[start angle=40, end angle=140, radius=2.1cm];
                \filldraw[fill=gray!20, draw=black,rotate=-180] 
                (-1.61,-1.35) arc[start angle=220, end angle=320, radius=2.1cm];

                % Draw the gauge boson lines
                \draw[decorate,decoration={snake},thick] (a) -- (v1);
                \draw[decorate,decoration={snake},thick] (b) -- (v1);
                \draw[decorate,decoration={snake},thick] (c) -- (v2);
                \draw[decorate,decoration={snake},thick] (d) -- (v2);
                
                % Draw the internal propagator 
                \draw[decorate,decoration={snake},thick] (v1) -- (v3);

                \draw[decorate,decoration={snake},thick] (v4) -- (v2);

                \draw[decorate,decoration={snake},thick] (v3) -- (v4);
    \end{tikzpicture}
\caption{The difference between Dirichlet and Neumann propagators reduces to a product of two three-point functions, glued by a ``two-point'' function along the boundary.}
\label{fig:NDdiff}
\end{figure}
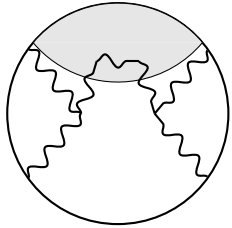

\paragraph{Neumann boundary conditions.} In the Feynman gauge we find the same terms as for Dirichlet boundary conditions, but the sign of the underlined ones is flipped as compared to the Dirichlet case. We have
\begin{align}
    \begin{aligned}
        4a\mathcal{W}^1_{s,\text{N}} &= -\frac{ag^2}{2EE_\text{L}E_\text{R}}\frac{\langle \bar{1}\bar{2}\rangle \langle \bar{3}4\rangle}{k_1k_2k_3}\Big(\langle \bar{1}4\rangle\langle\bar{2}4\rangle\langle \bar{3}\bar{4}\rangle-\frac{E}{4k}\big(\langle \bar{1}4 \rangle \langle \bar{2}|k\bar{k}|\bar{3}\rangle+\langle \bar{2}4 \rangle \langle \bar{1}|k\bar{k}|\bar{3}\rangle+\\
        &\qquad\qquad\qquad\qquad\qquad\qquad\qquad\qquad\qquad\qquad+\underline{\langle \bar{1}\bar{3} \rangle \langle 4|k\bar{k}|\bar{2}\rangle+\langle \bar{2}\bar{3} \rangle \langle 4|k\bar{k}|\bar{1}\rangle}\big)\Big) \,,\\
        4a\mathcal{W}^2_{s,\text{N}} &= \frac{ag^2}{8}\frac{\langle \bar{1}\bar{2}\rangle \langle \bar{3}4\rangle}{k_1k_2k_3k_4}\Big(\frac{\langle \bar{1}\bar{3} \rangle \langle \bar{2}4\rangle +\langle \bar{1}4 \rangle\langle \bar{2}\bar{3}\rangle}{E} +\underline{\frac{\langle \bar{1}\bar{2}|\bar{k}\bar{k}kk|\bar{3}4\rangle}{kE_\text{L}E_\text{R}}}\Big) \,.
    \end{aligned}
\end{align}
Therefore, it is easier to compute the difference
$4a\mathcal{W}_{s,\text{N}-\text{D}}\equiv 4a\big(\mathcal{W}_{s,\text{N}}-\mathcal{W}_{s,\text{D}}\big)$ between the Neumann and Dirichlet correlators
\begin{align}
    \begin{aligned}
        4a\mathcal{W}^1_{s,\text{N}-\text{D}} &= \frac{ag^2}{4kE_\text{L}E_\text{R}}\frac{\langle \bar{1}\bar{2}\rangle\langle \bar{3}4\rangle}{k_1k_2k_3}\Big(\langle \bar{1}\bar{3} \rangle \langle 4|k\bar{k}|\bar{2}\rangle+\langle\bar{2}\bar{3}\rangle\langle4|k\bar{k}|\bar{1}\rangle\Big) \,,\\
        4a\mathcal{W}^2_{s,\text{N}-\text{D}} &= \frac{ag^2}{4kE_\text{L}E_\text{R}}\frac{\langle \bar{1}\bar{2}\rangle\langle \bar{3}4\rangle}{k_1k_2k_3k_4}\langle \bar{1}\bar{2}|\bar{k}\bar{k}kk|\bar{3}4\rangle \,.
    \end{aligned}
\end{align}
The difference has a purely CFT interpretation \cite{Hartman:2006dy, Giombi:2011ya}: it comes from gluing two three-point functions along the boundary, see Figure \ref{fig:NDdiff}.

\paragraph{Mixed boundary conditions.} The propagator for mixed boundary conditions can be decomposed as in \eqref{Gs}, with the various components given in \eqref{CSmixed}. Accordingly, the correlator is decomposed as
\begin{align}
    \mathcal{W}=\mathcal{W}_\text{inh}+\mathcal{W}_\text{pure gauge}+\mathcal{W}_\gamma\,.
\end{align}
The respective components of the correlators $\mathcal{W}^1_s$ and $\mathcal{W}^2_s$ read\footnote{Note that the external lines have been normalized ``to 1'', i.e. the $\gamma$-dependent numerical factors are rescaled away. Therefore, $\mathcal{W}^{1,2}_{s,\gamma}$ is the genuine bulk difference between the Neumann and $\gamma$-mixed boundary conditions.}
\begin{align}
    \begin{aligned}
        4a\mathcal{W}^1_{s,\text{inh}} &= -\frac{ag^2}{2EE_\text{L}E_\text{R}}\frac{\langle \bar{1}\bar{2} \rangle \langle \bar{3}4\rangle}{k_1k_2k_3}\Big(\langle \bar{1}4\rangle \langle \bar{2} 4\rangle \langle \bar{3}4\rangle-\frac{E}{4k}\big(\langle \bar{1}4 \rangle\langle \bar{2}|k\bar{k}|\bar{3}\rangle+\langle \bar{2}4 \rangle\langle \bar{1}|k\bar{k}|\bar{3}\rangle\big)\Big) \,,\\
        4a\mathcal{W}_{s,\text{pure gauge}}^1 &=-\sigma\frac{ag^2}{8k^2E_\text{L}}\frac{\langle \bar{1}\bar{2}\rangle^2\langle \bar{3}4\rangle^2}{k_1k_2k_3}(k_1-k_2) \,,\\
        4a\mathcal{W}^1_{s,\gamma} &= -e^{-2i\gamma}\frac{ag^2}{8k^2E_\text{L}E_\text{R}}\frac{\langle \bar{1}\bar{2}\rangle \langle \bar{3}4\rangle}{k_1k_2k_3}\langle \bar{1}\bar{2}|\bar{k}\bar{k}kk|\bar{3}4\rangle\,,\\
        4a\mathcal{W}^2_{s,\text{inh}} &= \frac{ag^2}{8E}\frac{\langle \bar{1}\bar{2}\rangle \langle \bar{3}4\rangle}{k_1k_2k_3k_4}\big(\langle \bar{1}\bar{3}\rangle \langle \bar{2}4\rangle +\langle \bar{1}4\rangle \langle \bar{2}\bar{3}\rangle\big)\,,\\
        4a\mathcal{W}^2_{s,\text{pure gauge}}&= 0\,,\\
        4a\mathcal{W}^2_{s,\gamma} &= -e^{-2i\gamma}\frac{ag^2}{16kE_\text{L}E_\text{R}}\frac{\langle \bar{1}\bar{2} \rangle \langle \bar{3}4\rangle}{k_1k_2k_3k_4}\langle \bar{1}\bar{2}|\bar{k}\bar{k}kk|\bar{3}4\rangle\,.
    \end{aligned}
\end{align}
One observes that the sum of the diagrams of the second type vanishes in the self-dual limit: $\mathcal{W}^2_{s,\text{inh}}+\mathcal{W}^2_{s,\text{pure gauge}}+\mathcal{W}^2_{s,\gamma}+\mathcal{W}_s^\text{contact}\Big|_{\gamma\rightarrow -i\infty}=0$. It is more clear if we just look at the $\Psi-\Psi$ propagator that does vanish in the self-dual limit. Also, the sum of the diagrams of the first type gives 
\begin{align}
    \begin{aligned}
        &4a\big(\mathcal{W}^1_{s,\text{inh}} + \mathcal{W}_{s,\text{pure gauge}} +\mathcal{W}^1_{s,\gamma}\big)\Big|_{\gamma\rightarrow-i\infty} =\\
        &=-\frac{ag^2}{2EE_\text{L}E_\text{R}}\frac{\langle \bar{1}\bar{2} \rangle \langle \bar{3}4\rangle}{k_1k_2k_3}\Big(\langle \bar{1}4\rangle \langle \bar{2} 4\rangle \langle \bar{3}4\rangle-\frac{E}{4k}\big(\langle \bar{1}4 \rangle\langle \bar{2}|k\bar{k}|\bar{3}\rangle+\langle \bar{2}4 \rangle\langle \bar{1}|k\bar{k}|\bar{3}\rangle\big)\Big)+\\
        &-\sigma\frac{ag^2}{8k^2E_\text{L}}\frac{\langle \bar{1}\bar{2}\rangle^2\langle \bar{3}4\rangle^2}{k_1k_2k_3}(k_1-k_2) \,.
    \end{aligned}
\end{align}
Note that the $\gamma$-dependent contribution vanishes in the limit. This sum agrees with the SDYM result \eqref{SDYM4ptLorenzAx}, but one needs to massage it slightly with the help of $\epsilon_{AB}=\frac{1}{2k}(k_A\bar{k}_B-\bar{k}_Ak_B)$ to find
\begin{align}\notag
    \begin{aligned}
        \frac{1}{2}\big(\langle \bar{1}4\rangle \langle \bar{2}|k\bar{k}|\bar{3}\rangle + \langle \bar{2}4\rangle \langle \bar{1}|k\bar{k}|\bar{3}\rangle\big) &= -\bar{1}^A\bar{2}^A\bar{3}^{B'}4^B\epsilon_{AB}k_A\bar{k}_{B'} = -\frac{1}{2k}\langle \bar{1}\bar{2}|kk\bar{k}\bar{k}|\bar{3}4\rangle -\frac{E_\text{R}}{2k}\langle \bar{1}\bar{2}\rangle \langle \bar{3}4\rangle (k_1-k_2) \,.
    \end{aligned}
\end{align}
Therefore, the sum of the non-topological diagrams reproduces the SDYM correlator \eqref{SDYM4ptLorenzAx} in the self-dual limit.

\paragraph{Gauge dependence.} Gauge dependence resides entirely in
$\mathcal{W}_s^1$. Indeed, what appears in $\mathcal{W}_s^2$ is a gauge-invariant two-point function $\langle F F\rangle$ and the contact diagram $\mathcal{W}^{\text{contact}}_s$ does not have a bulk-to-bulk propagator. We are going to compute the difference between the Feynman and axial gauge correlators. The $\gamma$-dependent part of the bulk-to-bulk propagator is the same in all gauges. Therefore, the gauge dependence can be computed for just the Neumann case. To simplify the computation one can notice that the Dirichlet correlators are gauge independent and the jump between Dirichlet and Neumann inside the propagator is simple. The final result is that 
\begin{align}\label{gaugedep}
    \Delta\mathcal{W}_{s,\gamma}=\Delta\mathcal{W}^1_{s,\text{N}} = -\frac{g^2}{16E_\text{L}}\frac{\langle \bar{1}\bar{2} \rangle^2\langle \bar{3}4\rangle^2}{k_1k_2k_3}\frac{k_1-k_2}{k^2} \,,
\end{align}
which is the same as found in pure SDYM, see \eqref{SDYMaxialcorr}. Therefore, the gauge dependence of the SDYM result for $\sigma=-1$ does not have anything to do with SDYM per se. It is the usual gauge dependence in the Neumann case since the dual is a gauge field whose correlators are, in general, gauge dependent. For the SDYM case, there is a way to make this gauge variation vanish, as discussed in Section \ref{subsec:SDYMaxial}.

\begin{figure}[h!]
\centering
    \begin{tikzpicture}[scale=0.7]
        \begin{scope}[xshift=-6cm,rotate=180]
            \node at (-3,0) {$=$};
        
                    % Define the boundary circle
                \draw[thick] (0,0) circle (2.1cm);
                
                % Define the vertices inside the bulk
                \coordinate (v1) at (-1.4,0);
                \coordinate (v2) at (0.7,0);
                
                % Define the external legs on the boundary
                \coordinate (a) at (-1.7,1.23);
                \coordinate (b) at (-1.7,-1.23);
                \coordinate (c) at (1.7,1.2);
                \coordinate (d) at (1.7,-1.2);

                \filldraw[fill=gray!20, draw=black,rotate=-90] 
                    (1.61,-1.35) arc[start angle=30, end angle=150, radius=1.855cm];
                \filldraw[fill=gray!20, draw=black,rotate=-90] 
                (-1.61,-1.35) arc[start angle=220, end angle=320, radius=2.1cm];
        
                        % Draw parallel momentum arrows
                    \draw[-{Latex},thick,rotate=-90] (-0.9,-1.4) -- (-0.3,-1.2) node[midway,xshift=-0.2 cm,yshift=-0.2 cm] {\( k_3 \)};
                    \draw[-{Latex},thick,rotate=-90] (0.9,-1.4) -- (0.3,-1.2) node[midway,xshift=-0.15 cm,yshift=0.25 cm] {\( k_4 \)};
                
                % Draw the gauge boson lines
                \draw[decorate,decoration={snake},thick] (a) -- (v1);
                \draw[decorate,decoration={snake},thick] (b) -- (v1);
                \draw[decorate,decoration={snake},thick] (c) -- (v2);
                \draw[decorate,decoration={snake},thick] (d) -- (v2);
                
                % Draw the internal propagator 
                \draw[decorate,decoration={snake},thick] (v1) -- (v2);
                
                % Draw parallel momentum arrows
                \draw[{Latex}-,thick] (0.8,0.6) -- (1.2,1) node[midway,xshift=6pt,yshift=-6pt] {\( k_2 \)};
                \draw[{Latex}-,thick] (0.8,-0.6) -- (1.2,-1) node[midway,xshift=6pt,yshift=6pt] {\( k_1 \)};
                \draw[-{Latex},thick] (0.3,-0.3) -- (-0.3,-0.3)node[midway,above] {\( k \)};
                
                % Add labels for fields next to endpoints
                \node[xshift=4pt,yshift=-4pt] at (a) {\( \epsilon^+_3 \)};
                \node[xshift=5pt,yshift=5pt] at (b) {\( \epsilon^-_4 \)};
                \node[xshift=-5pt,yshift=-5pt] at (c) {\( \epsilon^+_2 \)};
                \node[xshift=-5pt,yshift=5pt] at (d) {\( \epsilon^+_1 \)};

                \fill (v2) circle (5pt);
        \end{scope}

                \begin{scope}[rotate=180]
                    % Define the boundary circle
                \draw[thick] (0,0) circle (2.1cm);
                
                % Define the vertices inside the bulk
                \coordinate (v1) at (-1.4,0);
                \coordinate (v2) at (0.7,0);
                
                % Define the external legs on the boundary
                \coordinate (a) at (-1.7,1.23);
                \coordinate (b) at (-1.7,-1.23);
                \coordinate (c) at (1.7,1.2);
                \coordinate (d) at (1.7,-1.2);

                \filldraw[fill=gray!20, draw=black,rotate=-90] 
                    (1.61,-1.35) arc[start angle=30, end angle=150, radius=1.855cm];
                \filldraw[fill=gray!20, draw=black,rotate=-90] 
                (-1.61,-1.35) arc[start angle=220, end angle=320, radius=2.1cm];
        
                        % Draw parallel momentum arrows
                    \draw[-{Latex},thick,rotate=-90] (-0.9,-1.4) -- (-0.3,-1.2) node[midway,xshift=-0.2 cm,yshift=-0.2 cm] {\( k_3 \)};
                    \draw[-{Latex},thick,rotate=-90] (0.9,-1.4) -- (0.3,-1.2) node[midway,xshift=-0.15 cm,yshift=0.25 cm] {\( k_4 \)};
                
                % Draw the gauge boson lines
                \draw[decorate,decoration={snake},thick] (a) -- (v1);
                \draw[decorate,decoration={snake},thick] (b) -- (v1);
                \draw[decorate,decoration={snake},thick] (c) -- (v2);
                \draw[decorate,decoration={snake},thick] (d) -- (v2);
                
                % Draw the internal propagator 
                \draw[decorate,decoration={snake},thick] (v1) -- (v2);
                
                % Draw parallel momentum arrows
                \draw[{Latex}-,thick] (0.8,0.6) -- (1.2,1) node[midway,xshift=6pt,yshift=-6pt] {\( k_2 \)};
                \draw[{Latex}-,thick] (0.8,-0.6) -- (1.2,-1) node[midway,xshift=6pt,yshift=6pt] {\( k_1 \)};
                \draw[-{Latex},thick] (0.3,-0.3) -- (-0.3,-0.3)node[midway,above] {\( k \)};
                
                % Add labels for fields next to endpoints
                \node[xshift=4pt,yshift=-4pt] at (a) {\( \epsilon^+_3 \)};
                \node[xshift=5pt,yshift=5pt] at (b) {\( \epsilon^-_4 \)};
                \node[xshift=-5pt,yshift=-5pt] at (c) {\( \epsilon^+_2 \)};
                \node[xshift=-5pt,yshift=5pt] at (d) {\( \epsilon^+_1 \)};
        
                \fill[blue] (0.5,0) circle (5pt);
                \end{scope}
    \end{tikzpicture}
\caption{The derivative from the SD-vertex in $\mathcal{T}_s^\text{R}$ can only act on the internal line.}
\label{fig:TRDecomp}
\end{figure}
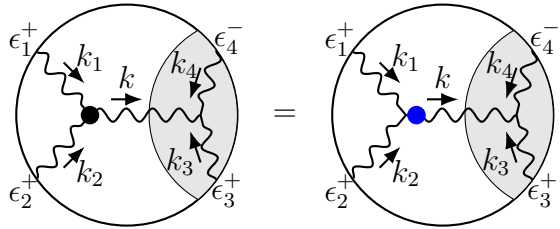

\paragraph{Topological diagrams.} Topological diagrams are relevant for $\gamma\neq0$. There are three topological diagrams: $\mathcal{T}_s^{\text{L}}$ and $\mathcal{T}_s^{\text{R}}$ and $\mathcal{T}_s^{\text{L,R}}$. The left one subdivides into two in accordance with where the derivative of the SD-vertex can land, see Figure \ref{fig:TLDecomp} 
\begin{align} \label{TDecomp}
    \mathcal{T}_s^{\text{L}} = \mathcal{T}_s^{\text{L},1}+\mathcal{T}_s^{\text{L},2} \,.
\end{align}
The actual expressions to compute are
\begin{align}\notag
    \begin{aligned}
        \mathcal{T}_s^{\text{L},1} = 2g^2\int \Phi_+^{AA'}(k_1,0)\Phi\fdud{+}{A}{A'}(k_2,0) \langle \Phi_{AA}(-k,0) \Phi_{BB'}(k,z') \rangle \Phi\fudu{+}{B,}{B'}(k_3,z') \Psi_-^{BB}(k_4,z') \,,\\
        \mathcal{T}_s^{\text{L},2} = -2g^2\int \Phi^{+AA'}(k_1,0)\Phi\fdud{+}{A}{A'}(k_2,0) \langle \Phi_{AA}(-k,0) \Psi_{BB'}(k,z') \rangle \Phi_+^{B,B'}(k_3,z') \Phi\fdud{-}{B}{B'}(k_4,z')\,.
    \end{aligned}
\end{align}
For the right diagram the derivative has a unique place, see Figure \ref{fig:TRDecomp}, and the expression is
\begin{align}\notag
    \mathcal{T}_s^\text{R} = -2g^2\int \Phi_+^{AA'}(k_1,z)\Phi\fdud{+}{A}{A'}(k_2,z) \langle \Psi_{AA}(-k,z) \Phi_{BB}(k,0) \rangle \Phi_+^{BB'}(k_3,0) \Phi\fdud{-}{B}{B'}(k_4,0) \,.
\end{align}
The left-right diagram resides on the boundary and does not have any derivatives
\begin{align}\notag
    \mathcal{T}_s^{\text{L,R}} = 4g^2\Phi_+^{AA'}(k_1,0)\Phi\fdud{+}{A}{A'}(k_2,0)\langle \Phi_{AA}(-k,0)\Phi_{BB}(k,0) \rangle\Phi_+^{BB'}(k_3,0)\Phi\fdud{-}{B}{B'}(k_4,0)\,. 
\end{align}

\begin{figure}[h!]
\centering
    \begin{tikzpicture}[scale=0.7]
        \begin{scope}[xshift=-12cm]
            % Define the boundary circle
        \draw[thick] (0,0) circle (2.1cm);
        
        % Define the vertices inside the bulk
        \coordinate (v1) at (-1.4,0);
        \coordinate (v2) at (0.7,0);
        
        % Define the external legs on the boundary
        \coordinate (a) at (-1.7,1.23);
        \coordinate (b) at (-1.7,-1.23);
        \coordinate (c) at (1.7,1.2);
        \coordinate (d) at (1.7,-1.2);

        \filldraw[fill=gray!20, draw=black,rotate=-90] 
                    (1.61,-1.35) arc[start angle=30, end angle=150, radius=1.855cm];
                \filldraw[fill=gray!20, draw=black,rotate=-90] 
                (-1.61,-1.35) arc[start angle=220, end angle=320, radius=2.1cm];

                % Draw parallel momentum arrows
            \draw[-{Latex},thick,rotate=-90] (-0.9,-1.4) -- (-0.3,-1.2) node[midway,xshift=0.2 cm,yshift=0.2 cm] {\( k_1 \)};
            \draw[-{Latex},thick,rotate=-90] (0.9,-1.4) -- (0.3,-1.2) node[midway,xshift=0.15 cm,yshift=-0.25 cm] {\( k_2 \)};
        
        % Draw the gauge boson lines
        \draw[decorate,decoration={snake},thick] (a) -- (v1);
        \draw[decorate,decoration={snake},thick] (b) -- (v1);
        \draw[decorate,decoration={snake},thick] (c) -- (v2);
        \draw[decorate,decoration={snake},thick] (d) -- (v2);
        
        % Draw the internal propagator 
        \draw[decorate,decoration={snake},thick] (v1) -- (v2);
        
        % Draw parallel momentum arrows
        \draw[{Latex}-,thick] (0.8,0.6) -- (1.2,1.0) node[midway,xshift=-6pt,yshift=7pt] {\( k_4 \)};
        \draw[{Latex}-,thick] (0.8,-0.6) -- (1.2,-1.0) node[midway,xshift=-4pt,yshift=-6pt] {\( k_3 \)};
        \draw[-{Latex},thick] (-0.3,0.3) -- (0.3,0.3) node[midway,above] {\( k \)};
        
        % Add labels for fields next to endpoints
        \node[xshift=-5pt,yshift=5pt] at (a) {\( \epsilon^+_1 \)};
        \node[xshift=-5pt,yshift=-6pt] at (b) {\( \epsilon^+_2 \)};
        \node[xshift=5pt,yshift=5pt] at (c) {\( \epsilon^-_4 \)};
        \node[xshift=4pt,yshift=-4pt] at (d) {\( \epsilon^+_3 \)};

        \fill (v2) circle (5pt);
            \node at (3,0) {$=$};
        \end{scope}

            \begin{scope}[yshift=0cm]
            \begin{scope}[xshift=-6cm]
                    % Define the boundary circle
            \draw[thick] (0,0) circle (2.1cm);
            
            % Define the vertices inside the bulk
            \coordinate (v1) at (-1.4,0);
            \coordinate (v2) at (0.7,0);
            
            % Define the external legs on the boundary
            \coordinate (a) at (-1.7,1.23);
            \coordinate (b) at (-1.7,-1.23);
            \coordinate (c) at (1.7,1.2);
            \coordinate (d) at (1.7,-1.2);

            \filldraw[fill=gray!20, draw=black,rotate=-90] 
                    (1.61,-1.35) arc[start angle=30, end angle=150, radius=1.855cm];
                \filldraw[fill=gray!20, draw=black,rotate=-90] 
                (-1.61,-1.35) arc[start angle=220, end angle=320, radius=2.1cm];
    
                    % Draw parallel momentum arrows
                \draw[-{Latex},thick,rotate=-90] (-0.9,-1.4) -- (-0.3,-1.2) node[midway,xshift=0.2 cm,yshift=0.2 cm] {\( k_1 \)};
                \draw[-{Latex},thick,rotate=-90] (0.9,-1.4) -- (0.3,-1.2) node[midway,xshift=0.15 cm,yshift=-0.25 cm] {\( k_2 \)};
            
            % Draw the gauge boson lines
            \draw[decorate,decoration={snake},thick] (a) -- (v1);
            \draw[decorate,decoration={snake},thick] (b) -- (v1);
            \draw[decorate,decoration={snake},thick] (c) -- (v2);
            \draw[decorate,decoration={snake},thick] (d) -- (v2);
            
            % Draw the internal propagator 
            \draw[decorate,decoration={snake},thick] (v1) -- (v2);
            
            % Draw parallel momentum arrows
            \draw[{Latex}-,thick] (0.8,0.6) -- (1.2,1.0) node[midway,xshift=-6pt,yshift=7pt] {\( k_4 \)};
        \draw[{Latex}-,thick] (0.8,-0.6) -- (1.2,-1.0) node[midway,xshift=-4pt,yshift=-6pt] {\( k_3 \)};
        \draw[-{Latex},thick] (-0.3,0.3) -- (0.3,0.3) node[midway,above] {\( k \)};
            
            % Add labels for fields next to endpoints
        \node[xshift=-5pt,yshift=5pt] at (a) {\( \epsilon^+_1 \)};
        \node[xshift=-5pt,yshift=-6pt] at (b) {\( \epsilon^+_2 \)};
        \node[xshift=5pt,yshift=5pt] at (c) {\( \epsilon^-_4 \)};
        \node[xshift=4pt,yshift=-4pt] at (d) {\( \epsilon^+_3 \)};

            \fill[blue] (0.9,0.2) circle (5pt);
                \node at (3,0) {$+$};
            \end{scope}
            \begin{scope}
                % Define the boundary circle
        \draw[thick] (0,0) circle (2.1cm);
        
        % Define the vertices inside the bulk
        \coordinate (v1) at (-1.4,0);
        \coordinate (v2) at (0.7,0);
        
        % Define the external legs on the boundary
        \coordinate (a) at (-1.7,1.23);
        \coordinate (b) at (-1.7,-1.23);
        \coordinate (c) at (1.7,1.2);
        \coordinate (d) at (1.7,-1.2);

        \filldraw[fill=gray!20, draw=black,rotate=-90] 
                    (1.61,-1.35) arc[start angle=30, end angle=150, radius=1.855cm];
                \filldraw[fill=gray!20, draw=black,rotate=-90] 
                (-1.61,-1.35) arc[start angle=220, end angle=320, radius=2.1cm];

                % Draw parallel momentum arrows
            \draw[-{Latex},thick,rotate=-90] (-0.9,-1.4) -- (-0.3,-1.2) node[midway,xshift=0.2 cm,yshift=0.2 cm] {\( k_1 \)};
            \draw[-{Latex},thick,rotate=-90] (0.9,-1.4) -- (0.3,-1.2) node[midway,xshift=0.15 cm,yshift=-0.25 cm] {\( k_2 \)};
        
        % Draw the gauge boson lines
        \draw[decorate,decoration={snake},thick] (a) -- (v1);
        \draw[decorate,decoration={snake},thick] (b) -- (v1);
        \draw[decorate,decoration={snake},thick] (c) -- (v2);
        \draw[decorate,decoration={snake},thick] (d) -- (v2);
        
        % Draw the internal propagator 
        \draw[decorate,decoration={snake},thick] (v1) -- (v2);
        
        % Draw parallel momentum arrows
        \draw[{Latex}-,thick] (0.8,0.6) -- (1.2,1.0) node[midway,xshift=-6pt,yshift=7pt] {\( k_4 \)};
        \draw[{Latex}-,thick] (0.8,-0.6) -- (1.2,-1.0) node[midway,xshift=-4pt,yshift=-6pt] {\( k_3 \)};
        \draw[-{Latex},thick] (-0.3,0.3) -- (0.3,0.3) node[midway,above] {\( k \)};
        
        % Add labels for fields next to endpoints
        \node[xshift=-5pt,yshift=5pt] at (a) {\( \epsilon^+_1 \)};
        \node[xshift=-5pt,yshift=-6pt] at (b) {\( \epsilon^+_2 \)};
        \node[xshift=5pt,yshift=5pt] at (c) {\( \epsilon^-_4 \)};
        \node[xshift=4pt,yshift=-4pt] at (d) {\( \epsilon^+_3 \)};

        \fill[blue] (0.5,0) circle (5pt);
            \end{scope}
            \end{scope}
        
    \end{tikzpicture}
\caption{The diagram $\mathcal{T}_s^{\text{L}}$ splits into two diagrams: $\mathcal{T}_s^{\text{L},1}+\mathcal{T}_s^{\text{L},1}$.}
\label{fig:TLDecomp}
\end{figure}
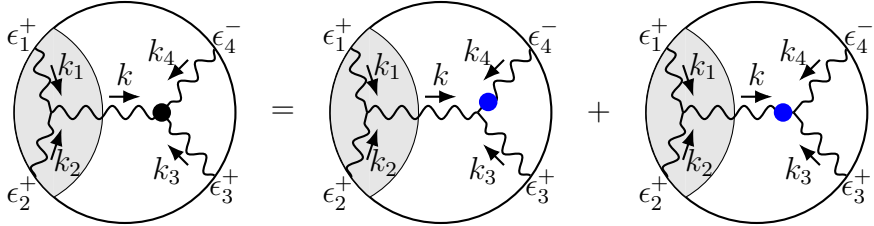

\paragraph{Neumann boundary conditions.} We proceed as for the non-topological diagrams and first give the Neumann results
\begin{align}
    \begin{aligned}
        2(b-a)\mathcal{T}^{\text{L},1}_{s,\text{N}} &= \frac{(b-a)g^2}{4kE_\text{R}}\frac{\langle \bar{1}\bar{2}\rangle \langle \bar{3}4 \rangle}{k_1k_2k_3} \Big(\langle \bar{1}\bar{3} \rangle \langle \bar{2}4 \rangle + \langle \bar{1}4 \rangle \langle \bar{2}\bar{3}\rangle\Big) \,,\\
        2(b-a)\mathcal{T}^{\text{L},2}_{s,\text{N}} &= \frac{(b-a)g^2}{16kE_\text{R}} \frac{\langle \bar{1}\bar{2} \rangle \langle \bar{3}4 \rangle}{k_1k_2k_3k_4}\Big(\langle \bar{1}\bar{3}\rangle \langle \bar{2}|\bar{k}k|4\rangle + \langle \bar{1}4\rangle \langle \bar{2}|\bar{k}k|\bar{3}\rangle + \langle \bar{2}\bar{3}\rangle \langle \bar{1}|\bar{k}k|4\rangle + \langle \bar{2}4\rangle \langle \bar{1}|\bar{k}k|\bar{3}\rangle\Big) \,, \\
        2(b-a)\mathcal{T}^{\text{R}}_{s,\text{N}} &= \frac{(b-a)g^2}{16kE_\text{L}} \frac{\langle \bar{1}\bar{2} \rangle \langle \bar{3}4 \rangle}{k_1k_2k_3k_4}\Big(\langle \bar{1}\bar{3}\rangle \langle \bar{2}|\bar{k}k|4\rangle + \langle \bar{1}4\rangle \langle \bar{2}|\bar{k}k|\bar{3}\rangle + \langle \bar{2}\bar{3}\rangle \langle \bar{1}|\bar{k}k|4\rangle + \langle \bar{2}4\rangle \langle \bar{1}|\bar{k}k|\bar{3}\rangle\Big) \,,\\
        2(b-a)\mathcal{T}^{\text{L,R}}_{s,\text{N}} &= \frac{(b-a)g^2}{8k} \frac{\langle \bar{1}\bar{2} \rangle \langle \bar{3}4 \rangle}{k_1k_2k_3k_4} \Big(\langle \bar{1}\bar{3} \rangle \langle \bar{2}4\rangle + \langle \bar{1}4 \rangle\langle \bar{2}\bar{3}\rangle\Big) \,.
    \end{aligned}
\end{align}

\paragraph{Mixed boundary conditions.} The components of the correlators for mixed boundary conditions are
\begin{align}
    \begin{aligned}
        2(b-a)\mathcal{T}^{\text{L},1}_{\text{inh}} &= \frac{(b-a)g^2}{8kE_\text{R}}\frac{\langle \bar{1}\bar{2} \rangle \langle \bar{3}4\rangle}{k_1k_2k_3}\big(\langle \bar{1}\bar{3}\rangle \langle \bar{2}4\rangle+\langle \bar{1}4\rangle \langle \bar{2}\bar{3} \rangle\big)\,,\\
        2(b-a)\mathcal{T}^{\text{L},1}_{\text{pure gauge}} &= -\sigma\frac{(b-a)g^2}{8k^3E_\text{R}}\frac{\langle \bar{1}\bar{2}\rangle^2\langle \bar{3}4\rangle^2}{k_1k_2k_3}(k_1-k_2)(k_3+k_4)\,,\\
        2(b-a)\mathcal{T}^{\text{L},1}_{\gamma} &= -\frac{(b-a)g^2}{16k^3E_\text{R}} \frac{\langle \bar{1}\bar{2} \rangle \langle \bar{3}4\rangle }{k_1k_2k_3} \Big(e^{2i\gamma}\langle \bar{1}\bar{2}|kk\bar{k}\bar{k}|\bar{3}4\rangle + e^{-2i\gamma}\langle \bar{1}\bar{2}|\bar{k}\bar{k}kk|\bar{3}4\rangle\Big) \,,\\
        2(b-a)\mathcal{T}^{\text{L},2}_{\text{inh}} &= \frac{(b-a)g^2}{32kE_\text{R}}\frac{\langle \bar{1}\bar{2}\rangle\langle \bar{3}4\rangle}{k_1k_2k_3k_4}\Big(\langle \bar{1}\bar{3} \rangle \langle \bar{2}|\bar{k}k|4\rangle + \langle \bar{1}4 \rangle \langle \bar{2}|\bar{k}k|\bar{3}\rangle + \langle \bar{2}\bar{3} \rangle \langle \bar{1}|\bar{k}k|4\rangle + \langle \bar{2}4 \rangle \langle \bar{1}|\bar{k}k|\bar{3}\rangle\Big)\,,\\
        2(b-a)\mathcal{T}^{\text{L},2}_{\text{pure gauge}} &= -\sigma \frac{(b-a)g^2}{16k^2E_\text{R}}\frac{\langle \bar{1}\bar{2}\rangle^2\langle \bar{3}4\rangle^2}{k_1k_2k_3k_4}(k_1-k_2)(k_3+k_4)\,,\\
        2(b-a)\mathcal{T}^{\text{L},2}_{\gamma} &= -e^{-2i\gamma}\frac{(b-a)g^2}{16k^2E_\text{R}} \frac{\langle \bar{1}\bar{2} \rangle \langle \bar{3}4 \rangle}{k_1k_2k_3k_4} \langle \bar{1}\bar{2}|\bar{k}\bar{k}kk|\bar{3}4\rangle\,,\\
        2(b-a)\mathcal{T}^\text{R}_\text{inh} &= \frac{(b-a)g^2}{32kE_\text{L}}\frac{\langle \bar{1}\bar{2}\rangle\langle \bar{3}4\rangle}{k_1k_2k_3k_4}\Big(\langle \bar{1}\bar{3} \rangle \langle \bar{2}|\bar{k}k|4\rangle + \langle \bar{1}4 \rangle \langle \bar{2}|\bar{k}k|\bar{3}\rangle + \langle \bar{2}\bar{3} \rangle \langle \bar{1}|\bar{k}k|4\rangle + \langle \bar{2}4 \rangle \langle \bar{1}|\bar{k}k|\bar{3}\rangle\Big)\,,\\
        2(b-a)\mathcal{T}^\text{R}_\text{pure gauge} &= -\sigma \frac{(b-a)g^2}{16k^2E_\text{L}}\frac{\langle \bar{1}\bar{2}\rangle^2\langle \bar{3}4\rangle^2}{k_1k_2k_3k_4}(k_1-k_2)(k_3+k_4) \,,\\
        2(b-a)\mathcal{T}^{\text{R}}_{\gamma} &= -e^{-2i\gamma}\frac{(b-a)g^2}{16k^2E_\text{L}} \frac{\langle \bar{1}\bar{2} \rangle \langle \bar{3}4 \rangle}{k_1k_2k_3k_4} \langle \bar{1}\bar{2}|\bar{k}\bar{k}kk|\bar{3}4\rangle\,,\\
        \frac{(b-a)^2}{a}\mathcal{T}^{\text{L,R}}_\text{inh} &= \frac{(b-a)^2g^2}{16ak}\frac{\langle \bar{1}\bar{2}\rangle \langle \bar{3}4\rangle}{k_1k_2k_3k_4}\big(\langle \bar{1}\bar{3} \rangle \langle \bar{2}4\rangle + \langle \bar{1}4\rangle \langle \bar{2}\bar{3} \rangle)\,,\\
        \frac{(b-a)^2}{a}\mathcal{T}^\text{L,R}_{\text{pure gauge}} &= -\sigma \frac{(b-a)^2g^2}{32ak^3}\frac{\langle \bar{1}\bar{2}\rangle^2\langle \bar{3}4\rangle^2}{k_1k_2k_3k_4}(k_1-k_2)(k_3+k_4)\,,\\
        \frac{(b-a)^2}{a}\mathcal{T}^{\text{L,R}}_{s,\gamma} &= -\frac{(b-a)^2g^2}{32ak^3} \frac{\langle \bar{1}\bar{2} \rangle \langle \bar{3}4 \rangle}{k_1k_2k_3k_4}\Big(e^{2i\gamma}\langle \bar{1}\bar{2}|kk\bar{k}\bar{k}|\bar{3}4\rangle + e^{-2i\gamma}\langle \bar{1}\bar{2}|\bar{k}\bar{k}kk|\bar{3}4\rangle\Big) \,.
    \end{aligned}
\end{align}%
The results for the topological diagrams seem gauge independent, at least whenever the external lines have definite helicity. One subtlety is that the diagram $\mathcal{T}^{\text{L},1}_{s,\gamma}$ has the $\langle\Phi\Phi\rangle$-propagator, which diverges in the self-dual limit. The diagram is multiplied by $2(b-a)$ which goes to zero at the same rate as $\mathcal{T}^{\text{L},1}_{s,\gamma}$ diverges. The finite leftover is
\begin{align}\label{extra}
    \lim_{\gamma\rightarrow-i\infty}2(b-a)\mathcal{T}^{\text{L},1}_{s,\gamma} &= -a\frac{g^2}{8k^3E_\text{R}} \frac{\langle \bar{1}\bar{2} \rangle \langle \bar{3}4\rangle }{k_1k_2k_3} \langle \bar{1}\bar{2}|kk\bar{k}\bar{k}|\bar{3}4\rangle \,.
\end{align}
Of course, one can argue that the limiting theory in the self-dual limit does not have any topological vertices to begin with. All other topological diagrams safely go to zero thanks to $(b-a)$ and $(b-a)^2$ prefactors. What makes $\mathcal{T}^{\text{L},1}_{s,\gamma}$ special is that it features the $\langle \Phi\Phi\rangle$-propagator since the derivative of the cYM vertex lands on the external line. Note that for genuine bulk diagrams we managed to massage them in such a way that no $\langle \Phi\Phi\rangle$-propagators appear, while $\langle \Psi\Phi\rangle$ and $\langle \Psi\Psi\rangle$ do not diverge in the SD-limit. The diagram can also be interpreted as a correction due to the nonlinearity of the mixed boundary conditions (due to $A^2$ in $F$). That \eqref{extra} is finite may indicate that within AdS/CFT the relation between parent theories and the self-dual truncations thereof can be more complicated than in flat space, where whole amplitudes (at a given loop order and with a fixed helicity structure), not just individual diagrams, can be assigned to SD-theories. In this lucky case, \eqref{extra} can be shown to cancel in the SD-limit against certain composite operators contributions \cite{Richard2026}, see also footnote \ref{compost}. Therefore, the limit from YM to SDYM is smooth for the four-point function we considered.

%%%%%%%%%%%%%%%%%%%%%%%%%%%%%%%%%%%%%%%%%%%%%%%%%%%%%%%%%%%%%
\section{Discussion and Conclusions}
\label{sec:discussion}
%%%%%%%%%%%%%%%%%%%%%%%%%%%%%%%%%%%%%%%%%%%%%%%%%%%%%%%%%%%%%
In the paper we made a proposal for the AdS/CFT duality that has self-dual theories on the bulk side of the duality. Self-dual theories are chiral, i.e. they discriminate fields by helicity. Therefore, it is not surprising that the AdS/CFT dictionary in the self-dual case also has rules that depend on helicity: $\Phi$ of SDYM couples to a positive helicity half $J_+$ of the current on the boundary and $\Psi$ couples to a negative helicity half $a_-$ of a gauge field on the boundary. The latter can be dualized to $J_-$ to build the complete current (the components that are independent for a conserved current). 

Concretely, we started from Yang--Mills theory with a theta-term to achieve the most general mixed boundary conditions. We computed two-, three- and four-point correlators in YM and SDYM with the goal to reveal the relation to SDYM. As a by-product, we have also computed three and four-point functions for Dirichlet, Neumann and mixed boundary conditions in Yang--Mills theory. The Dirichlet results in one or another form have already been available, see e.g. \cite{Raju:2011mp,Raju:2012zs,Raju:2012zr,Albayrak:2023kfk,Albayrak:2023jzl,Armstrong:2020woi}, where the closest one to us is \cite{Armstrong:2020woi,Chowdhury:2024dcy}. 

The relation between YM and SDYM is best seen via the Chalmers--Siegel ``parent action'' that reduces to both of them. However, the SD limit is subtle for many reasons, e.g. (a) the second order theory becomes a first order one, which changes the nature of the boundary problem/conditions/propagators; (b) Chalmers--Siegel's $\epsilon$ is not a genuine coupling constant; (c) YM with $F_{AB}=0$ imposed is essentially an empty theory in the bulk. What seems to make sense (as in flat space) is to compute the ``amplitudes'' that are shared by YM and SDYM. In the AdS/CFT setting this involves dialing the mixed boundary conditions to the self-dual point. 

Nicely, most of the diagrams that exist in YM can be either massaged to those of Chalmers--Siegel theory (where the vertex does not have any derivatives and coincides with the one of SDYM) or can be shown to vanish in the SD limit. It is quite obvious that Chalmers--Siegel's correlators have a smooth limit to those of SDYM. One important observation is that the ``right'' YM theory is the one with a theta-term fine-tuned to collapse the YM Lagrangian to cYM, i.e. to $(F_{AB})^2$, which can be resolved into the Chalmers--Siegel action. As in flat space, SDYM in (A)dS${}_4$ is a subsector of YM in (A)dS${}_4$. However, identifying the relation involves boundary conditions and a subtle self-dual limit.

All computations were done in Feynman/Lorenz and axial gauge. We showed that the leading energy pole gives a flat space amplitude. We could show directly that the Dirichlet results are gauge independent. For Neumann boundary conditions there is a gauge variation of the four-point function between Feynman and axial gauges. This variation can be attributed to the pure gauge terms present in the Feynman propagator that impose boundary conditions on the unphysical components via the method of images. When this propagator is extended to mixed boundary conditions it leads to the same variation as for the Neumann case. However, there is a discrete $\sigma=\pm1$ family of propagators that only differ by boundary conditions on the unphysical fields (Dirichlet vs. Neumann). For $\sigma=+1$, which inherits the boundary conditions from the Dirichlet problem, the four-point function is gauge independent. The same applies to the SDYM four-point function, i.e. at $\sigma=+1$ the Feynman and axial gauges yield the same result.

One should also discuss the relation to the light-cone gauge gauge \cite{Skvortsov:2018uru, Neiman:2023bkq, Neiman:2024vit, Chowdhury:2024dcy,Kozaki:2025jrj}, in particular, to  \cite{Skvortsov:2018uru,Chowdhury:2024dcy}. It is not easy to compare the light-cone gauge results to those in Feynman and axial gauges. First of all, the light-cone gauge can miss important boundary terms, which, as we have seen, affect the two- three- and four-point functions. For example, in the light-cone gauge, the free action for YM and SDYM is the same $\bar\phi \square \phi$, which clearly misses the effect of the Chern--Simons term. In addition, the covariant action for SDYM is a first order theory. It is also not clear how to impose mixed boundary conditions.\footnote{One option is to impose $a k \phi(0)+ b\pl_z \phi(0) =0$ mixed boundary conditions that are conformally invariant because we use the ``shadow'' $k \phi(0)$ instead of $\phi(0)$. } Since the Neumann/mixed boundary conditions' results are gauge dependent, there is an additional possible source of discrepancy. One more subtlety, see \cite{Chowdhury:2024dcy}, is that the light-cone gauge makes a choice of reference spinors that is not friendly with respect to the helicity basis natural for a more covariant treatment. Nevertheless, the light-cone approach is very efficient as all redundant degrees of freedom are gone and should capture the ``essence'' of the results.

Let us discuss a possible relation to the recent proposals \cite{Jain:2024bza, Aharony:2024nqs}, where the ``self-dual'' subsector of Chern--Simons matter theories was attempted to be isolated via a certain limit where the Chern--Simons level has to be taken to the complex plane. The case of SDYM should correspond to a small subsector of the (spin-one) global symmetry current's correlators in Chern--Simons matter theories with leftover global symmetry. Even though these models (with leftover nonabelian global symmetry) were not considered in \cite{Jain:2024bza, Aharony:2024nqs}, the basic conclusions should apply, e.g. in the ``chiral'' or ``self-dual'' subsector of Chern--Simons matter theories all correlators with total helicity negative (or positive for the opposite subsector) must vanish. This is true, of course, in the SDYM limit, except for somewhat ambiguous two-point functions. From the boundary vantage point the proposals of \cite{Jain:2024bza, Aharony:2024nqs} require the existence of nontrivial $\langle +++\rangle $-correlators, which are captured by a higher derivative vertex $\Tr \,\bar{F}^3$ and we have not included such a vertex. Also, there are some subtleties mentioned in \cite{Jain:2024bza, Aharony:2024nqs} regarding the proper adjustment of contact terms at the three-point level as well.

What can definitely be improved is the understanding of two-point functions. Even though they do not contain much information, it would still be important to settle the contact ambiguities present in there. It turns out that the entire two-point function appears to be a contact term with a free coefficient. As we noted, different approaches give either $\langle \pm \pm\rangle $ or $\langle \pm \mp\rangle$, which are mutually exclusive, whereas nonanalytic and contact terms get mixed in the helicity basis.

As an obvious extension of this work, one can consider higher point correlation functions and various other self-dual theories, e.g. self-dual gravity (SDGR) \cite{Krasnov:2016emc} and higher spin extensions of SDYM and SDGR \cite{Ponomarev:2017nrr, Krasnov:2021nsq}, which go all the way to chiral higher-spin gravity \cite{Metsaev:1991mt,Metsaev:1991nb, Ponomarev:2016lrm, Sharapov:2022awp, Sharapov:2022wpz}.\footnote{It was found recently that at least in flat space the variety of self-dual theories is much richer \cite{Serrani:2025owx}. } It would also be important to develop efficient twistor techniques to compute the AdS/CFT correlators, see e.g. \cite{CarrilloGonzalez:2025qjk}, since self-dual theories are most naturally formulated in twistor space, including higher spin extensions \cite{Tran:2021ukl, Herfray:2022prf, Tran:2022tft, Mason:2025pbz}.

Within AdS/CFT the pure (super)gravity approximation ``sees'' a universal part of the stress-tensor correlators, which is incomplete due to UV-divergences. The situation with self-dual theories, e.g. with SDGR, can be much better: since all SD-theories are UV-finite, they capture some truly universal part of the correlators. Such subsectors can be called self-dual CFTs as they are AdS/CFT dual to self-dual theories in the bulk. One possible definition, based on \cite{Jain:2024bza}, is that a self-dual CFT has nonvanishing correlators $\langle J_{
\lambda_1} ...\rangle \neq0$ whenever $\sum_{i=1}^n \lambda_i\geq n-2$ for all $n=3,4,...$ and $\langle J_{
\lambda_1} ...\rangle =0$, otherwise.

More generally, gauge theories in AdS, e.g. YM, SDYM, GR and SDGR, provide us with perfect examples of gauge theories on manifolds with boundaries where BV and BFV can be indispensable tools. For example, within the self-dual AdS/CFT correspondence we are led to properly set-up the boundary problem and to eventually compare non-gauge-invariant correlation functions in two quantum gauge theories. The latter aspects are practically absent for Dirichlet boundary conditions.  

At least when a theory is conformally invariant, e.g. SDYM and higher spin extensions of it, there can be an interesting correspondence between celestial and AdS/CFT structures, see e.g. \cite{Sheta:2025oep}. It would also be important to reconsider the vanishing of the leading piece of the AdS/CFT correlators in self-dual theories due to the recent discovery \cite{Guevara:2026qzd} that $(+...+-)$-amplitudes are supported on the collinear kinematics. This observation seems to have a higher-spin extension \cite{Ponomarev:2022atv, Ponomarev:2022qkx}, where chiral higher-spin gravity in flat space is dual to the flat space version of the free vector model, which is a celestial analog of the higher-spin/vector model AdS/CFT duality \cite{Klebanov:2002ja,Sezgin:2002rt,Leigh:2003gk}, one major difference being that both sides of the duality are chiral. We make a number of comments on the collinear limit of AdS/CFT correlators in Appendix \ref{app:collin}.

%%%%%%%%%%%%%%%%%%%%%%%%%%%%%%%%%%%%%%%%%%%%%%%%%%%%%%%%%%%%%
\section*{Acknowledgment}
%%%%%%%%%%%%%%%%%%%%%%%%%%%%%%%%%%%%%%%%%%%%%%%%%%%%%%%%%%%%%
We are indebted to Simone Giombi for many helpful and insightful discussions during the development of this work, which substantially improved the conceptual clarity of this work and clarified key aspects of the analysis. E.S. is grateful to Dhruva K.S., Sachin Jain, Euihun Joung, Ricardo Monteiro, Arthur Lipstein, Dmitry Ponomarev, Alexey Sharapov, Andrew Strominger, Xi Yin for useful discussions on the topics related to this paper. We would like to thank the anonymous Referee for a very useful question that lead us to introduce a more general family of propagators. The work of E. S. and R. van D. was partially supported by the European Research Council (ERC) under the European Union’s Horizon 2020 research and innovation programme (grant agreement No 101002551). E.S. is a research associate of the Fonds de la Recherche Scientifique – FNRS. 
%%%%%%%%%%%%%%%%%%%%%%%%%%%%%%%%%%%%%%%%%%%%%%%%%%%%%%%%%%%%%
\appendix 
%%%%%%%%%%%%%%%%%%%%%%%%%%%%%%%%%%%%%%%%%%%%%%%%%%%%%%%%%%%%%

%%%%%%%%%%%%%%%%%%%%%%%%%%%%%%%%%%%%%%%%%%%%%%%%%%%%%%%%%%%%%
\section{Spinor helicity and CFT}
\label{app:cft}
%%%%%%%%%%%%%%%%%%%%%%%%%%%%%%%%%%%%%%%%%%%%%%%%%%%%%%%%%%%%%
A two-point function of a conserved current $J_m$ in $3d$ CFT has two conformally-invariant structures:
\begin{align}
    \langle J_m(k) J_n(-k)\rangle& = a\,k\,\Pi^\text{e}_{mn}+b \,k\,\Pi^\text{o}_{mn} \,,
\end{align}
Here, $\Pi^\text{e}$ is parity-even and $\Pi^\text{o}$ is parity-odd: 
\begin{align}
     \Pi^\text{e}_{mn}&= \left(\delta_{mn}-\frac{k_m k_n}{k^2}\right) \,, & \Pi^\text{o}_{mn}&=\epsilon_{mn\lambda}k^\lambda /k \,.
\end{align}
In the two-component spinor language the same structures read
\begin{align}
    \Pi_{AA,BB}^\text{e}&=\epsilon^{AB}\epsilon^{AB}+\frac{ k^{AA}k^{BB}}{2k^2} \,, &  \Pi^\text{o}_{AA,BB}&= \frac{1}{k}\epsilon_{AB}k_{AB} \,.
\end{align}
The two-point function can be rewritten as 
\begin{align}
    \langle J_{AA}(k) J_{BB}(-k)\rangle& = a\,k\,\Pi_{AA,BB}^\text{e}+b \,k\, \Pi^\text{o}_{AA,BB} \,.
\end{align}
The same spinor-helicity language as discussed in Section \ref{sec:FG} can be employed on the CFT side. With the help of the polarization vectors $\epsilon_{\pm}^{AB}$ one can project two- and higher-point correlators onto components with specific helicities by contracting the currents as $J^\alpha =J(k)\cdot \epsilon^\alpha(k)$, where $\alpha=\pm$. The two-point functions in the helicity basis are
\begin{align}\label{twopointab}
       \langle J^+(k) J^+(-k)\rangle&=  k (a- b) \,, &
       \langle J^-(k) J^-(-k)\rangle&=  k (a+ b) \,.
\end{align}
One observes that adding the parity odd contribution, e.g. via a Chern--Simons term, can cancel one of the two helicity structures. The parity-even/odd projectors can be expressed as
\begin{align}
    \Pi^\text{e}_{AA,BB}&= \frac{1}{4k^2}(k_A k_A \brk_{B}\brk_{B}+\brk_A \brk_A k_{B} k_{B})\,,&
    \Pi^\text{o}_{AA,BB}&= \frac{1}{4k^2}(k_A k_A \brk_{B}\brk_{B}-\brk_A \brk_A k_{B} k_{B})   \,.
\end{align}
To bring the holographic correlation functions closer to amplitudes, it is convenient to define $1 \rangle = k^A$ etc. and contractions $\langle 1 2\rangle= k^A_1(k) k_A^2(-k)$ and express correlators in terms of such contractions. Note that $\epsilon_{\pm}$ do not scale with the momentum, but $k\rangle$ do. This way the two-point functions can be rewritten as 
\begin{align}
       \langle J^+(k) J^+(-k)\rangle&= \delta^{\alpha,\beta} \frac{\langle 1 2\rangle^2 }{k^2} (a- b)k \,,&
       \langle J^-(k) J^-(-k)\rangle&= \delta^{\alpha,\beta} \frac{\langle \bar 1 \bar 2\rangle^2 }{k^2} (a+ b)k \,.
\end{align}
Note that the two-point kinematics is somewhat degenerate since the operators have the same momentum up to a sign and, hence, $\langle 1 2\rangle= k^A(k) k_A(-k)=-k^A(k) \brk_A(k)=+2k$. In particular, we see that in the helicity basis there is not much difference between the parity-odd contact contribution and the parity-even one. 

%%%%%%%%%%%%%%%%%%%%%%%%%%%%%%%%%%%%%%%%%%%%%%%%%%%%%%%%%%%%%
\section{Collinear limit}
\label{app:collin}
%%%%%%%%%%%%%%%%%%%%%%%%%%%%%%%%%%%%%%%%%%%%%%%%%%%%%%%%%%%%%
In order to take $i\epsilon$'s in the propagators properly into account one has to return to Minkowski signature. However, at the level of three-point functions, the net effect is just a change of variables. Indeed, let us take \eqref{SDYM3pt} and set $q_1=q_2=r$
\begin{align}
    -\frac{2g}{E}\frac{\langle q_13 \rangle \langle q_23 \rangle}{\langle q_11 \rangle \langle q_22 \rangle} \langle \bar{1}\bar{2} \rangle &=-\frac{2g}{E} \frac{\langle r3 \rangle^2}{\langle r1 \rangle \langle r2 \rangle} \langle \bar{1}\bar{2} \rangle\,.
\end{align}
Now, on the pole $1/E$ the kinematics is that of the flat space, i.e. one has $\delta^4(\sum_i i_A \bar{i}_{A'})$. By choosing two auxiliary spinors $\mu$ and $\nu$, one can split the $\delta^4$ into two $\delta^2$:
\begin{align}
    \delta^4(\sum_i i_A \bar{i}_{A'})&= \langle \mu\nu\rangle^2 \delta^2(\sum_i \langle \mu i\rangle \bar{i}_{A'}) \delta^2(\sum_i \langle \nu i\rangle \bar{i}_{A'}) \,.
\end{align}
On choosing $\nu$ to be one of the $i$'s, e.g. $i=3$, one can massage the last $\delta$ into
\begin{align}
    \delta^2(\sum_i \langle \nu i\rangle \bar{i}_{A'}) &= |\langle \bar1 \bar 2\rangle|^{-1} \delta (\langle 1  3\rangle)\delta (\langle 2  3\rangle)\,.
\end{align}
Combining all essential factors we get on the energy pole with $\mu=r$ \cite{Guevara:2026qzd} (note that all corrections in \eqref{corrections3} die out in the limit)
\begin{align}
 \frac{\langle r3 \rangle^4}{\langle r1 \rangle \langle r2 \rangle} \sign\langle \bar{1}\bar{2} \rangle\, \delta (\langle 1  3\rangle)\,\delta (\langle 2  3\rangle) \,\delta^2(\sum_i \langle r i\rangle \bar{i}_{A'})\,.
\end{align}
In order to take the energy non-conservation into account, one can insert an additional $\delta(\sum k_i-E)$ to get the $4d$ conservation $\delta^4(\sum_i i_A \bar{i}_{A'}-\epsilon_{AA'}E)$. By splitting $\delta^4$ into two $\delta^2$ along $r_A$ and $3_A$, one finds for the measure
\begin{align}
    \langle r3 \rangle^2\frac{\sign\langle \bar{1}\bar{2} \rangle}{\langle \bar{1}\bar{2}\rangle}\, \delta \left(\langle 1  3\rangle - \tfrac{E\langle \bar 2  3\rangle}{\langle \bar1  \bar2\rangle}\right)\,\delta \left(\langle 2  3\rangle - \tfrac{E\langle 3  \bar 1\rangle}{\langle \bar 1  \bar 2\rangle}\right) \,\delta^2\left(\sum_i \langle r i\rangle \bar{i}_{A'} -r_{A'}E\right)\,,
\end{align}
which can now be combined with \eqref{corrections3} to get the complete three-point function. 

Starting from the five-point functions the collinear limit becomes highly nontrivial \cite{Guevara:2026qzd}. At the four-point level the collinear limit in flat space gets a simple contribution from the ``cut'' in accordance with $1/(p^2+i\epsilon)=\mathrm{PV}(1/p^2)-i \pi \delta (p^2)$. A careful analysis of the AdS-propagator is needed to recover the $i\epsilon$'s. A shortcut is to consider to the flat limit: 
\begin{align} \label{ieps}
    \int e^{-(k_1+k_2)z}\langle \Phi_{AA'}(-k,z)\Phi_{BB'}(k,z') \rangle e^{-(k_3+k_4)z'} = \frac{1}{E} \big(-\frac{\epsilon_{AB}\epsilon_{A'B'}}{E_\text{L}E_\text{R}-i\epsilon}\big) + \dots\,.
\end{align}
This leads to the following $\delta$-function,
\begin{align} \label{delta}
    \delta(E_\text{L}E_\text{R})=\delta(\langle 12 \rangle \langle \bar{1}\bar{2} \rangle - EE_\text{L})=\delta(\langle 34 \rangle \langle \bar{3}\bar{4} \rangle-EE_\text{R})=\frac{1}{|\langle \bar{3}\bar{4} \rangle|}\delta(\langle 34 \rangle - E\frac{E_\text{R}}{\langle \bar{3}\bar{4} \rangle}) \,.
\end{align}    
The momentum conservation can be treated as before. We first insert $E$ to restore the four-momentum conservation and write
\begin{align}
    \delta^4(\sum\limits_{i=0}^4 i_A\bar{i}_{A'}) = \delta^2(\sum\limits_{i=0}^4 \langle ri \rangle \bar{i}_{A'}-r_{A'}E)\delta^2(\sum\limits_{i=0}^4\langle4i\rangle \bar{i}_{A'}-4_{A'}E)\langle r4\rangle^2 \,.
\end{align}
The second delta leads to 
\begin{align}
    \delta^2(\sum\limits_{i=0}^4\langle4i\rangle \bar{i}_{A'}-4_{A'}E) = \delta\Big(\langle 14\rangle - \frac{E}{\langle \bar{1}\bar{2}\rangle}\big(E_\text{R}\frac{\langle \bar{2}\bar{3} \rangle}{\langle \bar{3}\bar{4} \rangle}+\langle \bar{2}4\rangle\big)\Big)\delta\Big(\langle 24 \rangle + \frac{E}{\langle \bar{1}\bar{2} \rangle} \big( E_\text{R}\frac{\langle \bar{1}\bar{3} \rangle}{\langle \bar{3}\bar{4} \rangle} +\langle \bar{1}4 \rangle\Big) \,.
\end{align}
Now, everything can be combined with the four-point functions in the main text.

%%%%%%%%%%%%%%%%%%%%%%%%%%%%%%%%%%%%%%%%%%%%%%%%%%%%%%%%%%%%%
\section{Propagators}
\label{app:propagators}
%%%%%%%%%%%%%%%%%%%%%%%%%%%%%%%%%%%%%%%%%%%%%%%%%%%%%%%%%%%%%
Propagators on constant curvature spacetimes have a long history, see e.g. \cite{Allen:1985wd,DHoker:1999bve, Raju:2010by, Moga:2025gdy}. Nevertheless, Charmers-Siegel theory and SDYM have not been discussed much in this context, as well as YM with mixed boundary conditions except for \cite{Chang:2012kt}. We use two gauges: Feynman gauge and axial gauge. The former is somewhat simpler and more covariant in the bulk, the latter is more handy in establishing the AdS/CFT dictionary. Below, we just list the results, see \cite{Richard2026} for details, but let us make a few general comments. A bulk-to-bulk propagator consists of two parts: an inhomogeneous one, which can be obtained by a Fourier transform of the flat space propagator, and a homogeneous one, which is needed to impose boundary conditions and regularity in the interior of $\text{AdS}_4$:
\begin{align}
    G(-k,z;k,z')&=G^\text{hom}(-k,z;k,z')+G^\text{inh}(-k,z;k,z') \,.
\end{align}
Bulk-to-bulk propagators in two different gauges are related by a ``gauge transformation'' and we prefer to relate everything to the Feynman gauge 
\begin{align} \label{gauge}
    \begin{aligned}
        G^{\text{some gauge}}_{AA',BB'}&(-k,z;k,z')= G^{\text{F}}_{AA',BB'}(-k,z;k,z')+\\
        &+\nabla_{k,z'}^{BB'} \xi_\text{L}^{AA'}(k,z,z')+\nabla_{-k,z}^{AA'} \xi_\text{R}^{BB'}(k,z,z')+\nabla_{-k,z}^{AA'} \nabla_{k,z'}^{BB'}\xi_{\text{c}}(k,z,z') \,.
    \end{aligned}
\end{align}
There is some redundancy in the last term, but it can be useful. Boundary conditions can always be imposed in terms of gauge-invariant quantities $\Psi_{AA}\equiv F_{AA}$, $\bar{\Psi}_{A'A'}\equiv\bar{F}_{A'A'}$,
\begin{align}\label{gaugeinvbc}
    \begin{aligned}
        e^{i\gamma} \langle \Psi^{AA}(-k,z\rightarrow 0) \Psi^{BB}(k,z')\rangle &=-e^{-i\gamma} \langle \bar\Psi^{AA}(-k,z\rightarrow 0) \Psi^{BB}(k,z')\rangle \,,\\
        e^{i\gamma} \langle \Psi^{AA}(-k,z\rightarrow 0) \bar\Psi^{BB}(k,z')\rangle &=-e^{-i\gamma} \langle \bar\Psi^{AA}(-k,z\rightarrow 0) \bar\Psi^{BB}(k,z')\rangle   \,.
    \end{aligned}
\end{align}
In case the gauge is not complete, one has to fix the pure gauge terms as well. 
A stronger version is sometimes possible where the bulk leg is kept in its original form of $\Phi_{A,A'}$,
\begin{align}\label{gaugenoninvbc}
       e^{i\gamma} \langle \Psi^{AA}(-k,z\rightarrow 0) \Phi^{B,B'}(k,z')\rangle &=-e^{-i\gamma} \langle \bar\Psi^{AA}(-k,z\rightarrow 0) \Phi^{B,B}(k,z')\rangle  \,.
\end{align}
By Feynman gauge we mean the one that leaves $\square$ as the kinetic operator in YM. In Chalmers--Siegel theory the $\langle \Psi \Phi\rangle$ propagator is obtained via the $\Psi$ equations of motion, i.e. as derivative of $\langle \Phi \Phi\rangle$-propagator. The latter has a smooth limit to SDYM. 

\subsection{YM: Feynman}

For Dirichlet/Neumann boundary conditions, the YM propagator in Feynman gauge is obtained using the method of images. Here it is important that the reflection operation also acts on the indices of the gauge field. The Dirichlet/Neumann propagator is
\begin{align} \label{D/N}
    \langle \Phi_{A,A'}(-k,z)\Phi_{B,B'}(k,z') \rangle_\text{D/N}= -\frac{\epsilon_{AB}\epsilon_{A'B'}}{2k} e^{-k|z-z'|}\pm e^{-k(z+z')}\frac1{2k}\left(\epsilon_{AB}\epsilon_{A'B'}-\epsilon_{AA'}\epsilon_{BB'}\right)\,,
\end{align}
where the $+$ is for Dirichlet. The first term is an inhomogeneous one $G^\text{inh}$ and the second term is a homogeneous one $G^\text{hom}$. Note that the image matrix acts both on the coordinates $x^\mu\equiv (x^i,z)$ and on the vector index of $\Phi_\mu\equiv \Phi_{A,A'}$. Therefore, $\Phi_i/\Phi_z$ obey Dirichlet/Neumann for the Dirichlet boundary conditions and the opposite for the Neumann ones, i.e. the four-vector $\Phi_\mu$ does not obey just Dirichlet or just Neumann conditions. 

For a general mixed boundary condition, a good starting point is the Neumann one since the CFT dual is always a gauge field except for the Dirichlet point. There is a part of the homogeneous component of the propagator that is fixed by the gauge-invariant boundary conditions \eqref{gaugeinvbc}:
\begin{align} \label{hom,gamma}
        G^{\text{homo},\gamma}_{AA',BB'}&= \frac{e^{-k(z+z')}}{2k} \Big(\cos (2\gamma)\Pi^\text{e}_{AA',BB'}+i\sin (2\gamma) \Pi^\text{o}_{AA',BB'}\Big)\,.
\end{align}
The pure gauge part of the homogeneous component can be fixed by \eqref{gaugenoninvbc}. Importantly, this pure gauge piece is $\gamma$-independent for $\gamma\neq0$. At $\gamma=\pi/2$ the pure gauge and the gauge-invariant homogeneous components conspire to give just the image term. Similarly, for the Dirichlet boundary conditions the gauge-invariant part of the homogeneous term is the same $G^{\text{homo},\gamma}$ at $\gamma=0$, but the pure gauge component has the opposite sign as compared the Neumann case. In other words,
\begin{align}
    G_{\text{D}}&= G^\text{inh}+ G^{\text{homo},\gamma=0}+\text{pure gauge}\,,\\
    G_{\text{N}}&= G^\text{inh}+ G^{\text{homo},\gamma=\pi/2}-\text{pure gauge}\,.
\end{align}
Note that $G^{\text{homo},\gamma=0}=-G^{\text{homo},\gamma=\pi/2}$ and the ``pure gauge'' term is the same in both cases above, which is important to reproduce \eqref{D/N}. The pure gauge homogeneous term affects only the boundary conditions for the unphysical $k_i \Phi^i$ and $\Phi_z$ components. Therefore, it makes sense to introduce a discrete parameter $\sigma=\pm 1$ to define a more general propagator as
\begin{align} \label{Gs}
    G&= G^\text{inh}+ G^{\text{homo},\gamma}+\sigma\times \text{pure gauge}\,.
\end{align}
For $\sigma=+1$ the unphysical components obey the boundary conditions implied by the Dirichlet problem on the physical ones, e.g. $\pl_z\Phi_z=k_i \Phi^i=0$ on the boundary. The latter implies that the ghost $c$ vanishes on the boundary.  Explicitly, $G^{\text{inh}}$ reads
\begin{align}
    G^\text{inh} = -\frac{\epsilon_{AB}\epsilon_{A'B'}}{2k}e^{-k|z-z'|} \,,
\end{align}
$G^{\text{homo},\gamma}$ is given in \eqref{hom,gamma} and the pure gauge term reads
\begin{align}
    \text{pure gauge} = -\frac{1}{4k}(\epsilon_{AA'}\epsilon_{BB'}+\frac{k_{AA'}k_{BB'}}{k^2})=\nabla_{AA'}\xi^\text{R}_{BB'}+\nabla_{BB'}\xi^\text{L}_{AA'}+\nabla_{AA'}\nabla_{BB'}\xi^\text{c}\,,
\end{align}
with
\begin{align}
    \xi^\text{R}_{BB'} &= -\frac{\epsilon_{BB'}}{4k^2}e^{-k(z+z')} \,, & \xi^\text{L}_{AA'} &=-\frac{\epsilon_{AA'}}{4k^2}e^{-k(z+z')} \,, & \xi^\text{c} &= \frac{1}{4k^3}e^{-k(z+z')}\,.
\end{align}

\subsection{YM: Axial}
The propagator in axial gauge can be found by Fourier transforming the flat space propagator to get the inhomogeneous term. The homogeneous one is fixed by regularity and by the gauge-invariant boundary conditions \eqref{gaugeinvbc} since the axial gauge is a complete gauge. 
The propagator in axial gauge can be related to the Feynman gauge propagator through \eqref{gauge}. We denote the difference between the Feynman gauge and axial gauge propagator as $\Delta\langle \Phi\Phi\rangle$. We note that the $\gamma$-dependent component of the propagator is the same for Feynman and axial gauge. Hence, the gauge variation of the propagators 
\begin{align}
    \Delta\langle \Phi_{A,A'}(-k,z)\Phi_{B,B'}(k,z')\rangle \equiv \langle \Phi_{A,A'}\Phi_{B,B'}\rangle^\text{Axial}-\langle \Phi_{A,A'}\Phi_{B,B'}\rangle^{\text{F}}\,.
\end{align}
does not depend on $\gamma$.
This is given by
\begin{align} \label{nablaxi}
    \begin{aligned}
        \Delta\langle \Phi_{A,A'}(-k,z)\Phi_{B,B'}(k,z')\rangle &= \nabla_{AA'}^{-k,z}\xi_{BB'}^\text{R}+\nabla_{BB'}^{k,z'}\xi^\text{L}_{AA'}+\nabla_{AA'}^{-k,z}\nabla_{BB'}^{k,z'}\xi^\text{c}\,,\\
        \xi^\text{L}_{AA'} &= \frac{\epsilon_{AA'}}{4k^2}\Big(\text{sign}(z-z')\big(1-e^{-k|z-z'|}\big)-\big(1-\sigma e^{-k(z+z')}\big)\Big) \,,\\
        \xi^\text{R}_{BB'} &= -\frac{\epsilon_{BB'}}{4k^2}\Big(\text{sign}(z-z')\big(1-e^{-k|z-z'|}\big)+\big(1-\sigma e^{-k(z+z')}\big)\Big) \,,\\
        \xi^\text{c} &= \frac{1}{4k^3}\Big(e^{-k|z-z'|}-\sigma e^{-k(z+z')}+k\big(|z-z'|-(z+z')\big)\Big) \,,
    \end{aligned}
\end{align}
where the $\sigma=+1$ is for Dirichlet and $\sigma=-1$ for Neumann.  The Dirichlet gauge parameters obey $\xi^\text{L}_{AA'}(z'=0)=\xi^\text{R}_{AA'}(z=0)=\xi^\text{c}(z=0)=\xi^\text{c}(z'=0)=0$, which proves crucial for the gauge independence of Dirichlet correlators. For Neumann boundary conditions, the gauge parameters do not vanish on the boundary. The axial gauge propagator with Dirichlet boundary conditions is known since \cite{Raju:2010by}.

\subsection{Chalmers--Siegel: Feynman}
Chalmers--Siegel theory is equivalent to cYM when evaluated on-shell. It is therefore not surprising that the $\langle\Phi\Phi\rangle$ propagator in the two formulations agree. In addition to $\langle \Phi\Phi \rangle$, Chalmers--Siegel contains a $\langle\Psi\Phi\rangle$ and $\langle\Psi\Psi\rangle$ propagator. The $\langle \Psi\Phi\rangle$ propagator is related to the $\langle \Phi \Phi \rangle$ one through
\begin{align}
    \langle \Psi_{AA}(-k,z)\Phi_{B,B'}(k,z')\rangle = \nabla^{-k,z}_{AA'}\langle \Phi\fdu{A,}{A'}(-k,z)\Phi_{B,B'}(k,z')\rangle.
\end{align}
For the Dirichlet and Neumann propagator we have
\begin{align} \label{PsiPhi}
    \begin{aligned}
        \langle \Psi_{AA}(-k,z)\Phi_{B,B'}(k,z')\rangle_\text{D} &= -\frac{\epsilon_{AB}}{2k}\big(k_{AB'}-\text{sign}(z-z')k\epsilon_{AB'}\big)e^{-k|z-z'|}+\frac{\epsilon_{AB'}}{2k}\bar{k}_Ak_Be^{-k(z+z')} \,,\\
        \langle \Psi_{AA}(-k,z)\Phi_{B,B'}(k,z')\rangle_\text{N} &= -\frac{\epsilon_{AB}}{2k}\big(k_{AB'}-\text{sign}(z-z')k\epsilon_{AB'}\big)e^{-k|z-z'|}-\frac{\epsilon_{AB'}}{2k}\bar{k}_Ak_Be^{-k(z+z')}
    \end{aligned}
\end{align}
The propagator for mixed boundary conditions is decomposed in a similar fashion as \eqref{Gs}:
\begin{align} \label{CSmixed}
    \begin{aligned}
        \langle \Psi_{AA}(-k,z)\Phi_{B,B'}(k,z')\rangle^{\text{inh}} &= -\frac{\epsilon_{AB}}{2k}\big(k_{AB'}-\text{sign}(z-z')\epsilon_{AB'}\big)e^{-k(z+z')} \,,\\
        \langle \Psi_{AA}(-k,z)\Phi_{B,B'}(k,z')\rangle^{\text{pure gauge}} &= \sigma\frac{k_{AA}}{4k^2}\nabla_{BB'}^{k,z'}e^{-k(z+z')}\,,\\
        \langle \Psi_{AA}(-k,z)\Phi_{B,B'}(k,z')\rangle^{\text{homo,}\gamma} &= -\frac{e^{-2i\gamma}}{4k^2}\bar{k}_A\bar{k}_Ak_Bk_{B'}e^{-k(z+z')} \,.
    \end{aligned}
\end{align}%
One crucial subtlety exists for the $\langle \Psi\Psi \rangle$ propagator. Inverting the kinetic term in the Chalmers--Siegel action demands
\begin{align} \label{eomPsi}
    \nabla\fdu{A}{A'}\langle \Psi^{AA}(-k,z) \Psi^{BB}(k,z')\rangle&=0 \,,
\end{align}
while taking derivatives on both legs of the $\langle \Phi\Phi \rangle$ propagator gives the two point function
\begin{align}
\begin{aligned}
    \langle F^{AA}(-k,z) F^{BB}(k,z')\rangle&=(\nabla_{-k,z})\fud{A}{A'} (\nabla_{k,z'})\fud{B}{B'}\langle \Phi^{AA'}(-k,z) \Phi^{BB'}(k,z')\rangle\,.
\end{aligned}
\end{align}
It consists of a term that satisfies \eqref{eomPsi} and another term containing a delta-function, which does not satisfy the homogeneous equation of motion. In other words, there is a difference between the propagator $\langle \Psi \Psi\rangle$ and the two-point function $\langle F F\rangle$:
\begin{align}
    \langle F F\rangle&= \langle F F\rangle^\text{inh}+\langle \Psi \Psi\rangle\,.
\end{align}
The actual $\langle \Psi\Psi \rangle$ propagators for Dirichlet/Neumann boundary conditions and $\langle F F\rangle^\text{inh}$ are
\begin{align}
    \begin{aligned}
    \langle F_{AA}(-k,z)F_{BB}(k,z')\rangle^\text{inh} &= -\delta(z-z')\epsilon_{AB}\epsilon_{AB}\,,\\
        \langle \Psi_{AA}(-k,z)\Psi_{BB}(k,z')\rangle_\text{D} &= \frac{\bar{k}_A\bar{k}_Ak_Bk_B}{2k}e^{-k(z+z')} \,,\\
        \langle \Psi_{AA}(-k,z)\Psi_{BB}(k,z')\rangle_\text{N} &= -\frac{\bar{k}_A\bar{k}_Ak_Bk_B}{2k}e^{-k(z+z')}\,,
    \end{aligned}
\end{align}
and for mixed boundary conditions we have
\begin{align}
    \begin{aligned}
        \langle \Psi_{AA}(-k,z)\Psi_{BB}(k,z')\rangle^\text{pure gauge} &= 0\,,\\
        \langle \Psi_{AA}(-k,z)\Psi_{BB}(k,z')\rangle^{\text{homo},\gamma} &= \frac{e^{-2i\gamma}}{2k}\bar{k}_A\bar{k}_Ak_Bk_Be^{-k(z+z')}\,.
    \end{aligned}
\end{align}

\subsection{Chalmers--Siegel: Axial}

The gauge variations of the $\langle \Psi \Phi \rangle$ and $\langle \Psi\Psi \rangle$ propagators read
\begin{align} \label{etaL}
    \begin{aligned} 
        \Delta &\langle \Psi_{AA}(-k,z)\Phi_{B,B'}(k,z')\rangle_\text{D/N} = \nabla_{BB'}^{k,z'}\eta^{\text{L}}_{AA}\,,\\
        \Delta &\langle \Psi_{AA}(-k,z)\Psi_{B,B'}(k,z')\rangle_\text{D/N} = 0\,,
    \end{aligned}
\end{align}
where $\eta^\text{L}_{AA} \equiv\nabla^{-k,z}_{AA'}\xi^\text{L}\fdu{A}{A'}$ and $\xi^\text{L}_{AA'}$ is defined in \eqref{nablaxi}. In the axial gauge the SDYM propagator is a smooth $\gamma\rightarrow-i\infty$ limit of the $\langle\Psi \Phi\rangle$-propagator in Chalmers--Siegel  theory.

\subsection{SDYM: Feynman}
\label{subsec:SDYMF}
The SDYM propagator can be obtained by taking the self-dual limit $\gamma\rightarrow-i\infty$ of the $\langle \Psi\Phi \rangle$ propagator in Chalmers--Siegel theory. Note that both the $\langle \Phi\Phi\rangle$ and $\langle \Psi \Psi \rangle$ propagators are absent in SDYM. This gives us
\begin{align}
        \langle \Psi_{AA}(-k,z)\Phi_{B,B'}(k,z')\rangle
        &=-\frac{\epsilon_{AB}}{2k}(k_{AB'}-\sign(z-z')k\epsilon_{AB'})e^{-k|z-z'|}+\sigma\frac{k_{AA}}{4k^2} \nabla_{BB'}^{k,z'}e^{-k(z+z')} \,.
\end{align}
One can notice something special at the SD-point: the homogeneous term is pure gauge. As for $\langle \Phi\Phi\rangle$-propagator in YM theory, the two values of $\sigma$ differ by the boundary conditions imposed on unphysical components $\Psi_k= k\cdot \Psi$, $\Phi_z$ and $\Phi_k=k \cdot \Phi$. What happens is that $\sigma=+1$ gives us the Dirichlet-like propagator (i.e. $\Phi_k=\pl_z \Phi_z=\pl_z \Psi_k=0$ at the boundary) and $\sigma=-1$ yields the Neumann-like propagator (i.e. $\pl_z\Phi_k=\Phi_z=\Psi_k=0$ at the boundary). The SDYM $\langle\Psi\Phi\rangle$ propagator in axial gauge is obtained by adding the gauge variation from \eqref{etaL} to the Feynman propagator. As discussed in the main text, the $\sigma=+1$ SDYM propagator gives a perfect agreement between Feynman and axial gauges.

\footnotesize
\providecommand{\href}[2]{#2}\begingroup\raggedright\endgroup

%%%%%%%%%%%%%%%%%%%%%%%%%%%%%%%%%%%%%%%%%%%%%%%%%%%%%%%%%%%%%
\end{document}